\documentclass[twocolumn,PRD,showpacs,nofootinbib,amsmath,amssymb]{revtex4}

\pdfoutput=1

\usepackage{graphicx}
\usepackage{dcolumn}
\usepackage{bm}
\usepackage{hyperref}
\usepackage{color}
\usepackage{rotate}
\usepackage{fancyhdr} 
\usepackage{cancel}
\usepackage{multirow}
\usepackage{ulem}

\def\be{\begin{equation}}
\def\ee{\end{equation}}
\def\bea{\begin{eqnarray}}
\def\eea{\end{eqnarray}}

\newcommand{\beq}{\begin{equation}}
\newcommand{\eeq}{\end{equation}}

\newcommand{\dnueff}{\Delta\nu_{\rm eff}}

\definecolor{darkred}{RGB}{175,0,0}
\definecolor{darkblue}{RGB}{14,0,185}

\newcommand{\refeq}[1]{Eq.~(\ref{eq:#1})}          
          
\newcommand{\reffig}[1]{Fig.~\ref{fig:#1}}          
\newcommand{\refsec}[1]{Sec.~\ref{sec:#1}}

\newcommand{\reftab}[1]{Tab.~\ref{tab:#1}}

\newcommand{\Msun}{M_\odot}

\newcommand{\fpbh}{{f_{\mathrm{PBH}}}}

\usepackage{aas_macros,braket}
\begin{document}
\title{Signatures of primordial black holes as seeds of supermassive black holes}
\author{Jos\'e Luis Bernal$^{1,2}$, Alvise Raccanelli$^{\star\let\thefootnote\relax\footnote{$^\star$Marie Sk\l{}odowska-Curie fellow},1}$, Licia Verde$^{1,3}$, Joseph Silk$^{4,5,6}$}
\affiliation{
$^1$Institut de Ci\`encies del Cosmos (ICCUB), Universitat de Barcelona (IEEC-UB), Mart\'{i} Franqu\`es 1, E08028 Barcelona, Spain~\\
$^2$Dept. de  F\' isica Qu\` antica i Astrof\' isica, Universitat de Barcelona, Mart\' i i Franqu\` es 1, E08028 Barcelona, Spain\\ 
$^3$ICREA, Pg. Llu\' is Companys 23, 08010 Barcelona, Spain\\
$^4$Department of Physics and Astronomy, Johns Hopkins University, 3400 N.\ Charles St., Baltimore, MD 21218~\\
$^5$Institut d'Astrophysique de Paris, UMR 7095, CNRS, UPMC Univ. Paris VI, 98 bis Boulevard Arago, 75014 Paris, France~\\
$^6$BIPAC, Department of Physics, University of Oxford, Keble Road, Oxford OX1 3RH, UK
     }

\date{\today}

\begin{abstract}
It is broadly accepted that Supermassive Black Holes (SMBHs) are located in the centers of most massive galaxies, although there is still no convincing scenario for the origin of their massive seeds.
It has been suggested that primordial black holes (PBHs) of masses $\gtrsim 10^{2} M_\odot$ may provide such seeds, which would grow to become SMBHs.
We suggest an observational test to constrain this hypothesis:
gas accretion around PBHs during the cosmic dark ages powers the emission of high energy photons which would modify the spin temperature as measured by 21cm Intensity Mapping (IM) observations.
We model and compute their contribution to the standard sky-averaged signal and power spectrum of 21cm IM, accounting for its substructure and angular dependence for the first time. 
If PBHs exist, the sky-averaged 21cm IM signal in absorption would be higher, while we expect an increase in the power spectrum for $\ell~\gtrsim 10^2-10^3$. 
We also forecast PBH detectability and measurement errors in the abundance and Eddington ratios for different fiducial parameter configurations  for various future experiments, ranging from SKA to a futuristic radio array on the dark side of the Moon. While the SKA could provide a detection, only a more ambitious  experiment  would provide accurate measurements.  
\end{abstract}

\maketitle

\section{Introduction}
The idea that density fluctuations can provide the seeds for galaxy formation via gravitational instability and leave detectable traces in the  Cosmic Microwave Background (CMB)~\cite{Silk:1967a, Silk:1967b}  introduced the concept that the graininess in the Universe would be the seeds around which galaxies form~\cite{Carr_graininess}. 
Now we know that Supermassive Black Holes (SMBHs) inhabit the centers of most galaxies (see~\cite{Kormendy_smbh} for a review). 
 Observations of quasars at $z\sim 6-7$  indicate that, even at these early times, there were SMBHs with masses of several $10^9\Msun$~\cite{Bañados_SMBH,Fan_quasar,Wu_quasar, Mortlock_quasar}. 
The existence of a population of Intermediate Mass Black Holes (IMBH) of masses around $10^2-10^6\Msun$ at  $z\sim 20-15$ would suffice~\cite{Smith_firstsmbh} to seed them.
The possible detection of a $\sim 10^5\Msun$ black hole in the Milky Way close to its center~\cite{oka_imbh1,oka_imbh2} may   provide evidence for such a relic and support the argument that Intermediate Mass Black Holes are the seeds of SMBHs. Besides, IMBHs may  inhabit the center of dwarf galaxies (e.g.,~\cite{silk_dwarf} and references therein).
 
The optimal conditions in the relevant parameter space of the mass of the black hole and the gas density around it, that lead to  fast growth of the black hole, were studied in~\cite{Pacucci_growth,Pacucci_growth2}. This happens if the combined effects of the angular momentum and radiation pressure are ineffective in stopping the stream of gas flowing from large scales towards the black hole. They find that this condition is fulfilled for $M\gtrsim 10^4-10^5~\Msun$ (where $M$ stands for the mass of the seed) and large gas densities, for which the growth of massive seeds up to SMBHs is  feasible. Even so, there is a limit on the maximum mass that the black hole can reach in an isolated halo, which depends on the  total mass of
the host halo and on  the radiative efficiency of the accretion \cite{Pacucci_limit}.
For smaller masses, the accretion is very inefficient, but  fast enough growth  can be achieved  via mergers.
 
However, the origin and formation mechanism of the massive seeds are still uncertain (see~\cite{Volonteri_smbhorigin, Latif_seeds} for a review). 
There are two main scenarios proposed to explain their origin: supercritical growth from stellar mass black holes formed from Population III stars and directly formed massive seeds at lower redshift. There are more exotic scenarios, such as IMBHs  formed by a subdominant component of the dark matter being dissipative~\cite{dAmico_dissDM}. 

According to the  first hypothesis, the seeds of SMBHs are remnants of Population III stars, formed with masses of tens of solar masses at $z\gtrsim 20$, which grow due to gas accretion and mergers~\cite{Madau_pop3,Tanaka_seedsaccretion,Li_bhmergers}. However, in order to reach masses such as those observed at $z\sim 6-7$~\cite{Madau_superedd, Lupi_superedd, Inayoshi_superedd}, the accretion needs to be supercritical over extended periods of time. Moreover, SMBH seeds growth is probably depressed due to the shallow gravitational potentials existing at those redshifts and the radiation pressure of the black hole emission. 
Indeed the recently discovered IMBHs in dwarfs are anorexic: apparently undermassive compared to the $M_{BH}-\sigma$ scaling relation~\cite{Baldassare_imbh,Mezcua_imbh}
Besides, cosmic X-ray background observations impose constraints on the growth of SMBHs, constraining the abundance of quasars with 
supercritical accretion~\cite{Salvaterra_xray} as well as of the abundance of miniquasars at high redshift \cite{Dijkstra04}. Therefore, this scenario alone is very unlikely to account for the present abundance of SMBHs.

On the other hand, SMBH seeds might also be formed due to the collapse of gas clouds which do not fragment or form ordinary stars, but directly form a massive black hole ($M\sim 10^5-10^6\Msun$) at lower redshifts ($z\lesssim 15$)~\cite{Bromm_dcbh, Begelman_dcbh, Lodato_dcbh, Choi_dcbh, Volonteri_evolseeds, Johnson_dcbh, Agarwal_dcbh, Shang_dcbh}. This kind of seed is  called a Direct Collapse Black Hole (DCBH). DCBHs may be realized if a metal-poor cloud is irradiated by non-ionizing ultraviolet light from nearby star-forming galaxies, which photodissociate molecular hydrogen and therefore prevent star formation. Hence, the gas can only cool via Lyman-$\alpha$ emission, which leads to a quasi-isothermal contraction without  fragmentation until the gravitational collapse and the formation of an IMBH  (see e.g.~\cite{Regan_dcbh}). 
Conveniently, the DCBH radiation is very efficient in preventing the formation of H$_2$. Therefore, a DCBH may trigger the formation of other DCBHs in a slowly-collapsing gas cloud more efficiently than galaxies \cite{Yue_dcbh}. 

Moreover, DCBHs are a good candidate for  explaining the large-scale power spectrum of the Near Infrared Background and its cross correlation with the cosmic X-ray background~\cite{Yue_dcbhNIRB}.
As DCBHs have a characteristic observational imprint \cite{Pacucci_dcbh}, it can be possible to identify these seeds in deep multi-wavelength surveys~\cite{Pacucci_seeds}.
Two promising candidates, whose infrared spectra require an exceptionally high star formation rate, were found at high redshift, with a predicted mass higher than $10^5\Msun$~\cite{Pacucci_seeds}. These candidates are likely to be formed by  direct collapse.
 
Nonetheless, the exact conditions and the probability of  obtaining DCBHs are still uncertain;  recent theoretical studies suggest that this mechanism might explain the abundance of the most luminous quasars at $z\sim 6-7$, but not the general population of SMBHs~\cite{Dijkstra_dcbh, Latif_dcbh,Habouzit_dcbh}. 

In summary, neither of these two scenarios individually  provide an entirely convincing explanation for 
 the origin of the seeds of SMBHs. However, massive seeds could have been formed much earlier. This third possibility (see \cite{Dolgov_smbh,Kohri_smbh, Clesse_pbh, Kawasaki_seeds,Khlopov_formmech,Dolgov_formmech} and references therein), much less explored in the literature, considers  Primordial Black Holes (PBHs) as the seeds which will grow to become SMBHs.
 If PBHs are formed with large enough masses, there is no need  for supercritical accretion, as is the case for Population III stars.

The idea of the existence of PBHs~\cite{Zeldovich_pbh} 
has recently regained popularity after they were  suggested to be the progenitors (and to make up a sizeable fraction of the dark matter, see e.g.~\cite{Bird_pbh}) of the stellar mass black holes ($\sim 30\Msun$) detected by LIGO+VIRGO Collaboration~\cite{abbott:ligo}.

Since then, a number of possible tests of the model have been performed with available data. They cover all of  the theoretically allowed range,  from the smallest masses constrained by black holes evaporation~\cite{Carr:evaporation}, to e.g. microlensing of stars~\cite{griest:keplerconstraint, niikura:microlensingconstraint}, to larger masses, constrained by e.g. X-ray and radio emission~\cite{gaggero}, wide-binaries disruption~\cite{quinn:widebinaryconstraint}, and accretion effects~\cite{Ali-Haimoud_PBH,
 poulin:cmbconstraint, Bernal_pbh}.
More innovative tests that can be performed in the future have also  been suggested, including using quantum gravity effects~\cite{Raccanelli:QG}, the lensing of fast radio bursts~\cite{Munoz:2016FRB}, the cross-correlation of gravitational waves with galaxy maps~\cite{raccanelli:cross, raccanelli:radio, Nishikawa:2017}, eccentricity of the binary orbits~\cite{cholis}, the black hole mass function~\cite{Kovetz:2017BHMF}, the gravitational wave mass spectrum~\cite{ Kovetz:2017GW}, merger rates~\cite{alihaimoud:pbhmergerrate} and the stochastic gravitational wave background~\cite{Mandic:2016, Cholis:2016, Nakama:2016,marti:gwpbhmergers}. \\

The mass range  required for PBHs to be the seeds of SMBHs is $\gtrsim10^{2} M_{\odot}$. In this range, the PBH abundance, $\fpbh =~\Omega_{\rm PBH}/\Omega_{\rm CDM}$, is strongly constrained by e.g., CMB observations~\cite{Ali-Haimoud_PBH,Bernal_pbh}, Ultra-Faint Dwarf Galaxies~\cite{brandt:ufdgconstraint} and wide binaries~\cite{quinn:widebinaryconstraint}. 
However, most of these constraints have been derived in the context of a model in which PBHs comprise most of the dark matter, and they assume a delta function in their mass distribution (see~\cite{bellomo:pbhemfconstraints} for updated constraints allowing for a wide mass distribution);
if, on the other hand, PBHs of these masses are only required to be the seeds of SMBHs and not a substantial part of the dark matter, the high-mass tail of the PBH mass distribution can have a very small $\fpbh$, satisfying all observational constraints.

Different scenarios for the SMBH seeds have different observational signatures. In this paper, we focus on 
 their imprints on 21 cm Intensity Mapping (IM). The term `IM' is sometimes dropped in the literature related to the emission from neutral hydrogen at large redshifts, in contrast with the studies of the emission lines from galaxies.  However, we maintain it for the sake of  clarity.
21 cm IM observations represent a promising future tool for cosmology (for a recent review, see~\cite{Kovetz:2017}). In particular, observations of spin temperature maps in the dark ages provide a direct window into the matter density fluctuations free of complications such as galaxy bias and most astrophysical processes. It can be thought of as a series of CMB-like screens, and therefore, besides the auto-correlation signal, one can also consider the ISW effect~\cite{Raccanelli_isw21cm} and lensing of 21 cm IM maps~\cite{Kovetz_lensing21cm}, including the possibility of  performing  tomographic analyses. 
It has recently been shown that 21 cm IM observations will give very powerful constraints on e.g., primordial non-gaussianity~\cite{Munoz_21cmbispec}, inflationary models~\cite{Munoz:2016, Pourtsidou:2016, Sekiguchi:2017}, scattering between dark matter and baryons~\cite{Munoz:2015}, statistical isotropy~\cite{Shiraishi_iso21cm} and annihilating and decaying dark matter \cite{Evoli_dcdm21,Valdes_anni-dm21cm}. 

21 cm IM will also be very powerful for  setting observational constraints on PBHs. 
Using the power spectrum originating from Poisson fluctuations, the authors of~\cite{Gong_pbh21cm} forecast constraints on $\fpbh$ based on future observations with SKA in the mass range $M\gtrsim 10^{-2}\Msun$. The abundance of PBHs  of much lighter masses can be constrained by  looking for the effects  of Hawking evaporation  on the Inter Galactic Medium (IGM) via  21 cm IM  of  the dark ages~\cite{Mack_lightpbh}. 
Minihalos 
have been also studied as interesting 21 cm emitters between reionization and the dark ages~\cite{Iliev_minihalo_02}, although they are hard to differentiate from the standard diffuse signal emanating from the IGM~\cite{Furlanetto06}.

The 21 cm IM sky-averaged signal can be also used to discriminate between the two main scenarios for the origin of SMBH seeds. Seeds formed from remnants of Popularion III stars dominate X-ray heating of the IGM and cause a rise in the 21 cm brightness temperature at $z\gtrsim 20$. An absence of such a signature might be due to the seeds being formed later, which 
would favor the DCBH scenario~\cite{Tanaka_smbh21cm}. However, such a signature could originate not only via seeds formed from Population III star remnants but also by PBHs. Besides, at these redshifts, the 21 cm IM signal is affected by a large number of astrophysical uncertainties and its dependence on redshift changes considerably with different assumptions~\cite{Cohen_21cm, Cohen_21cmPSparams},  making it difficult to identify the signal coming from the SMBH seeds. 

Here we study the scenario in which PBHs are the seeds of SMBHs. In order to avoid the astrophysical uncertainties mentioned above, we concentrate on the dark ages ($z\gtrsim 30$). The detection of a signal corresponding to the predictions reported in this work would be an indication that massive miniquasars were already present in the dark ages and the most likely explanation would be that these black holes are primordial, hence the most straightforward candidate to be the seeds of SMBHs.
In the standard scenario, during the so-called dark ages, the cosmic time previous to the formation of the first stars, there is no astrophysical feedback which contaminates the 21 cm IM signal and haloes are still not formed, so observations are free from galaxy bias and non-linearities in the clustering. Hence, the main uncertainties are only coming from the PBH sector. However, other exotic energy injections, such as that sourced in dark matter annihilation, might also heat up the IGM \cite{Valdes_anni-dm21cm,Evoli_dcdm21}. Nonetheless, we expect such signature to be distinguishable from the one of PBHs. A more quantitative evaluation of this issue will be presented
elsewhere.

We assume that the dark ages end at $z \sim 30$ (as it is standard convention), although in some scenarios star formation may start at earlier times and heat the IGM, hence changing both the sky-averaged  and  power spectrum of  21 cm IM (see e.g. \cite{Cohen_21cm,Cohen_21cmPSparams}). In such cases, the uncertainties in the standard signal at $z\sim 30$ would be larger and the identification of deviations as signatures of the presence of PBHs, more difficult.

We model the signature of massive PBHs,  with abundances required to explain the current SMBH population in the 21 cm IM signal. We compute 2-point statistics of the fluctuations accounting  explicitly for the temperature profiles around the PBHs in a comprehensive way, for the first time. We improve upon the work of  \cite{Iliev_minihalo_02,Furlanetto06,Sekiguchi:2017,Gong_pbh21cm} as we consider the  scale-dependence  of the PBH contribution to the  spectrum (and not only a rescaling of the amplitude of the standard 21 cm IM signal or only the  Poisson component).

 After characterizing the PBH contribution to the standard signal, we forecast the detectability with future experiments, ranging from the Square Kilometre Array (SKA,~\cite{SKA_IM}) to a futuristic radio array on the dark side of the moon~\cite{Silk:telescope}, which we refer to as the ``Lunar Radio Array'' (LRA).
 
This paper is structured as follows. First, we review the standard 21 cm IM sky-averaged signal and power spectrum coming from the IGM in the dark ages, as well as the instrumental noise, in~\refsec{background}. The effects of PBHs in the IGM and the spin temperature are characterized in~\refsec{PBH_effects}. Afterwards, the contribution to the 21 cm IM signal is modelled in~\refsec{PBH_global} and~\refsec{PBH_Cls} for the sky-averaged signal and the power spectrum, respectively. Finally, forecasts for different future experiments are presented in~\refsec{detect}. Discussions and conclusions can be found in~\refsec{Conclusions}. Throughout this paper, we assume the best fit values of the Planck 2015 TTTEEE+lowP power spectra~\cite{Planckparameterspaper} for the cosmological parameters.

\section{Standard signal}\label{sec:background}
We begin by reviewing the modelling of the standard 21 cm IM signal, i.e., without including the PBH contribution.

\subsection{Sky-averaged signal}
The optical depth of the IGM in the hyperfine transition is~\cite{Field_59}
\begin{equation}
\tau =~\dfrac{3c^3\hbar A_{10}x_Hn_H}{16k_BT_s\nu_0^2}~\dfrac{1}{H(z)+(1+z)\partial_rv_r},
\label{eq:tau}
\end{equation}
 where $c$ is the speed of light, $\nu_0 = 1420.4$ MHz is the rest-frame frequency of the hyperfine transition, $A_{10}=~2.85\times 10^{-15}$ s$^{-1}$ is the Einstein spontaneous emission rate coefficient for this transition, $T_s$ is the spin temperature of the gas, $H(z)$ is the Hubble parameter, $n_H= 8.6\times 10^{-6}\Omega_b h^2(1+z)^3$ cm$^{-3}$ is the hydrogen comoving number density~\cite{Planckparameterspaper}, $x_H$ is the neutral fraction of hydrogen, $k_B$ is the Boltzmann constant and $\partial_rv_r$ is the comoving gradient of the peculiar velocity along the Line of Sight (LoS). We define $T_{21}^{\rm obs}$ as the observed differential brightness temperature between the 21 cm emission and the CMB:
\begin{equation}
\begin{split}
& T_{21}^{\rm obs} =~\dfrac{T_s(z)-T_{\rm CMB}(z)}{1+z}\left(1-e^{-\tau}\right)~\approx ~\\
&~\approx  (27{\rm mK})(1+~\delta_b )x_H\left(1-\frac{T_{CMB}}{T_s}\right)\left(\frac{\Omega_bh^2}{0.023}\right)~\times\\ \times &~\left(\frac{1+z}{10}\frac{0.15}{\Omega_mh^2}\right)^{0.5}\frac{1}{1+(1+z)\frac{\partial_r v_r}{H(z)}},
\end{split}
\label{eq:T21}
\end{equation}
where $\delta_b$ is the local baryon overdensity, $h=H_0/100$ is the reduced Hubble constant and $\Omega_m$ and $\Omega_b$ are the matter and baryon density parameters, respectively. 
Therefore, the sky-averaged 21 cm IM signal, $\bar{T}_{21}$, can be obtained from~\refeq{T21} by setting $\delta_b=0$ and $\partial_rv_r=0$. 
We will mostly refer to the observed brightness temperature rather than to the local one, $T_{21}^{\rm loc}=~T_{21}^{\rm obs}(1+z)$, throughout the paper, so we drop the superscript ``$\rm{obs}$'' for simplicity. 

Assuming that the background radiation includes only CMB photons, the spin temperature can be expressed  
 as~\cite{Field58}:
\begin{equation}
T_s =~\frac{T_\star + T_{CMB}(z) + y_{k}T_{k}(z)+y_\alpha T_\alpha}{1+y_{k}+y_{\alpha}}
\label{eq:Tspin}
\end{equation}
where $T_\star = 0.068 K$ is the temperature correspondent to the 21 cm transitions, $T_k$ is the mean kinetic temperature of the IGM and $y_{k}$ and $y_\alpha$ are the kinetic and Lyman-$\alpha$ coupling terms, respectively. We set $T_\alpha\approx T_k$, since it is a very good approximation when the medium is optically thick to Lyman-$\alpha$ photons~\cite{Field_59B}, as in the case of study. 
The kinetic coupling term is due to the increase in the kinetic temperature by X-ray photon collisions with the gas:
\begin{equation}
y_{k} =~\frac{T_\star}{A_{10}T_{k}}(C_H+C_e+C_p),
\end{equation}
where $C_i$ are the de-excitation rates due to neutral hydrogen, electrons and protons, respectively 
. We use the fitting formulas of 
\cite{Kuhlen06}:
\begin{equation}
C_H = 3.1\times 10^{-11}n_H(z)T_k^{0.357}\exp(-32/T_{k})\; {\rm s^{-1}},
\end{equation}
\begin{equation}
C_e = n_e\gamma_e = n_H(z)(1-x_H(z,r))\gamma_e\; {\rm s^{-1}},
\end{equation}
\begin{equation}
C_p = 3.2x_H(z,r)C_H,
\end{equation}
where the number densities are in cm$^{-3}$
 and $\log (\gamma_e/{\rm cm^3/s}) = -9.607+0.5\log T_{k}~\exp(-(\log T_{k})^{4.5}/1800)$ if $T_{k}\leq 10^4$ K, otherwise, $\gamma_e=\gamma_e (T_{k}=10^4)$.

The coupling with the  Lyman-$\alpha$ photons is described by the Wouthusyen-Field effect~\cite{Field58}. It depends on Lyman-$\alpha$ photons intensity, $\tilde{J}_0$, given by:
\begin{equation}
\tilde{J}_0 =~\frac{\phi_\alpha c}{4\pi H(z)\nu_\alpha}n_Hx_H\int_{E_0}^\infty~\sigma(E)\mathcal{N}(E)dE
\end{equation}
where $\nu_\alpha$ is the frequency of the Lyman-$\alpha$ transition, $\phi_\alpha$ is the fraction of the absorbed energy that goes into kinetic excitation of Lyman-$\alpha$, $\mathcal{N}$ is the number of photons per unit area per unit time and $\sigma$ is the absorption cross-section. We use the parametrization of~\cite{Shull85}, given by $\phi_\alpha = 0.48\left( 1-x_e^{0.27}\right)^{1.52}$. 
Finally, the coupling term can be expressed as:
\begin{equation}
y_\alpha =~\frac{16\pi^2T_\star e^2f_{12}\tilde{J}_0}{27A_{10}T_km_ec}
\label{eq:yalpha}
\end{equation} 
where $f_{12}=0.416$ is the oscillator strength of the Lyman-$\alpha$ oscillator. 

\subsection{Fluctuations}\label{sec:standard_fluctuations}
The optical depth and the spin temperature of a hydrogen cloud depend on its density and velocity divergence. Small anisotropies in these two quantities create fluctuations in $T_{21}$. 
The 21 cm IM fluctuations power spectrum in the dark ages was computed in~\cite{Loeb_21cm}, and in~\cite{Bharadwaj_jlpm2} including the local velocity term. At the precision level we need in this work, given the uncertainties and assumptions in the modeling of the PBH contribution (see~\refsec{PBH_effects},~\refsec{PBH_global} and~\refsec{PBH_Cls}), it suffices to limit our computations to linear order. We follow the  formalism developed in ~\cite{Ali-Haimoud_21cmpert}, which includes the effects due to supersonic relative velocities between baryons and dark matter~\cite{Tseliakovich_supersonic}. This effect has been shown to help the formation of DCBHs at large redshifts \cite{Hirano_streaming}, but it does not play a major role in the population of SMBHs at $z\sim 6$~\cite{Tanaka_streaming}.  We refer the interested reader to~\cite{Pillepich_21cm, Lewis_21cm} for a more detailed description of the 21 cm IM fluctuations, extending the formalism to higher order and including fluctuations in other quantities, such as the ionized fraction.

Let us define $\delta_v\equiv -(1+z)~\partial_r v_r/H(z)$. Then, at linear order, the  fluctuations in the 21 cm IM signal can be expressed as:
\begin{eqnarray}
\delta T_{21}({\bf x})
&=& ~\alpha(z)~\delta_b({\bf x}) +~\bar T_{21}(z)~\delta_v({\bf x}),
\label{eq:T21_3D}
\end{eqnarray}
where $\alpha(z)=\mathrm d T_{21}/\mathrm d~\delta_b$, including gas temperature fluctuations.  
The observed $\delta T_{21}$ in a direction $\hat n$ on the sky and at a certain frequency $\nu$ is given by
\begin{eqnarray}
\delta T_{21}(\hat{n},~\nu) =~\int_0^{\infty} dx W_\nu(x)~\delta T_{21}(x,\hat{n}) ~,~\label{eq:T21_2D}
\end{eqnarray}
where $W_\nu(x)$ is the window function selecting the information at a certain frequency band centered in $\nu$ and $x$ is the comoving distance along the LoS. This $W_\nu(x)$ is a narrow function peaked at $x(z)$ which depends on the experiment. Here we assume a Gaussian function of width $\Delta\nu$. In Fourier space, 
 assuming that the baryons have caught up the dark matter and $\delta_b\propto (1+z)^{-1}$,   $\delta_v({\bf k}, z)=\mu^2\delta_b({\bf k}, z)$ at linear order, with $\mu=(\hat k~\cdot~\hat n)$. We can, therefore, define the transfer function of $\delta T_{21}$ as:
\begin{equation}
\begin{aligned}
& {\cal T}_{\ell}(k,\nu) =~\\ 
& =~\int_0^{\infty} dx W_\nu(x)~\left [\bar{T}_{21}(z) J_\ell(kx) +~\alpha(z) j_\ell(kx)~\right],
\end{aligned}
\label{eq:T_transfer}
\end{equation}
where $j_\ell$ is the spherical Bessel function with index $\ell$, and we have defined $J_\ell (kx)~\equiv -~\partial^2 j_\ell(kx) / (\partial kx)^2$, which can be written in terms of $j_\ell$, and $j_{\ell\pm2}$\footnote{$J_\ell(y)=\frac{-\ell(\ell-1)}{4\ell^2-1}j_{\ell-2}(y)+\frac{2\ell^2+2\ell-1}{4\ell^2+4\ell-3}j_\ell(y)+\frac{-(\ell+2)(\ell+1)}{(2\ell+1)(2\ell+3)}j_{\ell+2}$}~\cite{Bharadwaj_jlpm2}. Given this, we can easily compute the 21 cm IM angular power spectrum at a certain frequency $\nu$ as:
\be
C_{\ell}(\nu) =~\frac{2}{\pi}
\int_0^\infty k^2 dk P_m(k) 
{\cal T}_{\ell}^2(k,\nu) \,
\label{eq:Cell} ,
\ee
where $P_m(k)$ is the (isotropic) matter power spectrum.
For computational efficiency, we will employ the flat-sky approximation~\cite{Limber:1953} (for a pedagogical treatment, see e.g.~\cite{Hu_cmblensing_flatsky, White_flatsky}) for $\ell\geq 10^3$.

\subsection{Instrumental Noise}
Although in the cosmic-variance limit 
 the only source of noise is the variance arising by having a limited number of measurements of the power spectrum $C_\ell$, when considering an interferometer looking at the dark ages at a given frequency $\nu$, there is an additional noise power spectrum~\cite{Jaffe_noise,Knox_inflation,Kesden_gw,Zaldarriaga_21cm}:
\be
\ell^2 C_\ell^N =~\dfrac{(2\pi)^3 T_{\rm sys}^2(\nu)}{\Delta\nu~\,t_o f_{\rm cover}^2}\left(~\dfrac{\ell}{\ell_{\rm cover}(\nu)}\right)^2,
\ee
where  $t_o$ is the total time of observation, $\ell_{\rm cover}(\nu)\equiv 2\pi D_{\rm base}/\lambda (\nu)$ is the maximum multipole observable, $D_{\rm base}$ being the largest baseline of the interferometer, $f_{\rm cover}$ is the fraction of such baseline covered with antennas, and the amplitude $T_{\rm sys}$ is the system temperature, which we assume to be the synchrotron temperature of the observed sky:
\be
T_{\rm sys}(\nu) = 295~\left(~\dfrac{\nu}{150~\,~\rm MHz}\right)^{-2.62}~\rm K
\ee
found from extrapolating to lower frequencies the results of Ref.~\cite{Mozdzen_21cmskytemp}.

Therefore, the final uncertainty in the measurement of the $C_\ell$ at the required multipole $\ell$ is:
\begin{equation}
\sigma_{C_\ell} =~\sqrt{\frac{2\left( C_\ell+C_\ell^N\right) ^2}{f_{\rm sky}\left( 2\ell +1\right)}},
\label{eq:sigma}
\end{equation}
where $f_{\rm sky}$ is the fraction of the sky observed by the experiment.

\section{Effects of PBHs on the 21cm IM signal}
\label{sec:PBH_effects}
The presence of PBHs affect the gas spin temperature: the PBH accretion triggers the emission of high-energy photons which heat and ionize the gas around the PBH. In this work, we present for the first time a computation of the 2-point statistics of the fluctuations accounting for the whole scale-dependence of the temperature profiles around the PBHs, focusing on linear perturbations in the dark ages.

An accreting PBH builds up a classical Bondi profile (i.e., $r^{-3/2}$) around it. However, overdensities during the dark ages are still small and haloes are not formed yet. Therefore, as a first approach, we consider that there is no density profile in the gas around the PBH nor velocity inhomogeneities ($\delta_b(r)=0$ and $\partial_rv_r=0$, respectively). Regarding the interaction between radiation and gas, we neglect radiative transfer effects (and limit ourselves to integrate over the frequency, as in \refeq{xh}). Although these two effects might be  relevant in some parameter configurations, 
 they are competing: the former tends to reduce the volume affected by the PBH radiation, while radiative transfer increases the mean free path of high energy photons, hence increasing the distance to which X-rays can propagate and so the region heated by the PBH. 
  While a more careful treatment will be needed, especially for comparison with observations and  to assess their effective relative
importance, here, for this initial exploration and signal-to-noise estimate,  we assume that they compensate.

We assume that all processes are in equilibrium, given that their timescales are much smaller than the Hubble timescale. The steady-state approximation is very precise for masses $M\lesssim 3\times 10^4\Msun$~\cite{Ricotti_accretion}, but it breaks down for larger masses. Therefore, we limit our exploration to $M\leq 10^4\Msun$. To explore a suitable mass range we consider three representative cases: $M = 10^4\Msun$, $10^3\Msun$ and $10^2\Msun$. Given the slow growth of the PBHs at $z\gtrsim 30$, we assume that the PBH mass at different redshifts is the same when we perform a tomography analysis. Finally, we consider for simplicity that all PBHs have the same mass. This is an unrealistic scenario, but constraints for monochromatic mass distributions  can be translated to any extended mass distribution using e.g.,  the methods proposed in~\cite{bellomo:pbhemfconstraints,carr:comparison2}.

We explain below the formalism we use to compute the temperature profiles around a PBH  and show intermediate plots and results. Exact numerical calculations accounting for the time dependence can be found in~\cite{Thomas_mqcode}.

\subsection{Emission and neutral hydrogen fraction ($x_H(r)$)}
IMBH emission is usually modelled by the combination of three components: a ``multicolour  disk black body spectrum'' at low energies, a power-law spectrum from a surrounding ``hot corona''  at high energies and a small contribution from the reflected light from the corona by the gas around it. 
The  contribution to the total emission due to the reflected radiation is small, but the light emitted by the disk produces a rather hard spectrum peaking at $\sim 1$~KeV, as shown  in e.g., \cite{Tanaka_sed,Yue_dcbhNIRB} and references therein. 

 As the emission at low energies does not heat the gas around the PBH efficiently and sources at $z>22$ contribute only little to the Near Infrared Background~\cite{Yue_dcbhNIRB}, we assume that gas accretion around the PBH powers only X-ray emission. Moreover we assume, as commonly done, that the emission is spherically symmetric. 
Therefore, a bubble with 21 cm IM signal different from the sky-averaged value  is formed around the PBH.
Finally, we can safely assume that PBHs of the masses we consider do not affect cosmic reionization~\cite{Pacucci_spectrum}.

Following~\cite{Zaroubi07}, we assume that PBH accretion powers a miniquasar with a spherically symmetric power-law X-ray flux (limited to an energy range between 0.2 and 100 KeV). 
The difference between the heating of the gas by hard sources and those with a power-law spectrum may be significant (see e.g., \cite{Fialkov_sed,Pacucci_sed}). However, we show in the Appendix~\ref{sec:AppendixA} that the differences in the final angular power spectrum between a power-law spectrum and other more realistic choices (such as a piece-wise power-law \cite{Sazonov_spectrum} or including the emission from the disk as in \cite{Yue_dcbhNIRB}) are not significant with respect to the uncertainties in key parameters of the PBH population, i.e. their abundance, mass or Eddington ratio of the emission, as discussed below. Of course, in a refined application that goes beyond an initial
feasibility analysis such as this paper, all these affects must be correctly modelled.
 Then, the spectrum of the photon emission, $F(E)$, is given by:
\begin{equation}
F(E) =~\mathcal{A}(M\lambda)\;E^{-1} {\rm s}^{-1},
\label{eq:spectrum}
\end{equation}
where $\mathcal{A}$ is a normalization factor chosen to have a luminosity  $L =~\lambda L_{\rm Edd}$, where $\lambda$ is the Eddington ratio and $L_{\rm Edd}$ is the Eddington luminosity:
\begin{equation}
\mathcal{A}(M\lambda) =~\frac{\lambda L_{Edd}(M)}{\int_{E_{\rm range}}{E^{-1}}dE}\;{\rm keV/s},
\label{eq:A_Ml}
\end{equation}
\begin{equation}
L_{Edd}(M) = 8.614\times 10^{46} (M/M_\odot)\;{\rm keV/s}.
\label{eq:Ledd}
\end{equation}
Combining~\refeq{A_Ml} and~\refeq{Ledd}, it is easy to notice that $\lambda$ and $M$ are degenerate when computing the emission of the PBH, since $\mathcal{A}\propto ~\lambda M \equiv\mathcal{M}$.  
As explained below, relevant quantities, as $x_H$ or $T_{21}$, only depend on the redshift and the intensity of the emission. Therefore, in order to illustrate how these quantities depend on both $\lambda$ and $M$, we will show them in terms of $\mathcal{M}$ .

The spectrum of~\refeq{spectrum} translates into number of photons per unit area per unit time at a comoving distance $r$ from the source:
\begin{equation}
\mathcal{N}(E,r) = e^{-\tau(E,r)}\frac{\mathcal{A}(\mathcal{M})E^{-1}}{4\pi r^2}\; {\rm cm^{-2}s^{-1}} \,,
\label{eq:N_Er}
\end{equation}
where
\begin{equation}
\tau (E,r) =~\int_0^r n_H(z)x_H(r)\sigma(E) dr.
\end{equation}

We use the fitting formula of~\cite{Zdziarski_sigma} to compute the absorption cross section taking into account the contribution from helium and hydrogen atoms:
\begin{equation}
\sigma(E) = 4.25\times 10^{-21}\left(\frac{E}{0.25 {\rm keV}}\right)^{-p}\; {\rm cm^{2}},
\end{equation}
with $p = 2.65$ if $E<0.25$ keV and $p=3.30$ if $E>0.25$ keV. 
The emitted photons ionize the surrounding gas at a rate per hydrogen atom, $\Gamma$, as a function of  the  comoving distance $r$,  given by:
\begin{equation}
\Gamma(r) =~\int_{E_0}^\infty~\sigma(E)\mathcal{N}(E,r)(1+\frac{E}{E_0}\phi(E,x_e))\frac{dE}{E}\,,
\end{equation}
where $x_e(r)$=($1-x_H(r)$) is the ionized fraction,  and the term $\frac{E}{E_0}\phi(E,x_e)$ is introduced to take into account secondary ionizations. We apply the fitting formulas from~\cite{Dijkstra04} and~\cite{Shull85} for $E\leq 0.5$ KeV and $E>0.5$ KeV, respectively.

Therefore, the neutral fraction is determined by the equilibrium between ionization and recombination rates:
\begin{equation}
\alpha_H n_H^2(z)(1-x_H(r))^2 =~\Gamma(r)x_H(r)n_H(z),
\label{eq:xh}
\end{equation}
where $\alpha_H = 2.6\times 10^{-13} T_4^{-0.85}$ cm$^3$/s is the recombination cross-section to the second excited atomic level, with $T_4=T_k/10^4$ K. For this computation, we assume $T_4=1$ (as in~\cite{Zaroubi07}).

\begin{figure}
\centering
\includegraphics[width=\columnwidth]{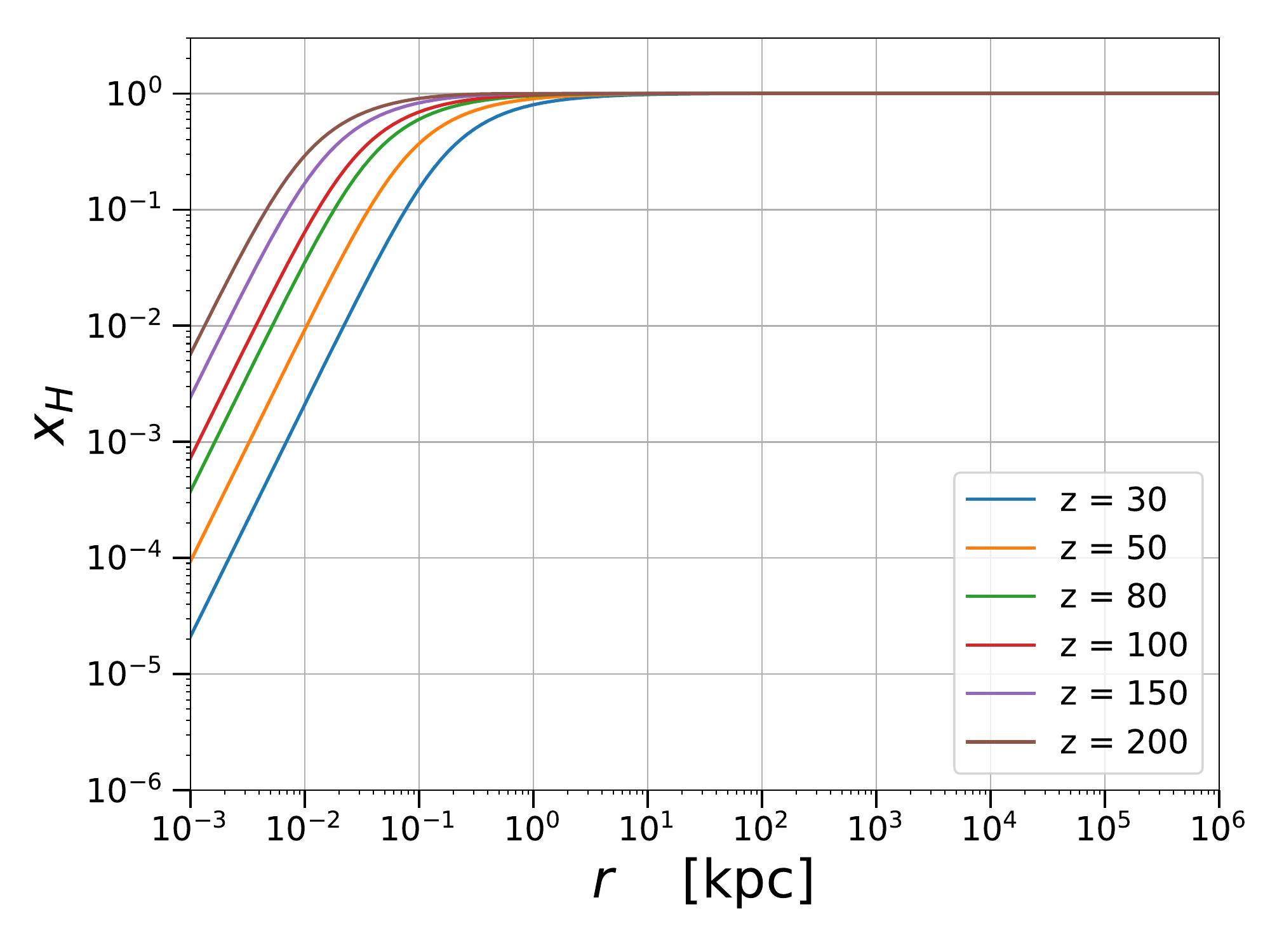}
\includegraphics[width=\columnwidth]{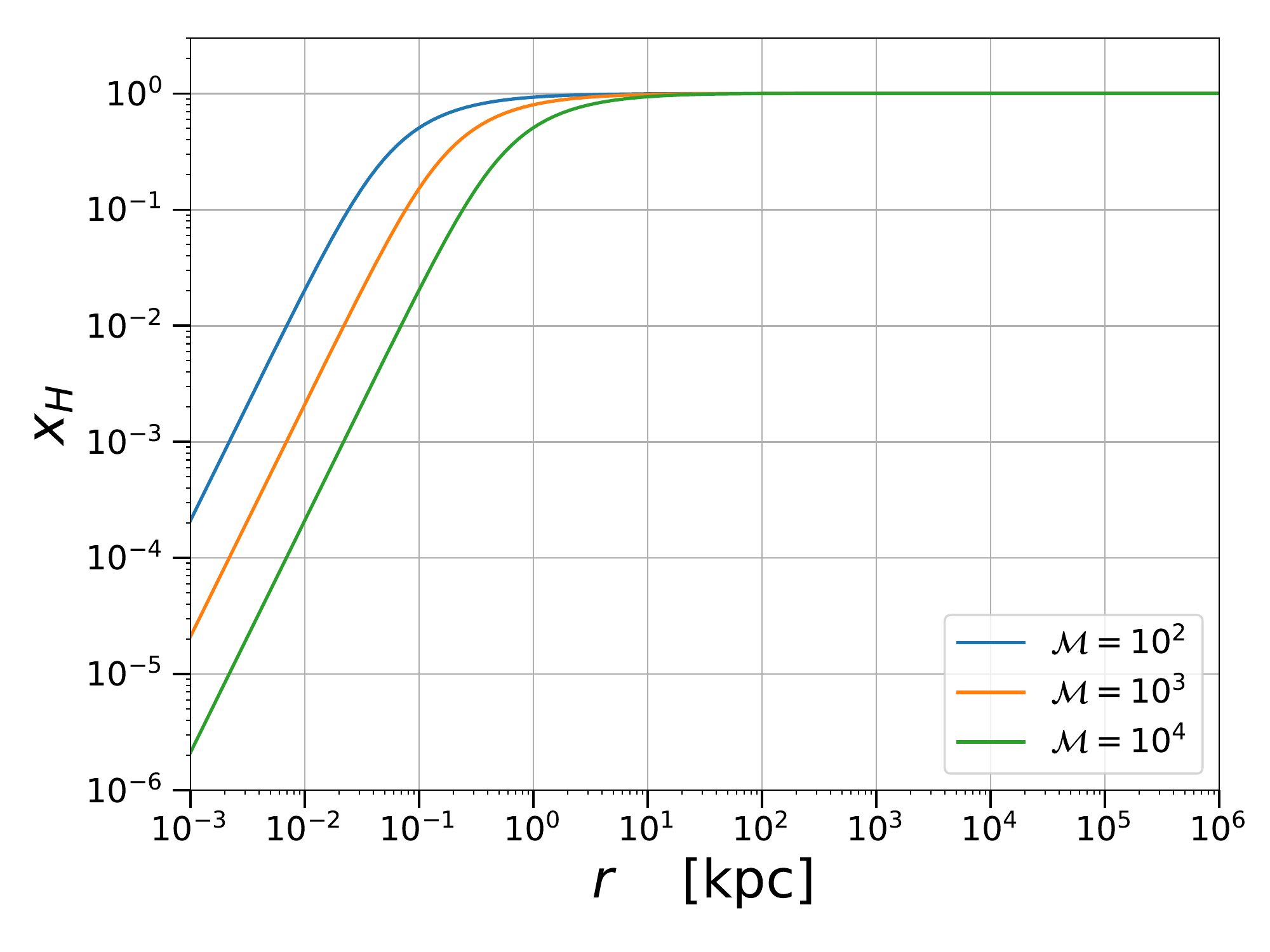}
\caption{\label{fig:xh}
Neutral hydrogen fraction profile $x_H$ for a PBH with $\mathcal{M}=100$ at various redshifts (top) and for a PBH with various $\mathcal{M}$ values at $z=30$ (bottom).}
\end{figure}

The neutral fraction radial profile, $x_H(r)$, is shown in~\reffig{xh} for different redshifts and values of $\mathcal{M}$. With increasing redshift, the hydrogen density increases; in a given volume at fixed photon flux, there are more atoms to ionize, hence the  size of the ionized region decreases.

On the other hand, for  increasing masses or Eddington ratios (i.e., larger $\mathcal{M}$), as the PBH emission is more intense,  the ionized region becomes larger.

\subsection{Kinetic temperature}
In addition to being ionized, the gas around the PBH is heated by the photons emitted by the miniquasar and cooled by the interaction with the CMB and the expansion of the Universe. The miniquasar heating affects the kinetic temperature,
hence $T_k$ varies with the distance to the PBH. The heating rate per unit volume per unit time at a given comoving distance $r$ from the source is: 
\begin{equation}
\mathcal{H}_{\rm PBH} = f(x_e(r))n_H(z)x_H(r)\int_{E_0}^\infty~\sigma(E)\mathcal{N}(E,r)dE,
\end{equation}
where $f(x_e(r))$ is the fraction of the photon energy absorbed through collisional excitations. We use an extrapolation of the fitting formula of~\cite{Shull85}: $f=0.9771(1-(1-x_e^{0.2663})^{1.3163})$. As this fitting formula does not work well for a low-ionization medium (in reality $f$ never goes to $0$), we consider a floor  $f=0.15$ when $x_e\leq 10^{-4}$ \cite{Thomas_mqcode}. 

Since the gas is exposed to Compton cooling by CMB photons,
the heating  rate per unit volume per unit time due to Compton processes is:
\begin{equation}
\begin{split}
&~\mathcal{H}_{\rm Compton} =~\frac{32\pi^5~\sigma_Tc k_B^5n_e(z,r)T_{CMB}^4(z)}{15(hc)^3m_ec^2}\times~\\& ~\times(T_{CMB}(z)-T_k(r)),
\end{split}
\end{equation}
where $n_e(z,r)=n_H(z)x_e(r)$ is the number density of electrons. On the other hand, 
the adiabatic expansion cooling per unit volume per unit time  is $\mathcal{H}_{\rm exp}~=~-~ 3H(z)k_BT_k(r)n_H(z)(2-x_H(r))$. Then, in equilibrium, $\sum~\mathcal{H}_i = 0$.

Here, we do not consider Compton heating due to the emitted photons, because it is efficient only very close to the source~\cite{Thomas_mqcode}. Nonetheless, at those distances the hydrogen is totally ionized,  so there is no signal in 21 cm IM and the results do not change. Moreover, those scales are far beyond the reach of 21 cm IM power spectrum resolution. 

At large distances from the source, the gas is not affected by the PBH emission and its temperature is only determined by the adiabatic cooling due to the expansion of the Universe (there are no free electrons to scatter via Compton). Therefore, we need to set a contour condition by which 
$T_k(r\rightarrow~\infty) = T_k^0$, the mean kinetic temperature of the IGM (without PBHs, which we take from the output of HyRec~\cite{alihamoud:hyrec1,alihamoud:hyrec2}). We include this condition in our computation of $T_k$ by a adding $T_k^0x_H(r)$ to the obtained $T_k$. We will remove this contribution when computing $T_{21}$ of an isolated PBH.

\begin{figure}
\centering
\includegraphics[width=\columnwidth]{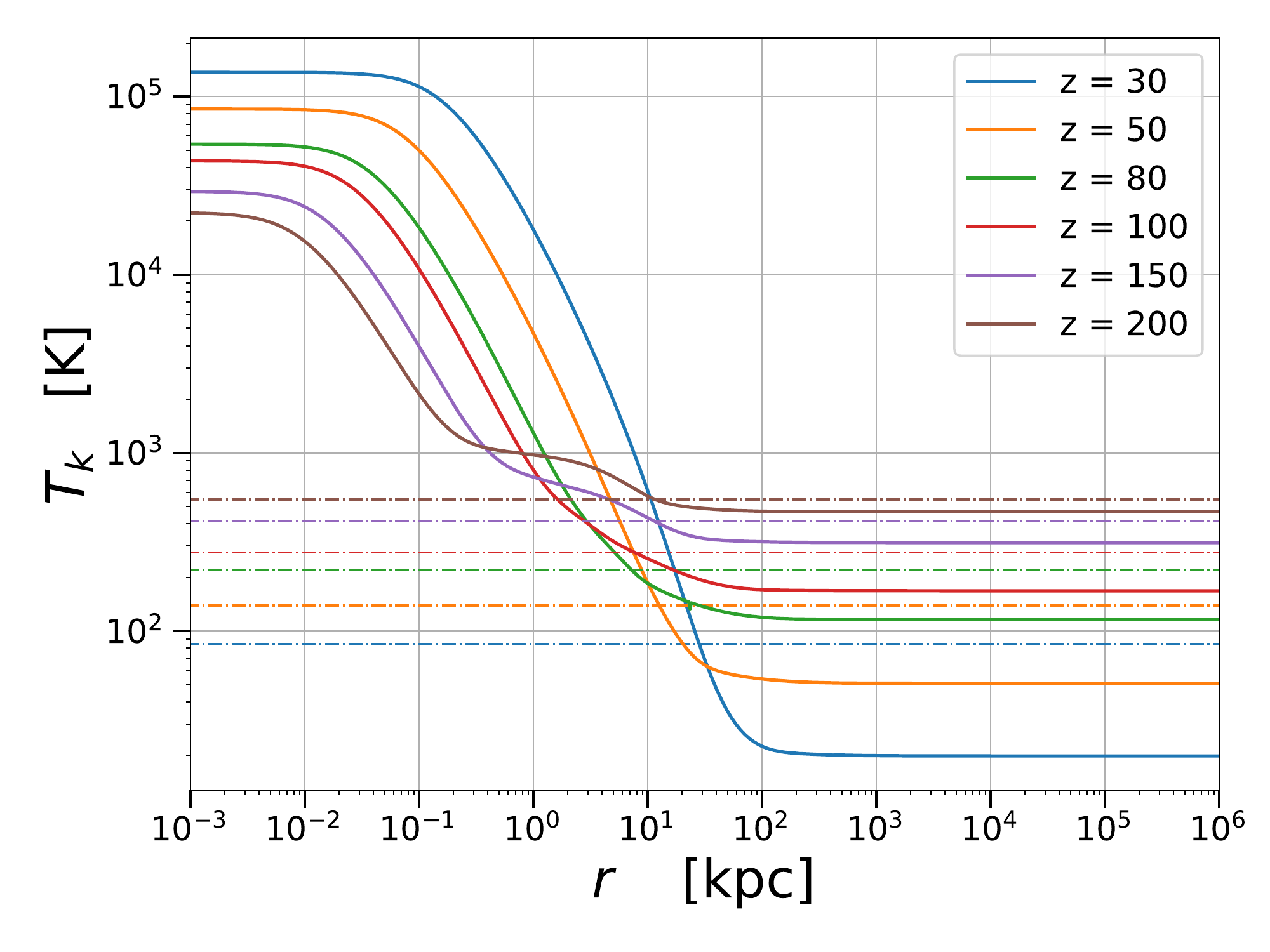}
\includegraphics[width=\columnwidth]{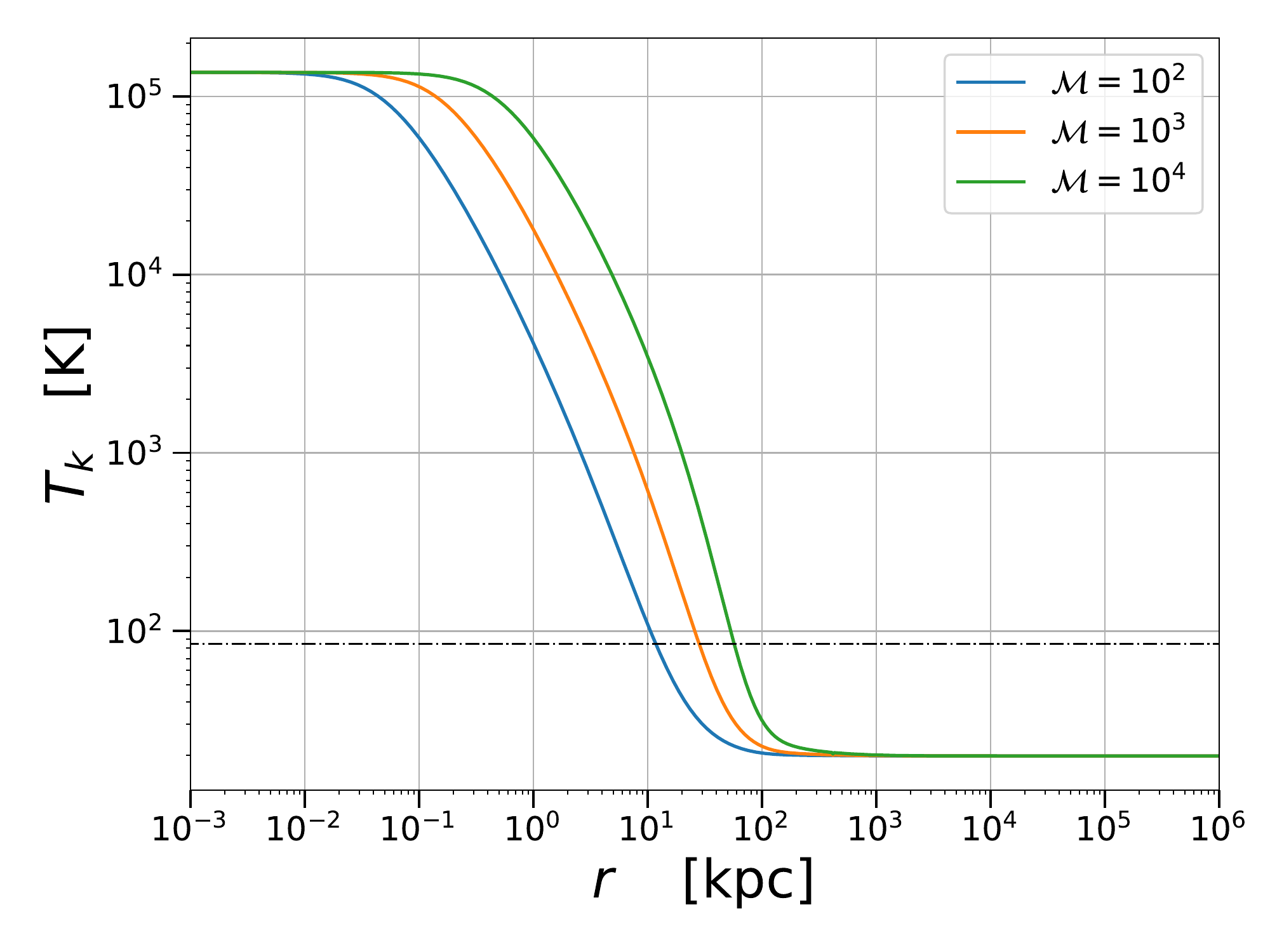}
\caption{\label{fig:Tk}
Kinetic temperature profile $T_k$ for a PBH with $\mathcal{M}=100$ at various redshifts (top) and for a PBH with various values of $\mathcal{M}$ at $z=30$ (bottom). The CMB temperature is shown in dot-dashed line for reference for each redshift in the upper panel and for $z=30$ in the lower. }
\end{figure}

We show gas temperature profiles as function of the comoving distance to the PBH in~\reffig{Tk} for different redshifts and values of $\mathcal{M}$. At large distances from the PBH, $T_k = T_k^0$, hence the gas temperature is lower at lower redshifts. In the inner regions, the heating due to the emission of the PBH is coupled only to the neutral hydrogen, but, as the number of photons decays exponentially with the distance, this heating is more efficient close to the PBH. In these regions, PBH heating dominates over Compton  and  adiabatic cooling, so $T_k$ needs to be high to reach equilibrium. If Compton heating due to the emitted photons were considered, $T_k$ at distances tending to 0 would be much higher. However, as stated before, this would not change the signal in 21 cm IM because the hydrogen is totally ionized in those regions. At intermediate distances, PBH heating loses efficiency and $T_k$ drops even below $T_{\rm CMB}$ until it reaches $T_k^0$. 

\subsection{Spin temperature and differential brightness temperature}
\begin{figure}
\centering
\includegraphics[width=\columnwidth]{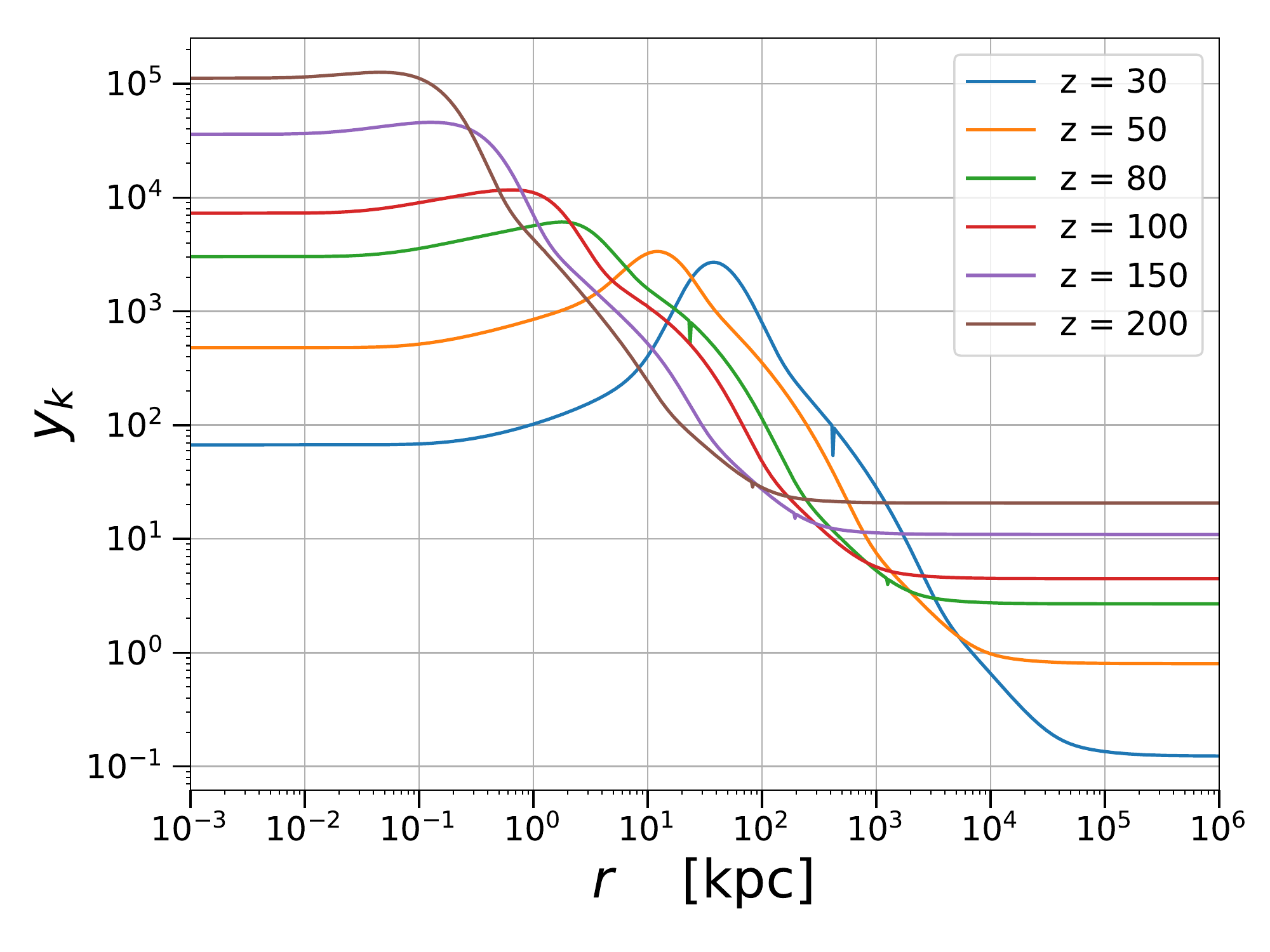}
\includegraphics[width=\columnwidth]{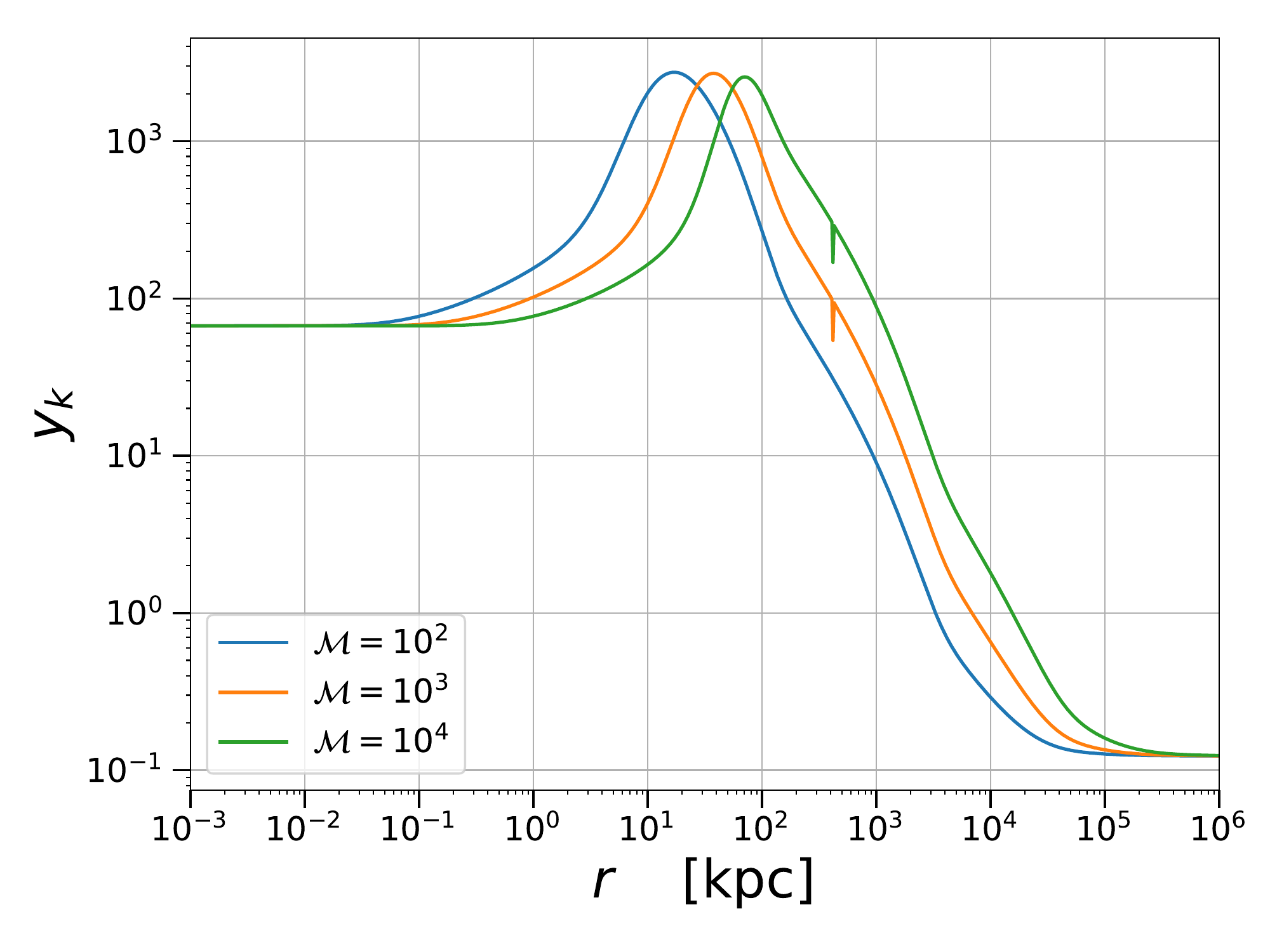}
\caption{\label{fig:yk} Radial profile of the kinetic coupling term $y_k$ of the spin temperature for a PBH with $\mathcal{M}=100$ at various redshifts (top) and for a PBH with various values of $\mathcal{M}$ at $z=30$ (bottom). }
\end{figure}

\begin{figure}
\centering
\includegraphics[width=\columnwidth]{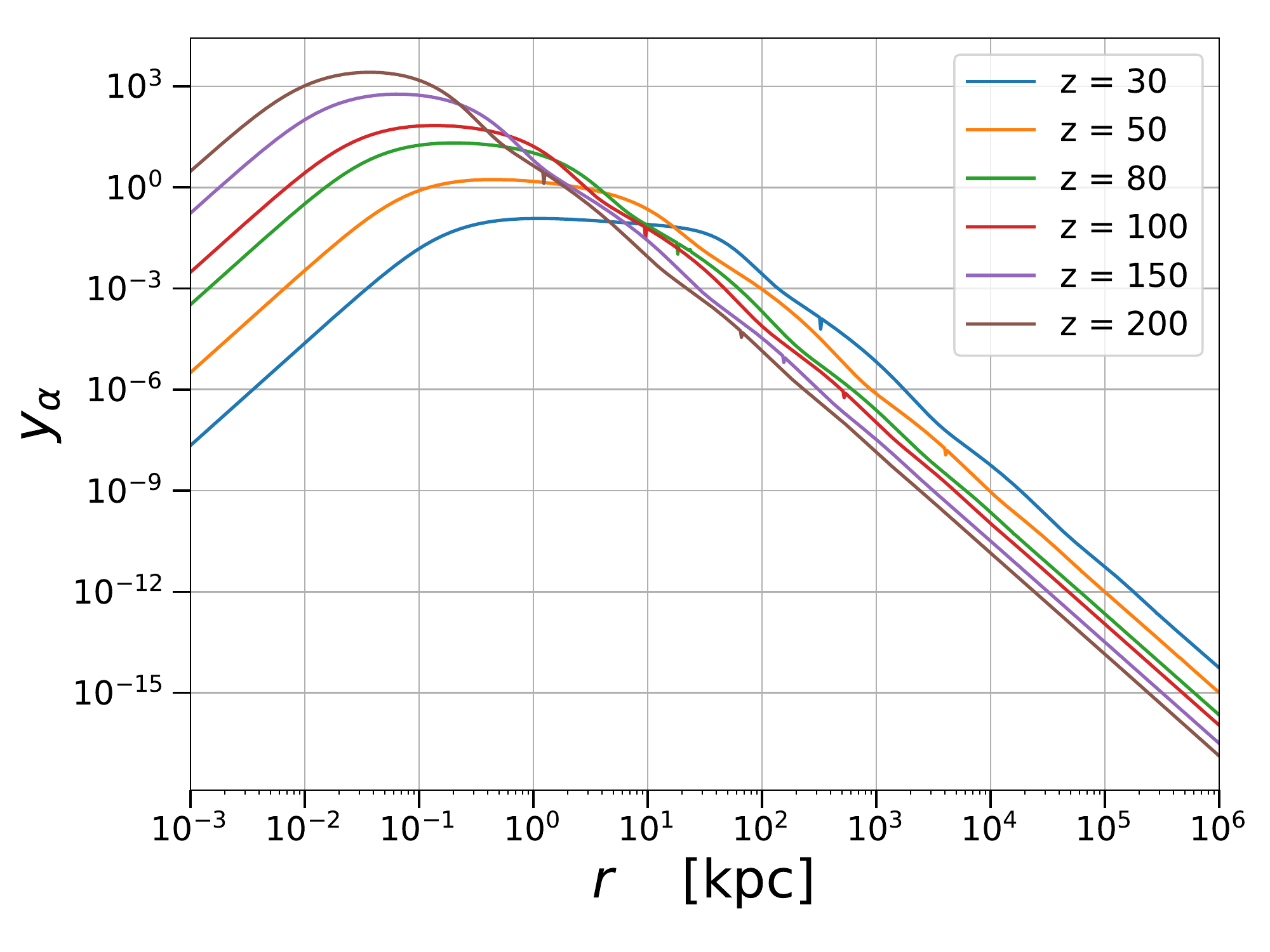}
\includegraphics[width=\columnwidth]{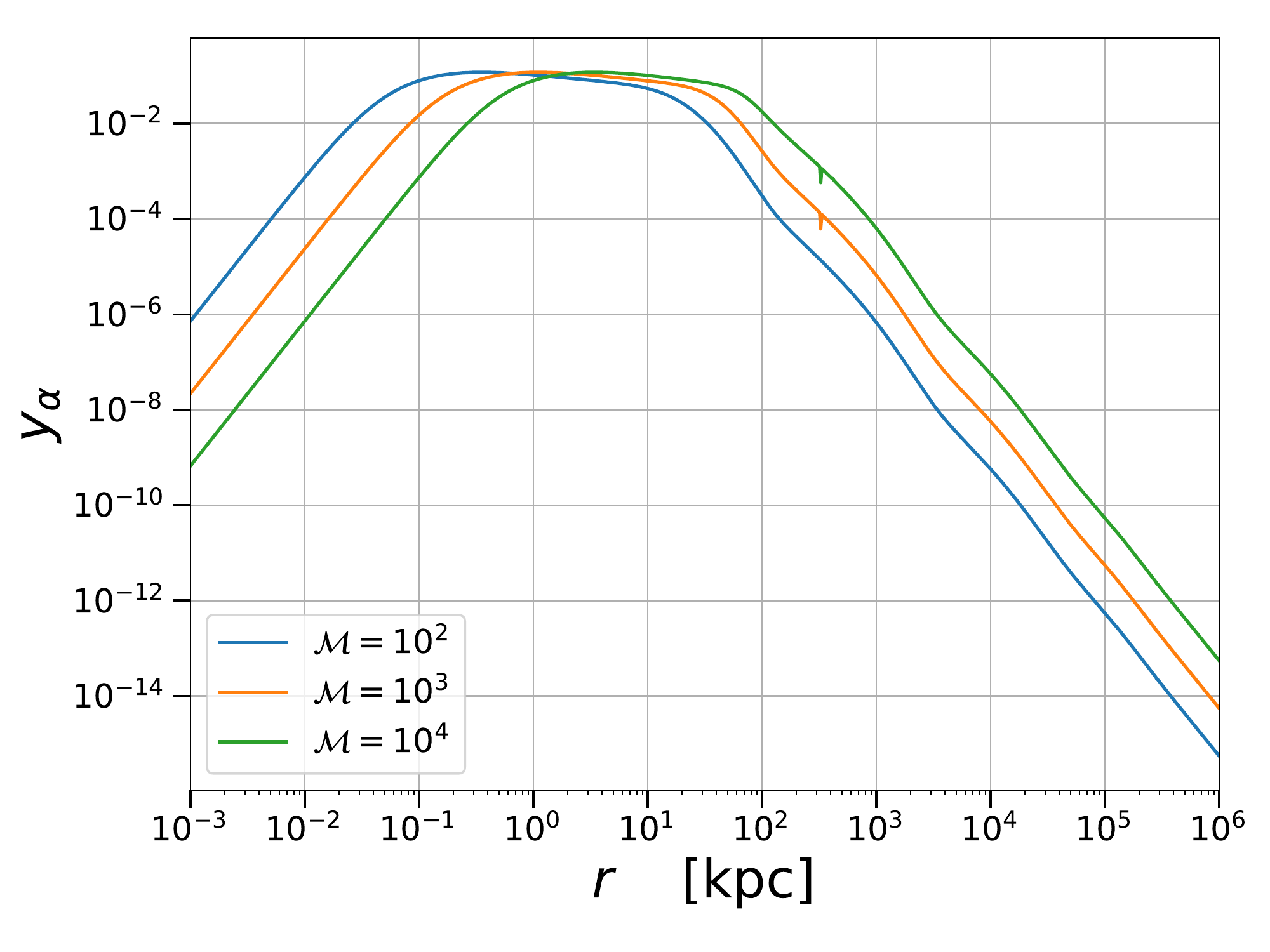}
\caption{\label{fig:yalpha} Radial profile of the coupling term of the spin temperature with photons due to the Wouthusyen-Field $y_\alpha$ of the spin temperature for a PBH with $\mathcal{M}=100$ at various redshifts (top) and for a PBH with various values of $\mathcal{M}$ at $z=30$ (bottom). }
\end{figure}

Once we have computed the ionization fraction and gas temperature profiles ($x_H(r)$ and $T_k(r)$), obtaining the spin and differential brightness temperature is straightforward using ~\refeq{Tspin} and~\refeq{T21}, respectively. $T_s$ may be driven whether by the collisional coupling or via the Wouthusyen-Field effect, whose weight is encoded in the coupling terms $y_k$ and $y_\alpha$ in~\refeq{Tspin}, respectively. We show radial profiles of $y_k$ and $y_\alpha$ in~\reffig{yk} and~\reffig{yalpha}, respectively, which make evident that $T_s$ is driven by collisional coupling in all the cases of study.

\begin{figure}
\centering
\includegraphics[width=\columnwidth]{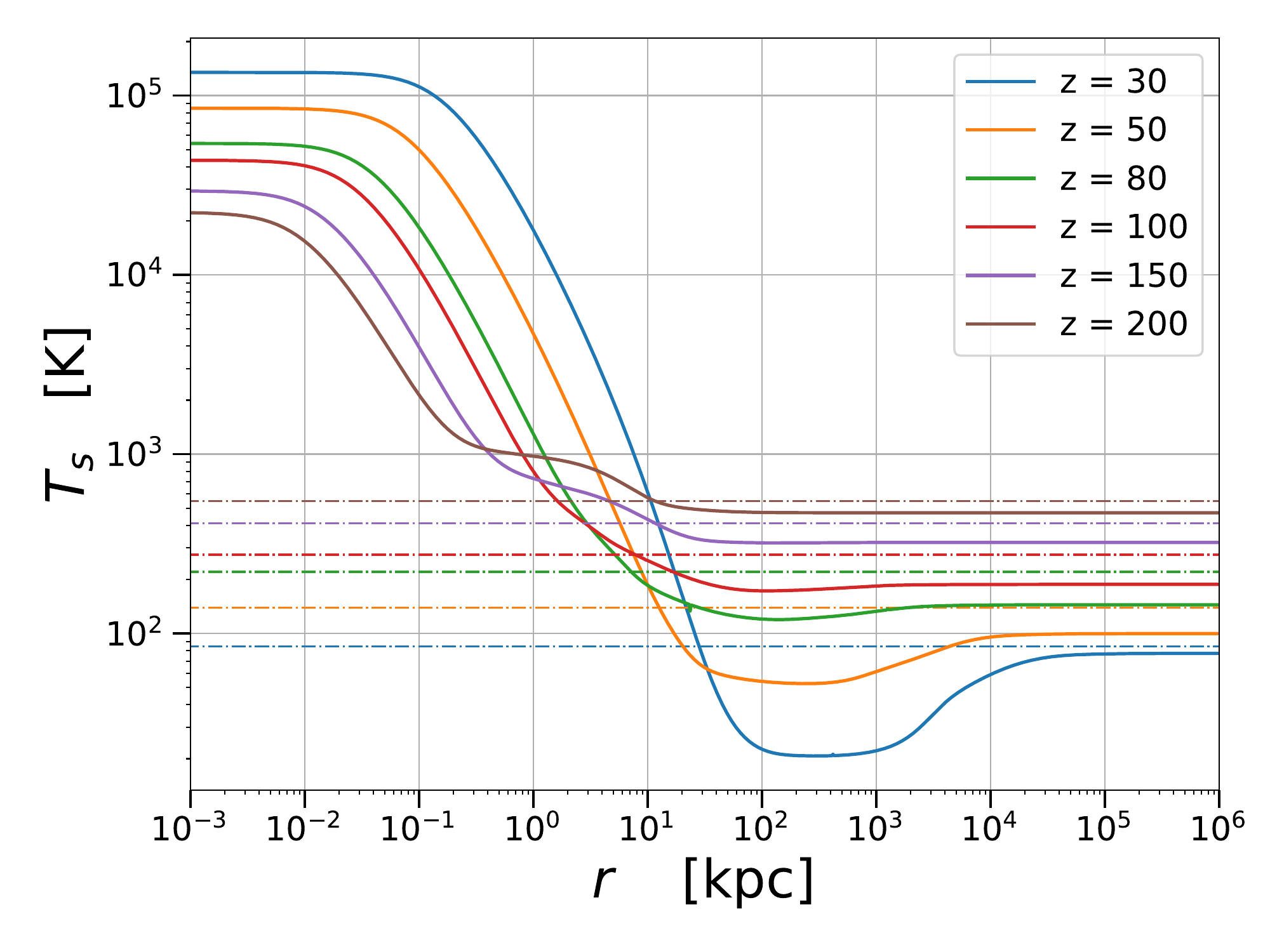}
\includegraphics[width=\columnwidth]{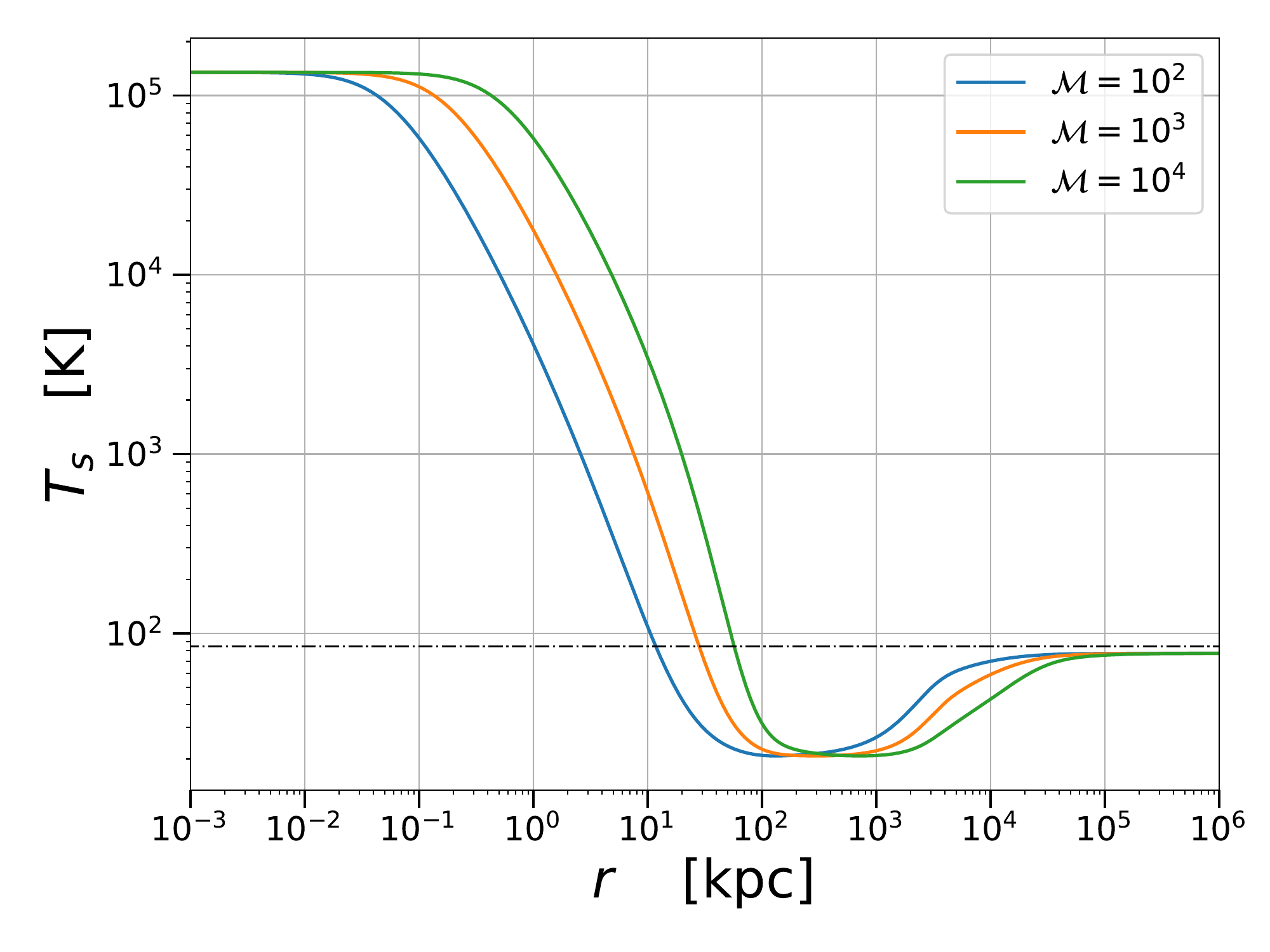}
\caption{\label{fig:Ts} Spin temperature profile $T_s$ for a PBH with $\mathcal{M}=100$ at various redshifts (top) and for a PBH with various values of $\mathcal{M}$ at $z=30$ (bottom). The CMB temperature is shown as a  dot-dashed line for reference for each redshift in the upper panel and for $z=30$ in the lower. }
\end{figure}

\begin{figure}
\centering
\includegraphics[width=\columnwidth]{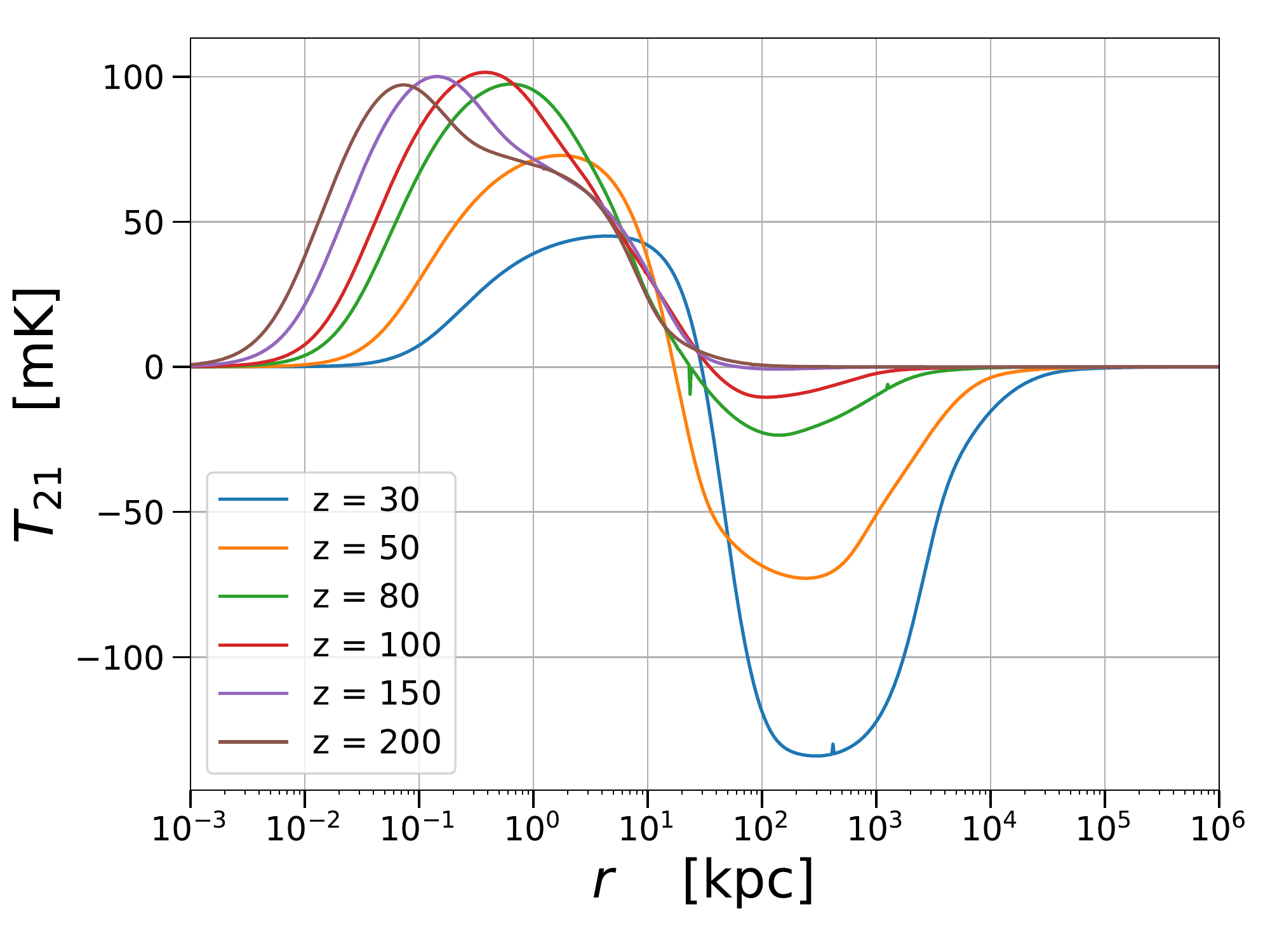}
\includegraphics[width=\columnwidth]{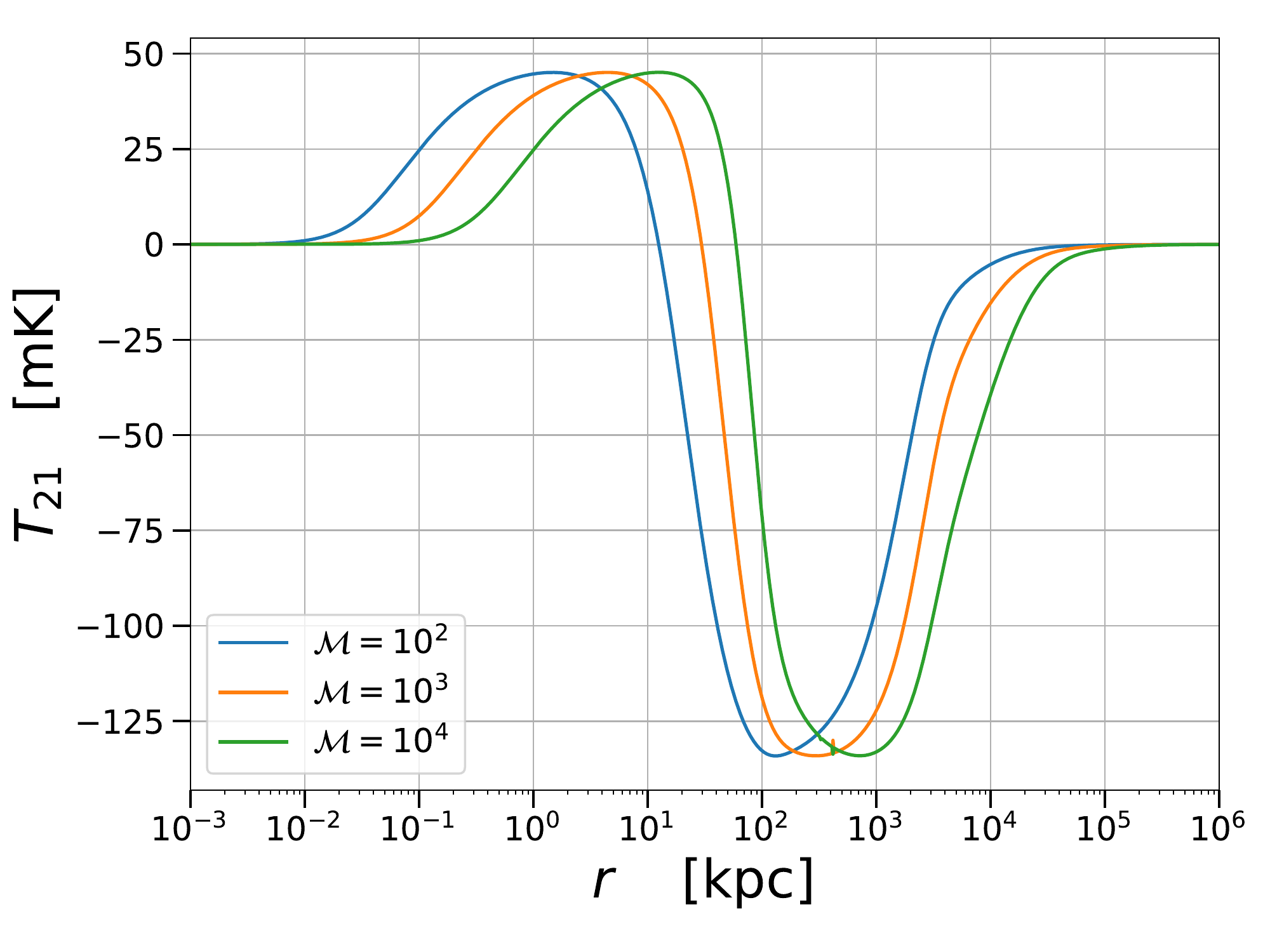}
\caption{\label{fig:dtb}
Differential brightness temperature profile $T_{21}$ for a PBH with $\mathcal{M}=100$ at various redshifts (top) and for a PBH with various values of $\mathcal{M}$ at $z=30$ (bottom). }
\end{figure}

Spin temperature profiles can be seen in~\reffig{Ts}. $T_s$ behaves qualitatively similar to $T_k$ until $T_k\approx T_s<T_{\rm CMB}$, where spin temperature coupling with CMB photons dominates and $T_s$ rises until $T_s\approx T_{\rm CMB}$, as can be seen in \reffig{yk}.

So far, we have applied the boundary condition that all quantities must match the standard values (i.e., without PBHs) when the distance to the PBH is large enough (e.g., $T_k(r\rightarrow~\infty)=T_k^0$). Nonetheless, we are interested on the isolated signal in 21 cm IM of a single PBH. Therefore, we subtract the contribution added due to the boundary condition in the same way that it was added before:
\begin{equation}
T_{21}(r)~\rightarrow T_{21}(r)-T_{21}^0x_H(r),
\label{eq:T21_substract}
\end{equation}
where $T_{21}^0$ is the sky-averaged $T_{21}$ without PBHs.

The $T_{21}(r)$ profile shown in~\reffig{dtb} can be explained as follows. In the inner part, $T_{21} = 0$ because all of the gas is ionized. The region with $T_{21}>0$ corresponds to the region where $T_k>T_{\rm CMB}$ and $x_H$ starts to grow; then, when $T_k$ drops because the PBH heating at those distances is less efficient,  $T_{21}$ drops to negative values. Finally, $T_{21}$ rises again due to the collisional and Lyman-$\alpha$ coupling of the photons to the source with the gas becomes totally inefficient and $T_s\rightarrow T_s^0$ so $T_{21}\rightarrow T_{21}^0$. Given that the PBH signal is isolated, at these distances, $T_{21}=0$. 

\begin{figure}
\centering
\includegraphics[width=\columnwidth]{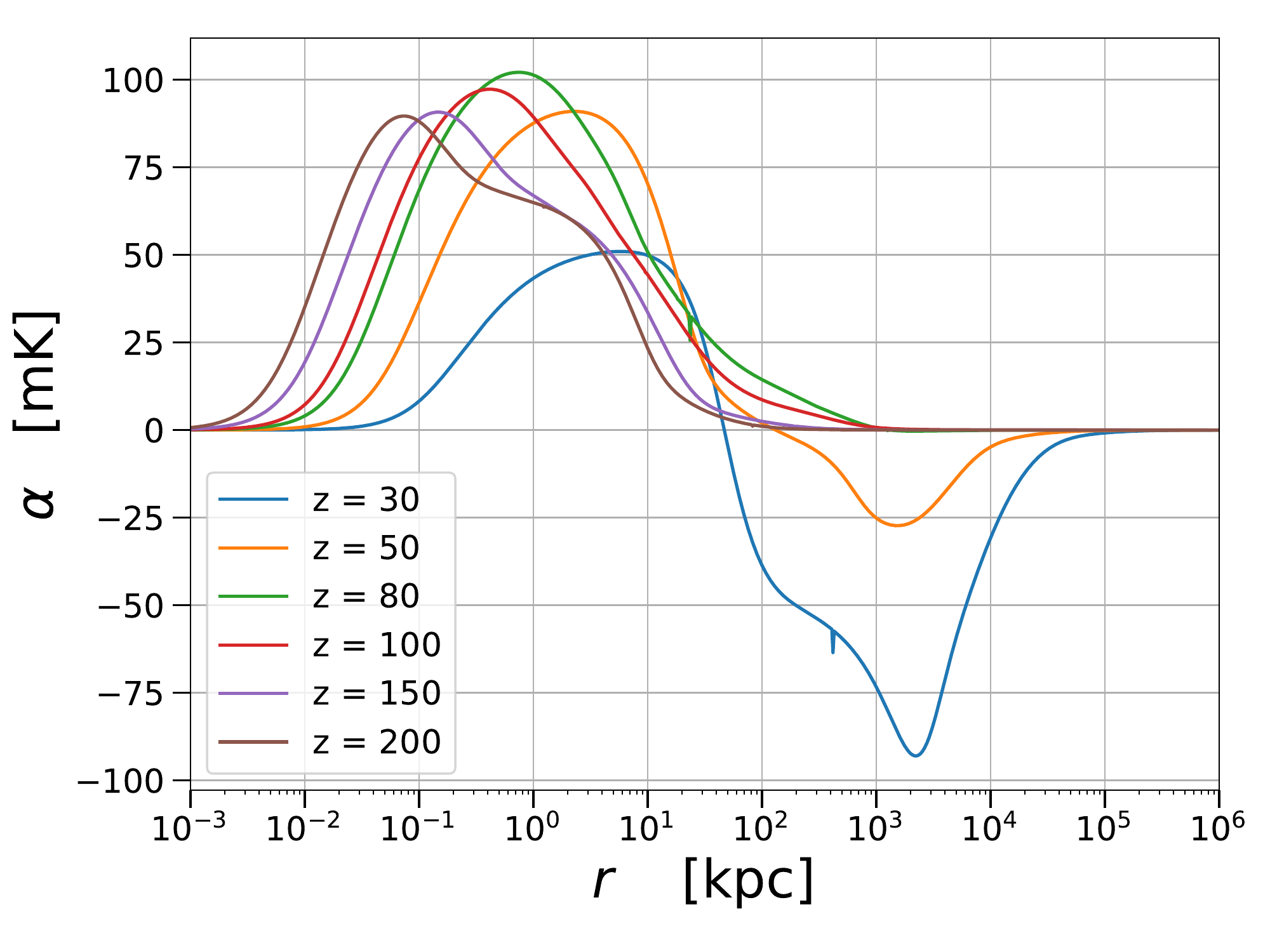}
\includegraphics[width=\columnwidth]{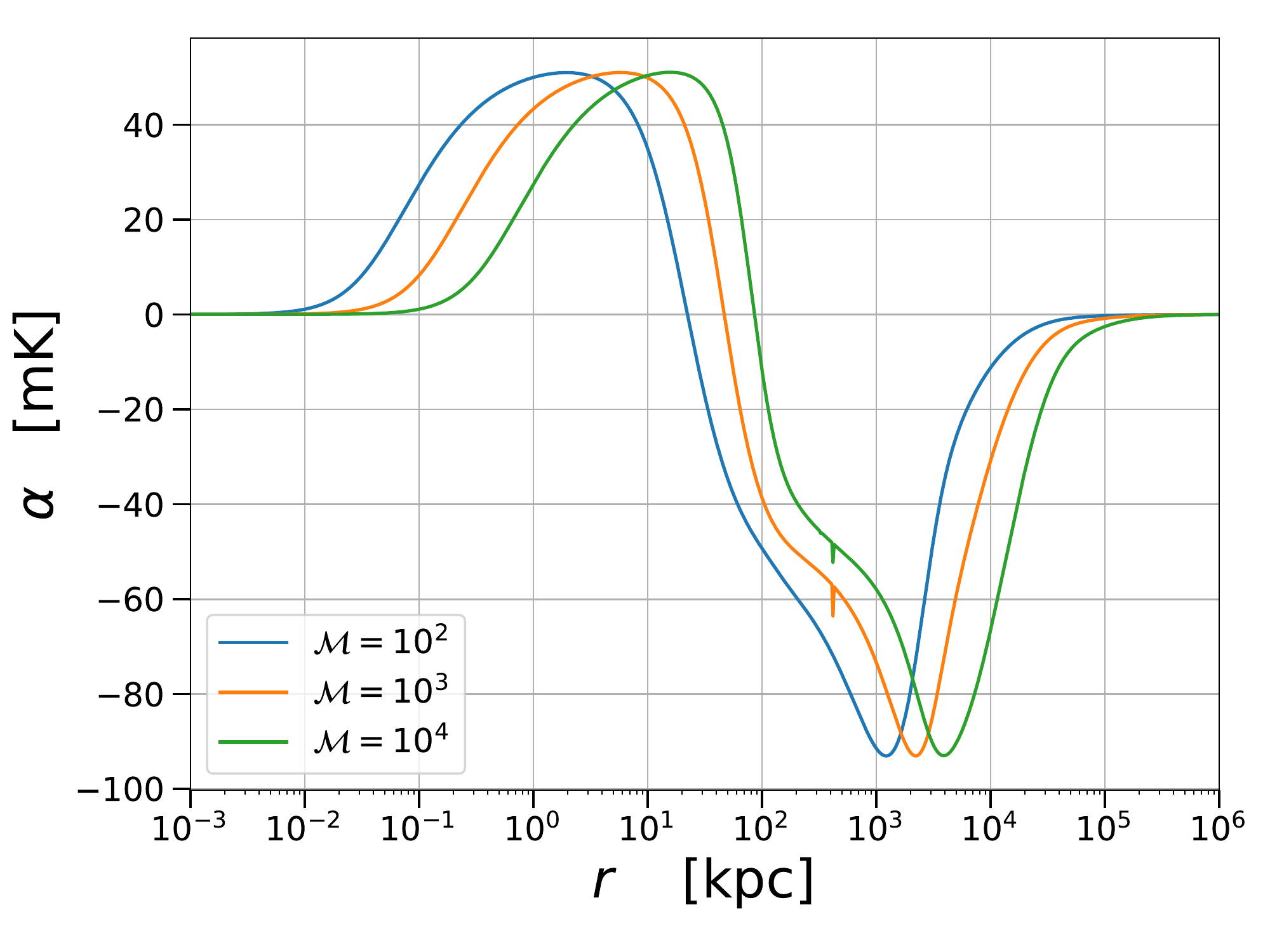}
\caption{\label{fig:alpha_r}
$\alpha$ profile for a PBH with $\mathcal{M}=100$ at various redshifts (top) and for a PBH with various values of $\mathcal{M}$ at $z=30$ (bottom). }
\end{figure}

In order to compute the fluctuations of $T_{21}$, we need to compute also $\alpha$ profiles as a function of distance to the PBH, for which we follow the analytic expressions of~\cite{Pillepich_21cm}. Such profiles can be seen in~\reffig{alpha_r}. 

\section{Contribution to the sky-averaged signal of 21 cm IM}\label{sec:PBH_global}
Considerations about the minimum seed  masses required  \cite{Pacucci_growth, Smith_firstsmbh},  number of galaxies in the universe hosting SMBH \cite{Conselice_numbergalaxies}, uncertainties on the accretion mechanisms, and  CMB observations constraints on the maximum allowed dark matter fraction in PBH \cite{Ali-Haimoud_PBH},  lead us to consider a range of  $10^{-8}<\Omega_{\rm PBH} <10^{-6}$~ \cite{Pacucci_spectrum}.
 
 Key parameters of the model are largely unknown:  SMBHs abundance and  Eddington ratio (which is a proxy for the radiative efficiency) and mass. We consider here some representative values. 

In addition to considering that all PBHs have the same mass, we also consider that all of them have the same Eddington ratio. This is an idealized case, since each kind of SMBH population (e.g. not active, type 1, type 2 and so on) has a different distribution of Eddington ratio (see e.g.~\cite{Shen_lambda, Kelly_lambda, Hopkins_lambda}). It is customary to consider that SMBHs are active if $\lambda\gtrsim 10^{-4}$, although this is an arbitrary limit, given that the Eddington ratio distribution is broad, and extends towards $\lambda<10^4$, as pointed by observations~\cite{Panessa_Plambda, Hopkins_Plambda, Babic_Plambda}. 

In any case, a characteristic value of the Eddington ratio is also largely unconstrained. Observational studies of X-ray selected SMBHs (which of course implies a selection bias favouring the most active luminous SMBHs) suggest large values of the Eddington ratio, i.e. $\lambda\sim 0.1$. 
 Nonetheless, one can consider that all SMBHs are active (not only those with $\lambda\gtrsim 10^{-4}$) and then, $\lambda$ can take values $\ll 10^{-4}$~\cite{Merloni_lambda}. Moreover, in~\cite{Ali-Haimoud_PBH}, the evolution of PBH accretion under the most conservative assumptions was studied in a cosmological context assuming spherical accretion, finding much lower and mass dependent Eddington ratios. Besides, we assume for simplicity a duty cycle of unity, so the Eddington ratio would be smaller to match more realistic cases with lower duty cycles but higher luminosity.

Taking all this into account, we prefer to consider different parameter configurations to account for different possibilities spanning a wide range in the parameter space. We consider all the possible combinations of three masses ($10^2$, $10^3$ and $10^4$ $\Msun$) and three abundances ($\Omega_{\rm PBH}= 10^{-8},\, 10^{-7}$ and $10^{-6}$). We also consider two possible scenarios with different choices of $\lambda$ for each combination of $M$ and $\Omega_{\rm PBH}$: one with large Eddington ratio ($\lambda\sim 0.1$ for astrophysical considerations) and another with small $\lambda$ (see ~\cite{Ali-Haimoud_PBH}).  If a disk is formed and the accretion is not spherical, values of $\lambda$ above this lower limit, but still below the astrophysical one,  are expected~\cite{poulin:cmbconstraint}.
 Following, ~\cite{Ali-Haimoud_PBH}, as the change of $\lambda$ with redshift for $z\lesssim 200$ is small, we consider it constant and we take $\lambda=10^{-4}$ for $M = 10^4\Msun$, $\lambda=10^{-7}$ for $M = 10^3\Msun$ and $\lambda=10^{-10}$ for $M = 10^2\Msun$. 

If we assume that there are PBHs present in the dark ages, their signal is superimposed to the standard one coming from the IGM and temperature fluctuations. We consider that the gas ``bubble'' around the PBH extends until the distance where $\lvert T_{21}\lvert <\Delta T$, which we set $\Delta T~=~1$~mK. This distance corresponds to the point in which $T_{21}$ (\reffig{dtb}) becomes flat, and refer to it as $r_{lim}$.

The differential flux per unit frequency received from the bubble can be expressed in terms of the differential brightness temperature as:
\begin{equation}
\delta~\mathcal{F}_\nu =~\frac{2~\nu_{\rm rec}^2}{c^2}k_bT_{21}\Delta\Omega_{\rm bubble},
\label{eq:diff_flux}
\end{equation}
where $\Delta\Omega_{\rm bubble} = A/\chi^2(z)$, being  $A=\pi r_{lim}^2$ the comoving cross section of the bubble, and $\chi(z)$, the comoving distance to us. 
Furthermore, the line-integrated differential flux, $\delta F$, can be obtained multiplying the differential flux evaluated in the desired frequency, $\nu^\prime$, by a redshift effective line width, $\Delta\nu_{\rm eff}=\left(\mathcal{F} (\nu)d\nu~\right)/\mathcal{F}(\nu^\prime)$. For an optically thin cloud, $\dnueff$ can be approximated by:
\begin{equation}
\dnueff =~\frac{\nu^\prime}{(1+z)}\sqrt{\frac{2k_BT_k}{m_Hc^2}}.
\end{equation}
As both our gas and differential brightness temperature have radial profiles, we use an effective surface average defined as:
\begin{equation}
\tilde{T}_{21} =~\frac{2\pi}{A}\int_0^{r_{lim}} T_{21}(r^\prime)\dnueff(r^\prime)r^\prime dr^\prime.
\end{equation}

The comoving number density of PBHs is:
\begin{equation}
\begin{split}
 n_{\rm PBH}(\Omega_{\rm PBH},M) = 1.256\times 10^{-2}~\times~\\ 
 \times~\left(\frac{\Omega_{\rm PBH}}{10^{-9}}\right)\left(\frac{M}{10^4M_\odot}\right)^{-1} {\rm Mpc^{-3}}.
\end{split}
\label{eq:npbh}
\end{equation}
As was discussed in the previous section, $\lambda$ and $M$ are  degenerate  when considering the signal of an individual PBH. However, when considering the entire population, as the comoving number density of PBHs (\refeq{npbh}) only depends on $\Omega_{\rm PBH}$ and $M$, this degeneracy is broken.
The average contribution of all the bubbles around the PBHs population to the differential flux per unit frequency is 
\begin{equation}
\langle\delta\mathcal{F}_\nu\rangle =~\frac{\Delta z\Delta\Omega_{\rm beam}}{\Delta~\nu}\frac{d^2V}{d\Omega dz}\delta F n_{\rm PBH}.
\end{equation}

Finally, taking into account that $\Delta\nu/\Delta z =~~\nu_0/(1+z)^2$ and defining the beam-averaged effective differential brightness temperature, $\langle T_{21}\rangle$, using $\langle~\delta\mathcal{F}_\nu\rangle = ~  2~\nu_{\rm rec}^2k_b\langle T_{21}\rangle\Delta\Omega_{\rm beam}/c^2$, we obtain~\cite{Iliev_minihalo_02}:
\begin{equation}
\langle T_{21}\rangle =~\frac{(1+z)^2}{\nu_0}\frac{c}{H(z)}n_{\rm PBH}\tilde{T}_{21} A.
\end{equation}

\begin{figure}
\centering
\includegraphics[width=\columnwidth]{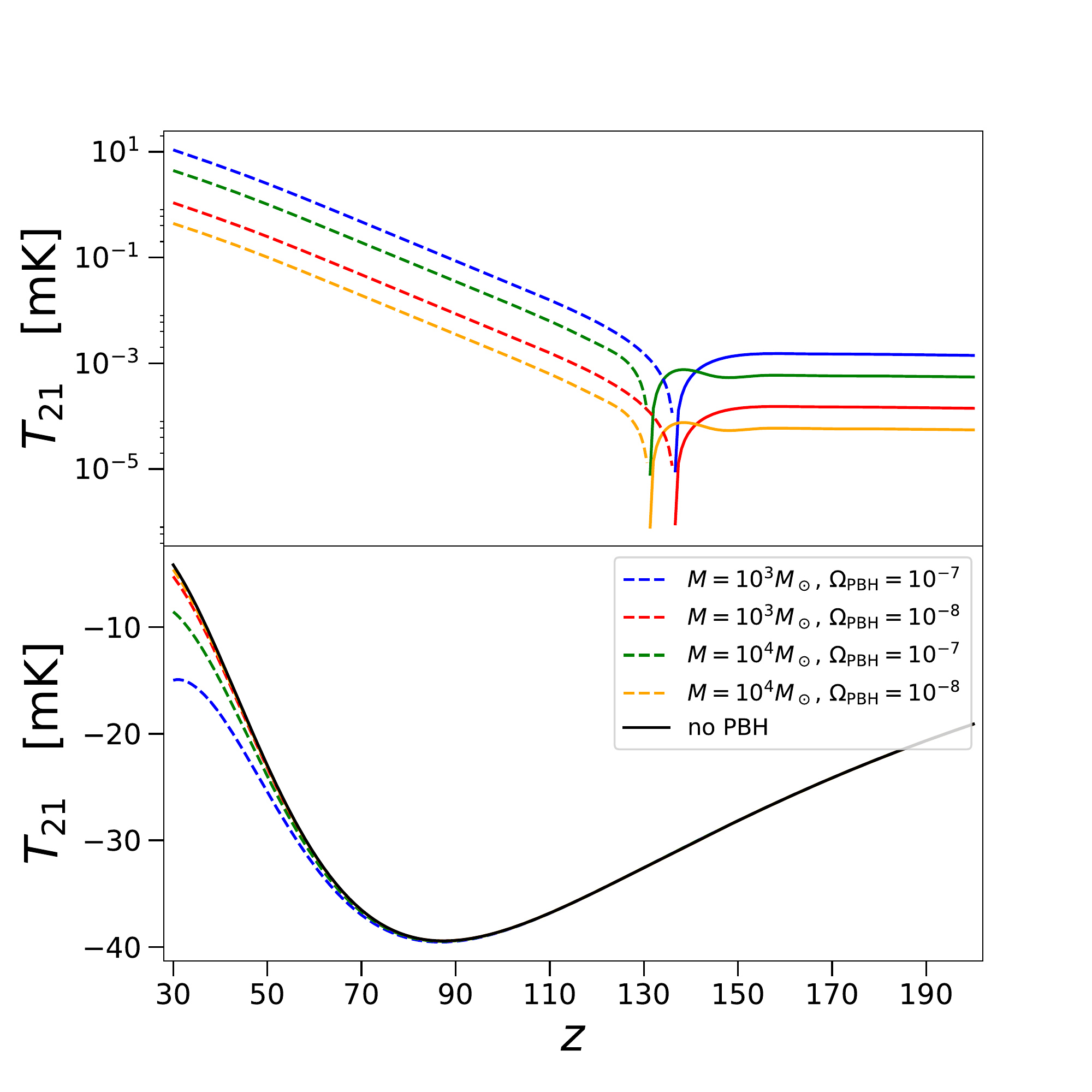}
\caption{\label{fig:T_z}
Sky-averaged differential brightness temperature as a function of redshift. Top panel: isolated contribution of PBHs assuming different masses and abundances (positive values are shown as solid lines while negative, in dashed lines). Bottom panel: Total signal.
}
\end{figure}

We show the evolution of the sky-averaged signal with redshift in~\reffig{T_z} for different cases with $\lambda = 0.1$. As can be seen, the contribution to the standard signal is positive (detected in emission) for $z\gtrsim 130$ (for which $T_{21}>0$ at any distance,~\reffig{dtb}), and negative (detected in absorption) for lower $z$. However, the contribution is only appreciable for $z\lesssim 50$. For the same values of $\mathcal{M}$, the PBH contribution is larger for larger $n_{\rm PBH}$, which is reasonable. On the other hand, for the same number density, the contribution is larger for larger intensity of the emission (i.e. larger $\mathcal{M}$). Therefore, the contribution of PBHs to the standard sky-averaged signal in the cases where $\lambda$ takes much smaller values will be negligible unless the number density is really high.

\section{Contribution to the angular power spectrum of 21 cm IM}
\label{sec:PBH_Cls}
In this section, we introduce how we compute the angular power spectrum of 21 cm IM, accounting for the first time for the emission of PBHs,  the temperature profiles around them, and thus the full scale dependence of their contribution to the 21cm IM signal.  
As reference, note that the corresponding scale for the multipole $\ell$ at redshift $z$ fulfills approximately $k\chi(z)=\ell$ (using the Limber approximation); therefore, at $z=30$, $\ell=10^3$ corresponds to $k\sim 0.09$ Mpc$^{-1}$ in a $\Lambda$CDM cosmology with the best fit parameters of Planck. 

The modelling of the PBH signal in the fluctuations of $T_{21}$ is similar to that of  the  Sunyaev-Zel'dovich effect fluctuations from clusters of galaxies~\cite{Komatsu99_SZ} or 21 cm IM from minihaloes before reionization~\cite{Furlanetto06}. In all these cases, there are extended sources tracing the peaks of the matter density field. In analogy, we use the halo model~\cite{Cole88} to characterize the $T_{21}$ power spectrum during the dark ages in the presence of PBHs.
A review of the  formalism of the halo model can be found in~\cite{Cooray_halomodel}. 
Given that we only consider a monochromatic PBH population, all the integrals in mass that appear in the halo model formalism, which are of the type $\int_{M_{\rm min}}^{M_{\rm max}}dM^\prime n_{\rm PBH}(M^\prime)\mathcal{G}(M^\prime)$, where $\mathcal{G}$ is a general function, 
simplify to
$ n_{\rm PBH}(M)\mathcal{G}(M)$.   

In the halo model, the power spectrum is the sum of two components: the correlation between points within the same halo or bubble is described by the `one-halo' term, while the correlation between points in separate halos/bubbles is encoded in the `two-halo' term. Hence $P_{\rm PBH}(k) = P^{\rm 1h}_{\rm PBH}+P^{\rm2h}_{\rm PBH}$. 
In the same way, one can express the angular power spectrum in multipole coefficients as $C_\ell^{\rm PBH} = C_\ell^{\rm PBH (1h)} +C_\ell^{\rm PBH (2h)} $.

We build on ~\refeq{T21_2D} to obtain the observed fluctuations of the 21 cm temperature fluctuations originated due only to the presence of PBHs in a direction $\hat{n}$ and in a frequency $\nu$:
\begin{equation}
\begin{split}
~\delta T_{21, {\rm PBH}}^\ell(\hat{n},~\nu) =~\int_0^{\infty} dx \left[ W_\nu(x) ~\alpha_{\rm PBH}(r)~\delta_b({\bf r}) + \right. ~\\
 + \left. T_{21, {\rm PBH}}(r)~\delta_v({\bf r}) \right],~\label{eq:T21_ell}
\end{split}
\end{equation}
where $T_{21}(r)$ and $\alpha(r)$ are the quantities obtained in~\refsec{PBH_effects}, $r=\sqrt{x^2+R^2}$ is the comoving distance to the center of the PBH and $R=\chi(z)/\ell$ is the comoving transverse distance to the center of the PBH.  By using $R=\chi(z)/\ell$, we assume a plane parallel approximation.  This is justified because for low $\ell$ (where the plane parallel approximation breaks down), $r\gg r_{\rm lim}$, hence $\delta T_{21, {\rm PBH}}({r})=0$. Once we have computed $\delta T_{21, {\rm PBH}}^\ell$, we obtain the transfer function for the 21 cm IM fluctuations due to PBHs, $\mathcal{T}_\ell^{\rm PBH}$, as in ~\refeq{T_transfer}.

 As the standard contribution in the linear regime without the PBHs comes from a continuum where there are no haloes, we consider that the one-halo term of the standard contribution vanishes. Therefore, we obtain the total angular power spectrum as the sum of the one-halo and two-halo terms, expressed as:
\begin{equation}
C_\ell^{\rm PBH (1h)} =~\frac{2}{\pi}n_{\rm PBH}~\int_0^\infty dk k^2~\left(\cal{T}_\ell^{\rm PBH}\right)^2,
\label{eq:Cells1h}
\end{equation}
\begin{equation}
C_\ell^{\rm PBH (2h)} =~\frac{2}{\pi}~\int _0^\infty dk k^2~\left(\mathcal{T}_\ell+n_{\rm PBH}b\mathcal{T}_\ell^{\rm PBH}\right)^2P_m(k),
\label{eq:Cells2h}
\end{equation}
where we assume that PBHs are completely correlated with the dark matter distribution and $b$ is a scale-independent bias. 
This is motivated by the following consideration.  If PBHs are the seeds of the SMBHs, they are  located at the centers of the potential wells  so galaxies will form around them. We  take the bias factor to be  approximately the mean value of the galaxy bias. In explicit calculations we assume   $b = 1.25$. 

Given that the formation of a PBH is a rare event and PBHs  spatial distribution is discrete, there is a Poissonian fluctuation in the number density of PBHs. Therefore, in addition to the standard matter power spectrum appearing in ~\refeq{Cells2h}, there is a Poissonian power spectrum contribution. 
These fluctuations behave like isocurvature modes, as the formation of compact objects at small scales does not affect immediately the curvature at large scales \cite{Afshordi_pbh}. The primordial power spectrum that describes them is:
\begin{equation}
P_{\rm PBH}^0 =~\frac{\fpbh^2}{n_{\rm PBH}}.
\end{equation} 
The isocurvature behaviour  is enclosed in the transfer function of isocurvature modes, which is scale-independent ($T_{\rm iso} =~\frac{3}{2}(1+z_{\rm eq})$, where $z_{\rm eq}$ is the redshift of matter-radiation equality, $1+z_{\rm eq}~\approx 3400$). Therefore, the power spectrum generated by the Poisson fluctuations is:
\begin{equation}
\begin{split}
P_{\rm Poisson}(z) =~\left(T_{\rm iso}D(z)\right)^2P_{\rm PBH}^0 
 =~\\
 =~\frac{9}{4}(1+z_{\rm eq})^2 D^2(z)\frac{f_{\rm PBH}^2}{n_{\rm PBH}},
\end{split}
\label{eq:Poisson}
\end{equation}
where $D(z)$ is the growth factor. The mass fraction, $f_{\rm PBH}$, appears because this contribution comes only from the fluctuation in number of PBHs and not all the matter. $P_{\rm Poisson}$ should be added to the two-halo term multiplied only by $\mathcal{T}_\ell^{\rm PBH}$. Nonetheless, given the ranges of $\fpbh$ we consider, the Poisson contribution is negligible at all scales. 
Only in studies exploring PBHs as a sizable fraction of the dark matter, where $\fpbh\sim 1$, it is found that the contribution of ~\refeq{Poisson} dominates at small scales. In fact, Afshordi et al. (2003)~\cite{Afshordi_pbh} and Kashilinsky (2016)~\cite{Kashilinsky16} propose to constrain the abundance of PBHs by looking for this scale independent contribution to the power spectrum in observations of the Ly$\alpha$ forest and the Cosmic Infrared
Background anisotropies, respectively. 

\begin{figure}[hbtp!]
\centering
\includegraphics[width=\columnwidth]{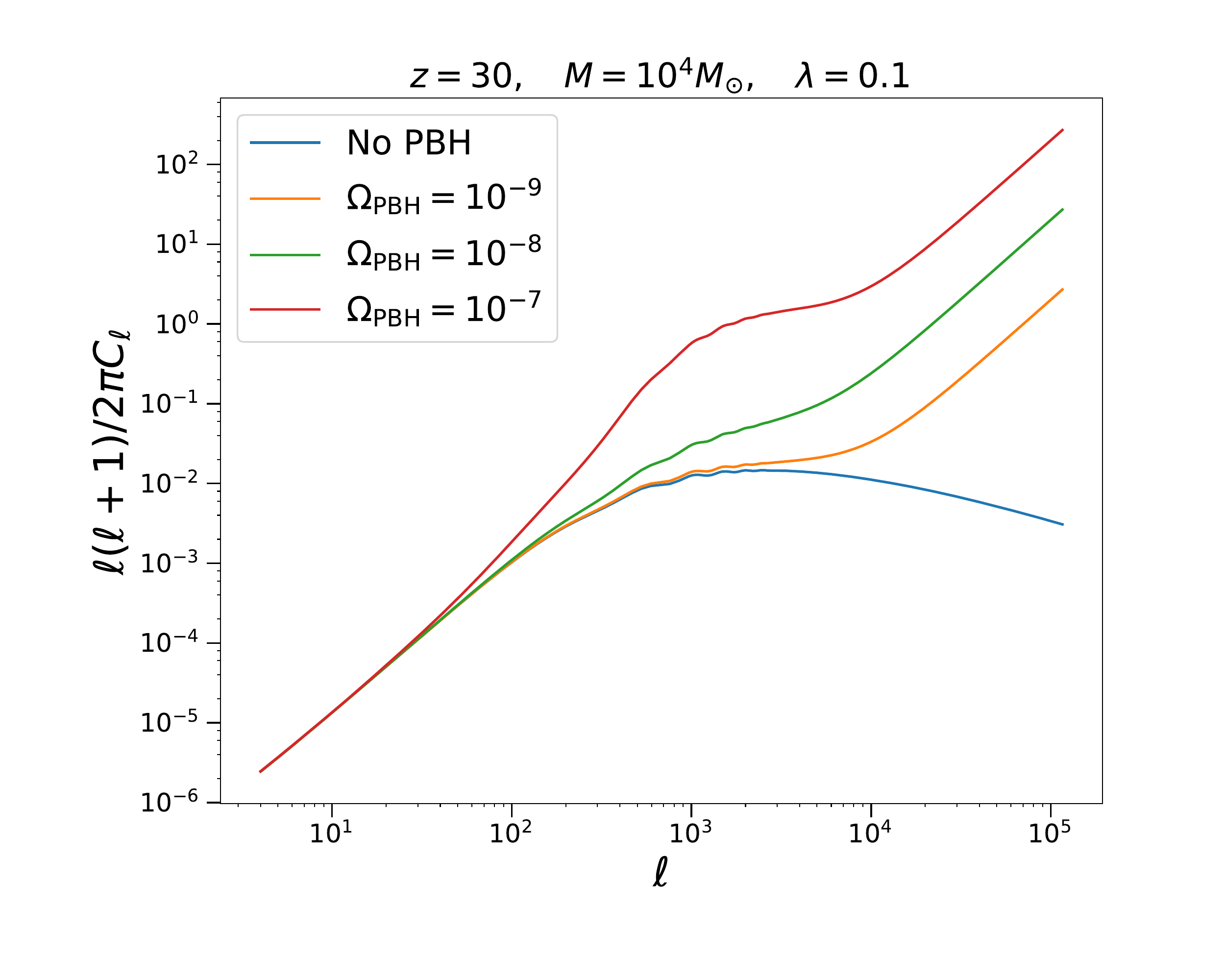}
\includegraphics[width=\columnwidth]{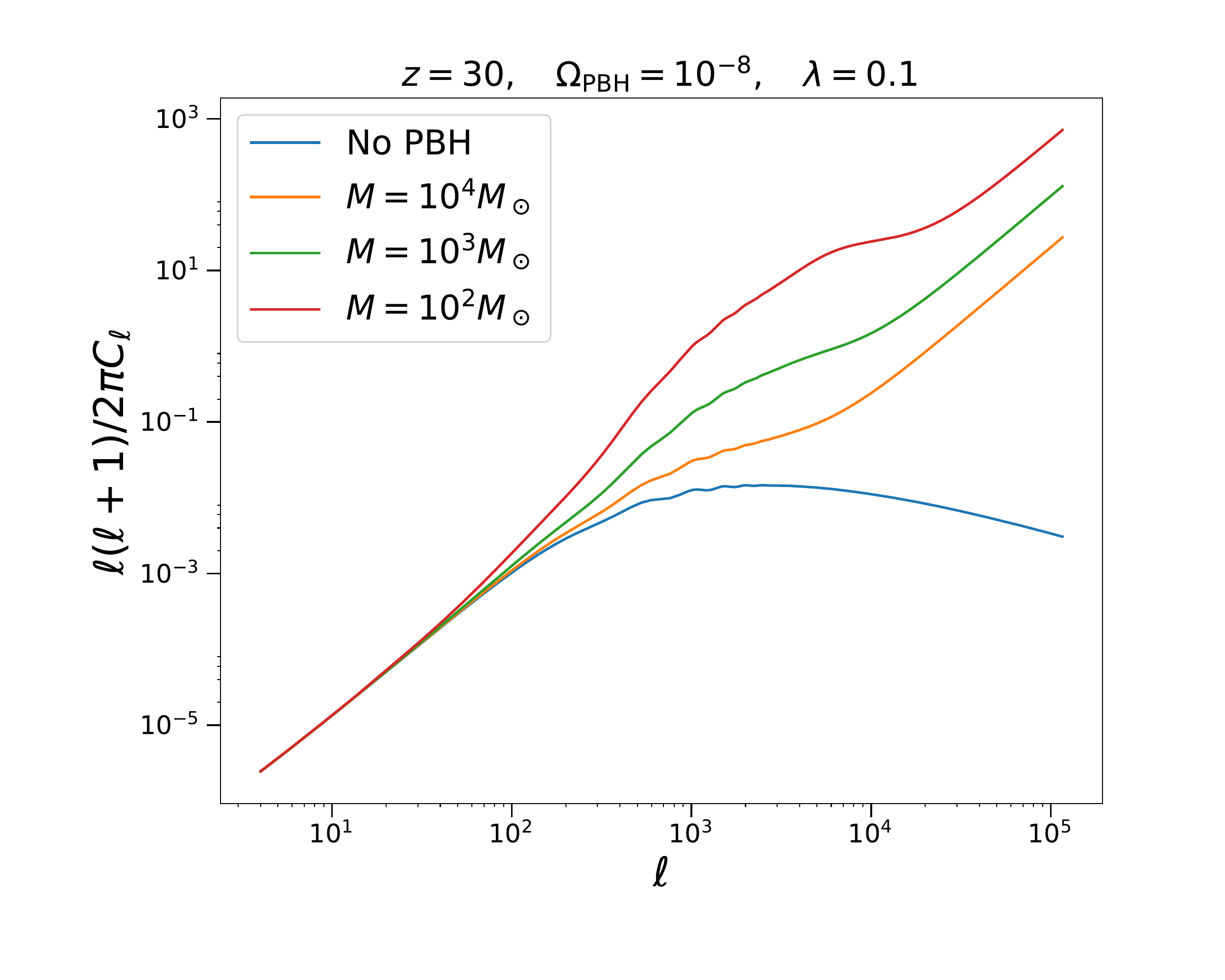}
\includegraphics[width=\columnwidth]{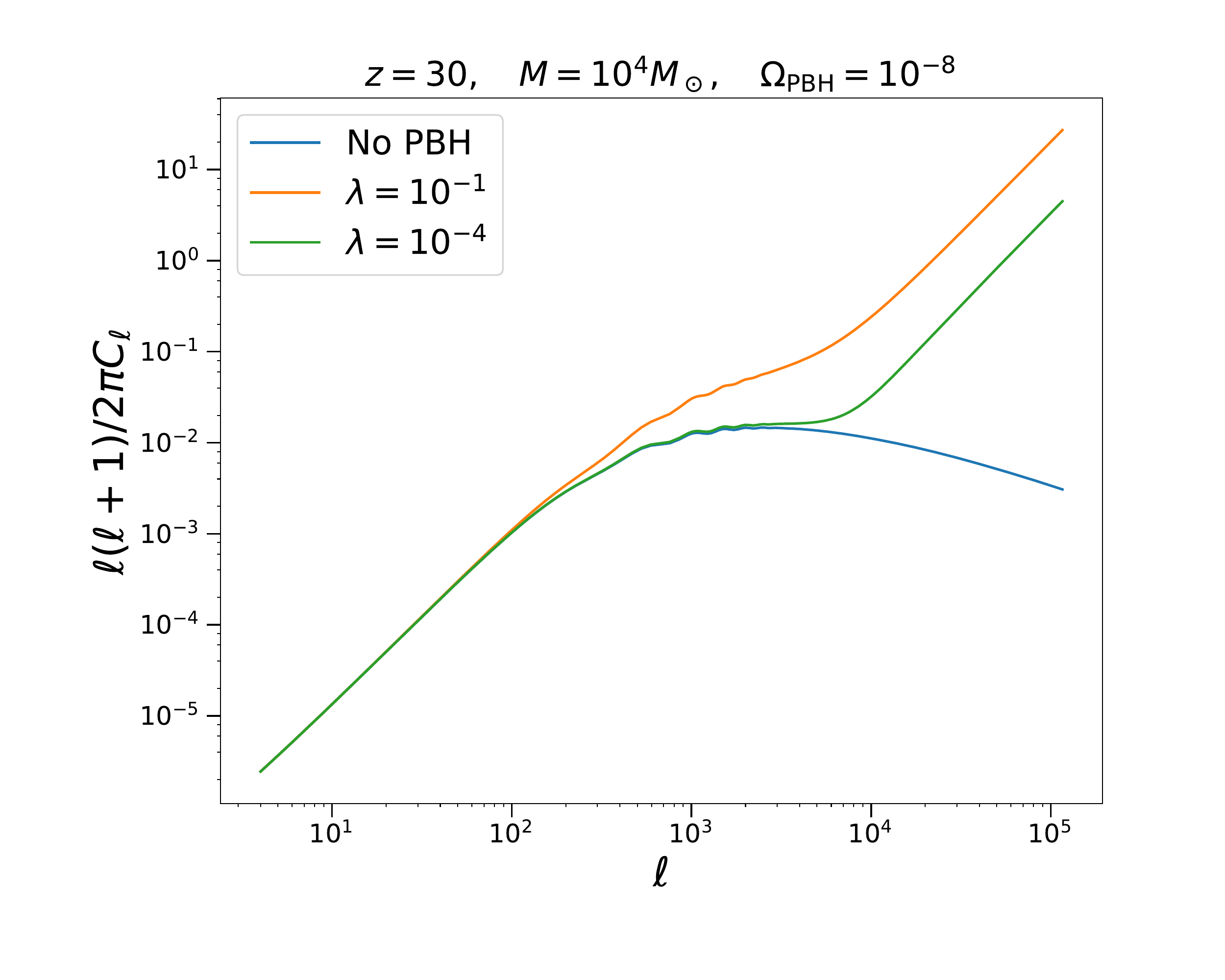}
\caption{\label{fig:cells_cases}
Angular power spectrum of the total signal in 21 cm IM at $z = 30$ varying the density parameter of PBH (top), the mass (middle) and the Eddington ratio (bottom). }
\end{figure}

\begin{figure}[hbtp!]
\centering
\includegraphics[width=\columnwidth]{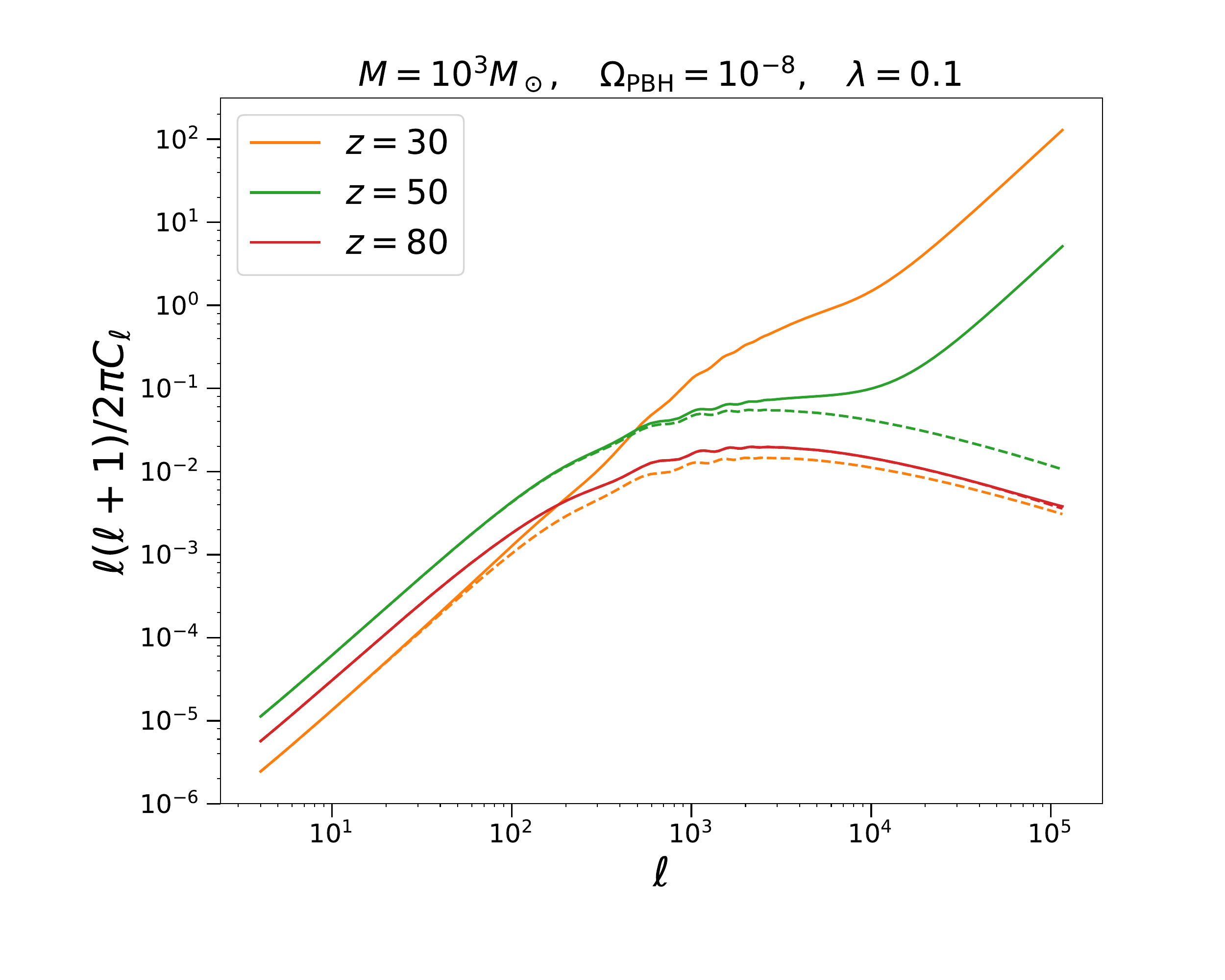}
\caption{\label{fig:cells_z}
Angular power spectrum comparing the total signal in 21 cm IM (solid lines) with the case where there are no PBHs (dashed lines) for $M=10^3\Msun$, $\Omega_{\rm PBH} = 10^{-8}$, $\lambda=0.1$ and $\Delta\nu = 1$ MHz at different redshifts.}
\end{figure}

Looking at~\refeq{T21_ell},~\refeq{Cells1h} and~\refeq{Cells2h} it is easy to notice that the angular power spectrum will depend only on two quantities related to PBHs: $n_{\rm PBH}$ and $\mathcal{T}_\ell^{\rm PBH}$ ($C_\ell$s also depend on other quantities not related with PBHs, such as the redshift). Therefore, although we do consider three parameters regarding PBHs, the relevant quantities are combinations of them: $\mathcal{M}=M\lambda$ and $n_{\rm PBH}\propto~\Omega_{\rm PBH}/M$. The former is needed to compute the size of the bubble around the PBH (i.e.,~$r_{lim}$). 
Essentially, varying $\mathcal{M}$ shifts the features related with PBHs to different multipole ranges  (via $\mathcal{T}_\ell^{\rm PBH}$). The latter is a rescaling of such contributions ($C_\ell^{\rm PBH(1h)}\propto n_{\rm PBH}$ and $C_\ell^{\rm PBH(2h)}\propto n_{\rm PBH}^2$). Therefore, varying $n_{\rm PBH}$ changes the amplitude of the PBH features. These two effects are relevant to determine at which scale the PBH contribution starts to dominate.  
 Thus, there is a degeneracy among the PBH parameters:
\begin{equation}
C_\ell (M,\lambda,\Omega_{\rm PBH}) = C_\ell (M/\beta,\lambda\beta,\Omega_{\rm PBH}/\beta),
\end{equation} 
where $\beta$ is an arbitrary positive constant. All these effects can be seen in~\reffig{cells_cases}. In most of the cases, the PBH effects modify the standard power spectrum at $\ell~\sim 10^2~-~10^3$, with a large variation at $\ell\sim 10^5$.

As can be seen in~\reffig{cells_z}, the PBH-induced deviation from the standard signal  decreases  with redshift because the size of the bubble decreases with  redshift (see~\reffig{dtb} and~\reffig{alpha_r}), so the multipole at which the deviation is appreciable at fixed $n_{\rm PBH}$ and $\mathcal{M}$ increases.

\section{Detectability}\label{sec:detect}
We have characterized the imprints of massive PBHs in both the sky averaged signal and the power spectrum of 21 cm IM. The sky averaged signal requires dedicated single dipole experiments, such as EDGES \cite{EDGES}, LEDA \cite{LEDA} or SARAS~\cite{SARAS}, to be measured. On the other hand, radio arrays as SKA aim to measure the fluctuations. As the PBH contribution on the sky-averaged signal is very small in most of the cases, we focus on the power spectrum and observations done with radio arrays.

We  study the detectability of the signal and  forecast constraints on massive PBH parameters assuming observations in the dark ages done with the SKA~\cite{SKA_IM} and a futuristic Earth-based experiment, similar to the SKA but with much larger baseline and $f_{\rm cover}$, which we refer to as ``SKA$_{\rm Adv}$''. Given that the atmosphere is opaque for  frequencies $\lesssim 45$ MHz, SKA will not be able to observe much further than $z\approx30$. Then, in order to observe well beyond the end of the dark ages ($z\gtrsim 30$) it will be necessary to observe from outside the Earth's atmosphere; a good candidate as a location for such observations is provided by the Moon~\cite{Jester_darkmoon, Burns_darkmoon}.  This is why we also consider three different realizations of a futuristic radio array on the dark side of the Moon, that we call the ``Lunar Radio Array'' (LRA)~\cite{Silk:telescope}. The relevant specifications of the experiments considered can be found in~\reftab{ST}.

	\begin{table}[hbtp!]
	\setlength\extrarowheight{3.5pt}
		\begin{tabular}{|c | c | c | c | c | c |}
			\hline
			Spec & SKA & SKA$_{\rm Adv}$ & LRA1 & LRA2 & LRA3~\\             
			\hline
			\hline
			$D_{\rm base}$ (km) & 6  & 100 & 30 & 100 & 300~\\             
			\hline
			$f_{\rm cover}$ & 0.02 & 0.2 & 0.1& 0.5 & 0.75~\\
			\hline			
			$t_{\rm obs}$ (years) & 5 & 10 & 5 & 5 & 5~\\
			\hline
			$l_{\rm cover}\frac{1+z}{31}$	& 5790 & 96515 & 28954 & 96515 & 289547~\\
			\hline
		\end{tabular}
		\caption{Instrument specifications for SKA, advanced SKA and three different realizations of the Lunar Radio Array.}
		\label{tab:ST}
	\end{table}

Because of its wide frequency coverage we consider that it will be possible to do tomography with LRA  beyond $z \sim 30$.
We follow the arguments introduced in~\cite{Munoz_21cmbispec} to determine  the redshift bins that can be considered independent when observing with $\Delta\nu =1$ MHz between $z=30$ and $z=200$. 

\begin{figure}[hbtp!]
\centering
\includegraphics[width=\columnwidth]{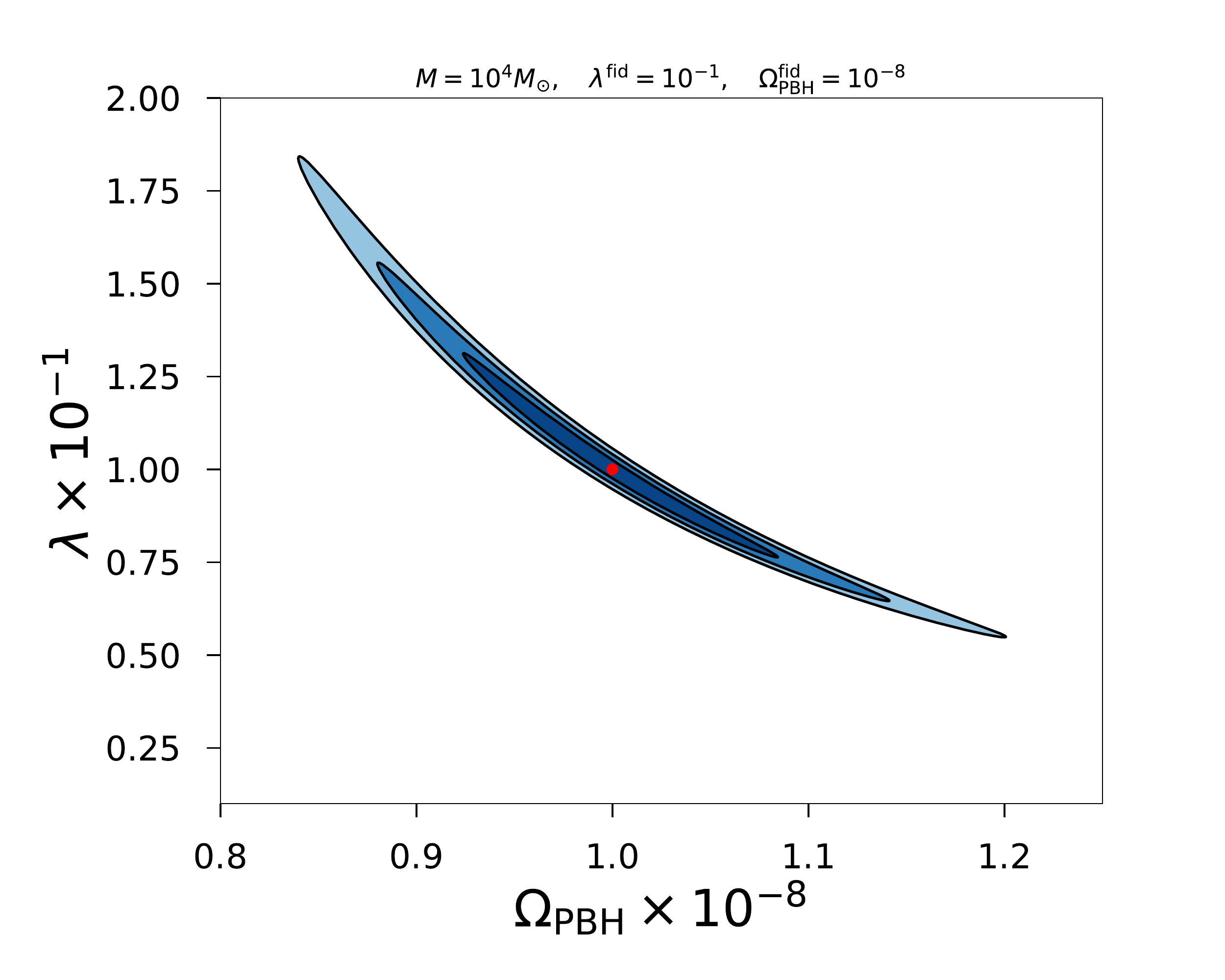}
\includegraphics[width=\columnwidth]{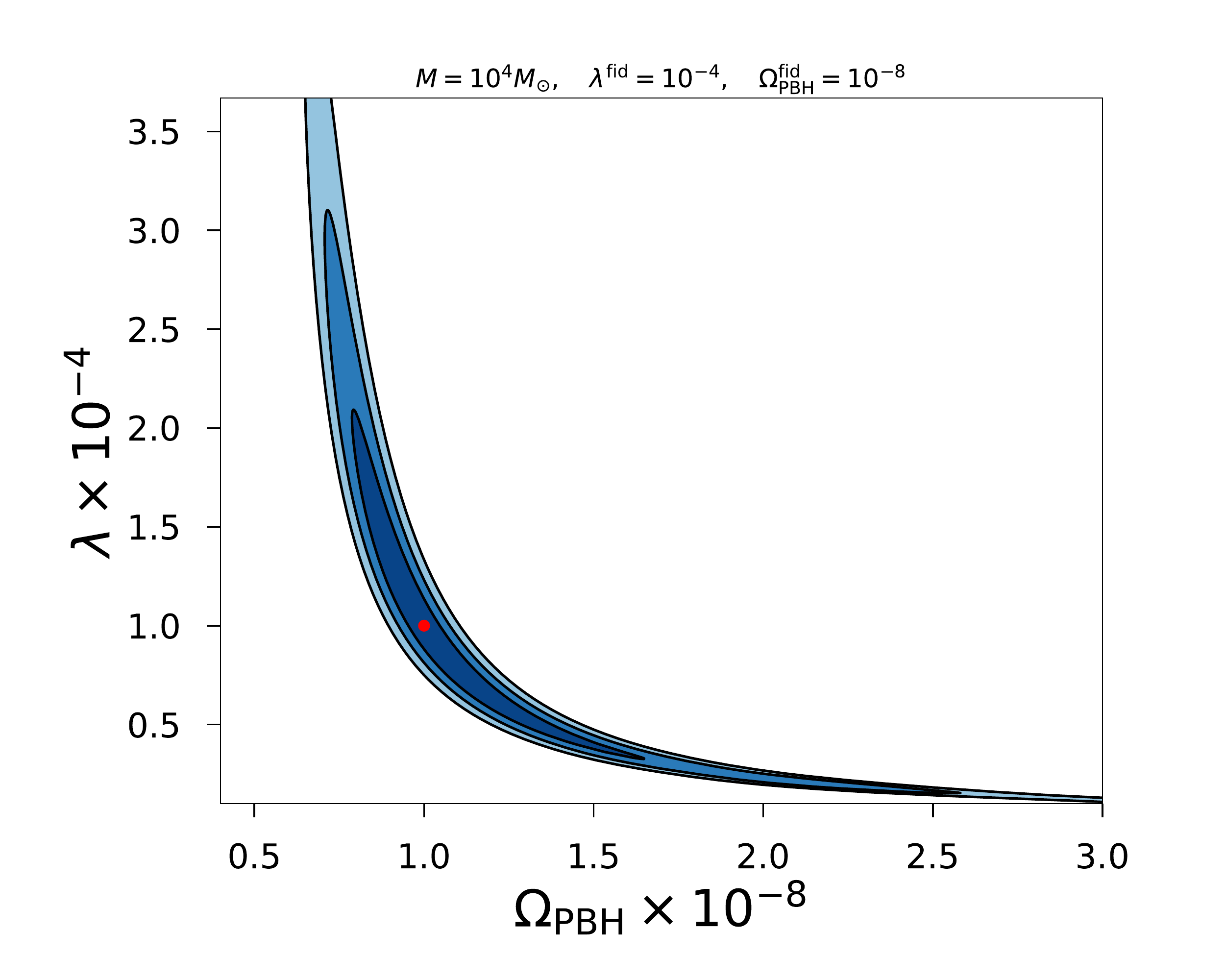}
\caption{\label{fig:Dchi2}
$1\sigma$, $2\sigma$ and $3\sigma$ confidence level forecasted constraints in the $\Omega_{\rm PBH}$-$\lambda$ plane from theoretical $\Delta\chi^2$ values for the fiducial cases of $M^{\rm fid}=10^4\Msun$, $\Omega_{\rm PBH}^{\rm fid}=10^8$ and $\lambda^{\rm fid} = 0.1$ (top) and $\lambda^{\rm fid}=10^{-4}$ (bottom), considering SKA$_{\rm Adv}$ in both cases. The fiducial case is marked with a red dot.}
\end{figure}

First of all, we compute the $\Delta~\chi^2$ considering SKA$_{\rm Adv}$ as a function of  $\Omega_{\rm PBH}$ and $\lambda$ for two fiducial cases ($M^{\rm fid}=10^4\Msun$, $\Omega_{\rm PBH}^{\rm fid}=10^8$ and $\lambda^{\rm fid} = 0.1$ (top) and $\lambda^{\rm fid}=10^{-4}$ (bottom)). The $1\sigma$, $2\sigma$ and $3\sigma$ contours  are shown in ~\reffig{Dchi2} where  the degeneracy among the parameters discussed in~\refsec{PBH_Cls} can be appreciated.

We also forecast the errors on $\Omega_{\rm PBH}$ ($\sigma_\Omega$) and $\lambda$ ($\sigma_\lambda$) using Fisher matrices  for all the fiducial cases we consider. The resulting forecasts are reported in~\reftab{forecast}. 
The Fisher forecasts obtained should be considered as a rough estimate, especially for low fiducial values for $\lambda$: \reffig{Dchi2} shows that the constant $\Delta \chi^2$ contours are not well described by ellipses, which is what the Fisher approach assumes.

	\begin{table*}[hbtp!]
	\setlength\extrarowheight{2.5pt}
		\begin{tabular}{| c |c |}
			\hline			$M_{\rm fid} (\Msun)$ & \begin{tabular}{m{.85cm} |m{14cm} @{}m{0pt}@{}} $\lambda_{\rm fid}$ &\centering Forecasted precision & \\ \end{tabular}      \\
			\hline
			\hline
			$10^4$	&  \begin{tabular}{m{.85cm}|p{14cm}}
									
									$10^{-1}$ & \raisebox{-0.5\height}{\includegraphics[width=13cm]{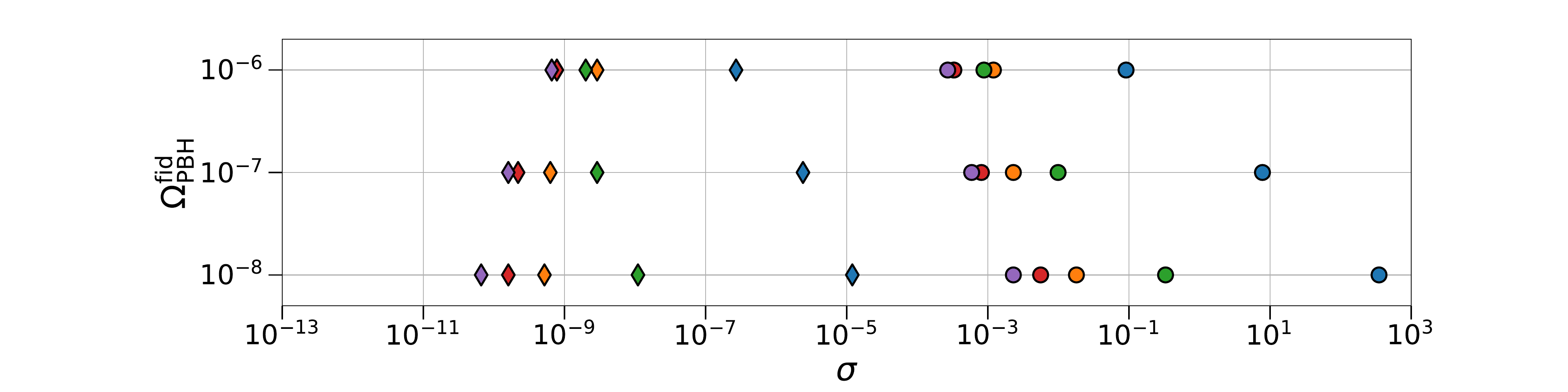}}\\ \hline
									$10^{-4}$ & \raisebox{-0.5\height}{\includegraphics[width=13cm]{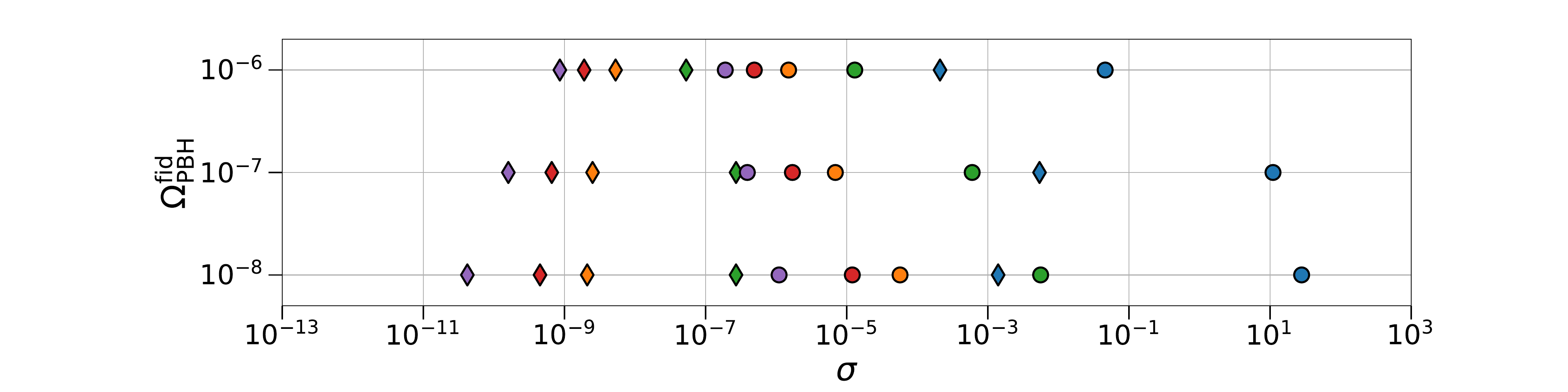}}\\
							\end{tabular}
							\\
			\hline
			
			\hline
			$10^3$	&  \begin{tabular}{m{.85cm}|p{14cm}}
									
									$10^{-1}$ & \raisebox{-0.5\height}{\includegraphics[width=13cm]{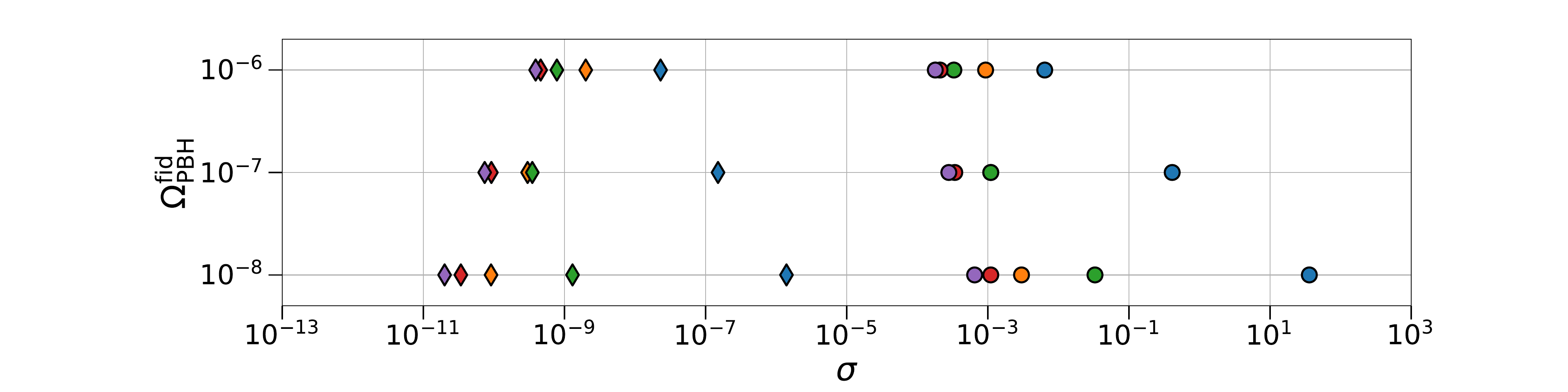}}\\ \hline
									$10^{-7}$ & \raisebox{-0.5\height}{\includegraphics[width=13cm]{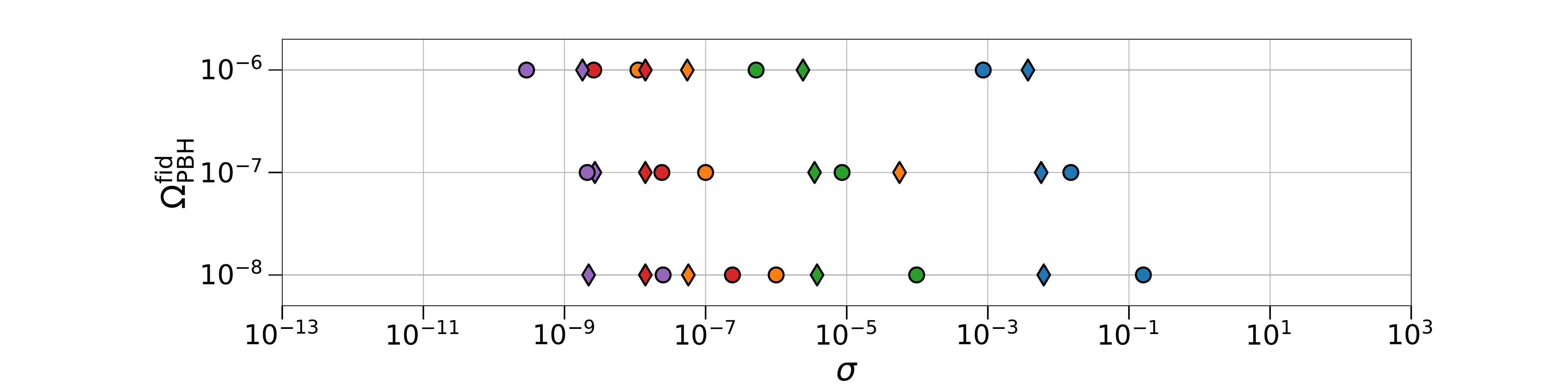}}\\
							\end{tabular}
							\\
			\hline
			
			\hline
			$10^2$	&  \begin{tabular}{m{.85cm}|p{14cm}}
									
									$10^{-1}$ & \raisebox{-0.5\height}{\includegraphics[width=13cm]{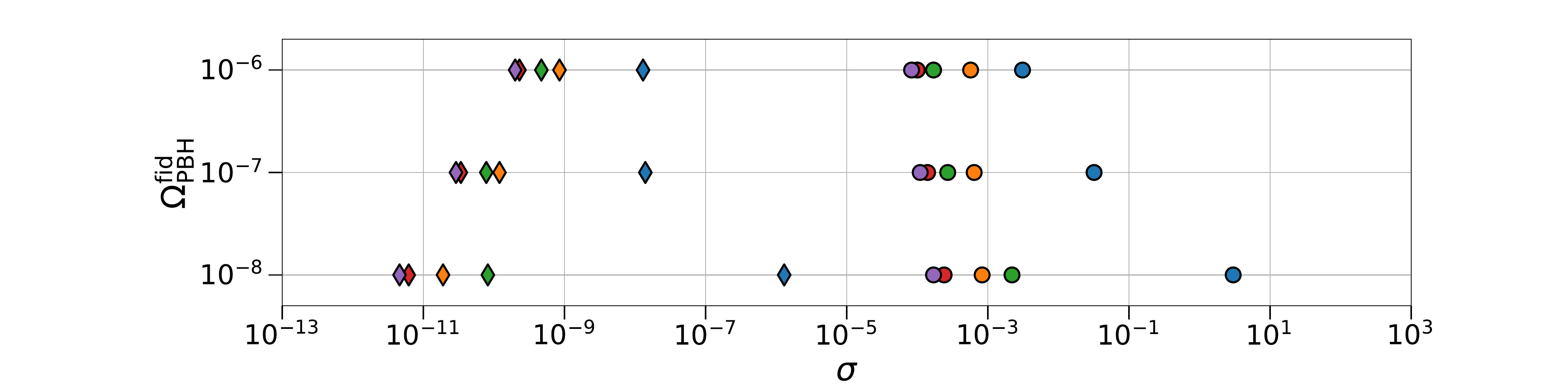}}\\ \hline
									$10^{-10}$ & \raisebox{-0.5\height}{\includegraphics[width=13cm]{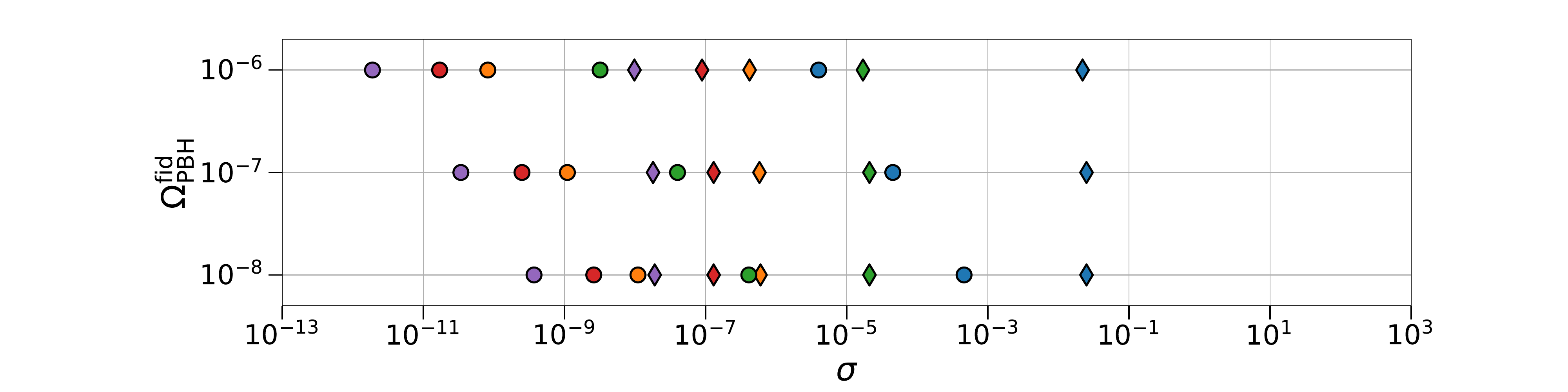}}\\
							\end{tabular}
							\\
			\hline
			 	
			\hline
		\end{tabular}
		\centering
		\includegraphics[scale=0.4]{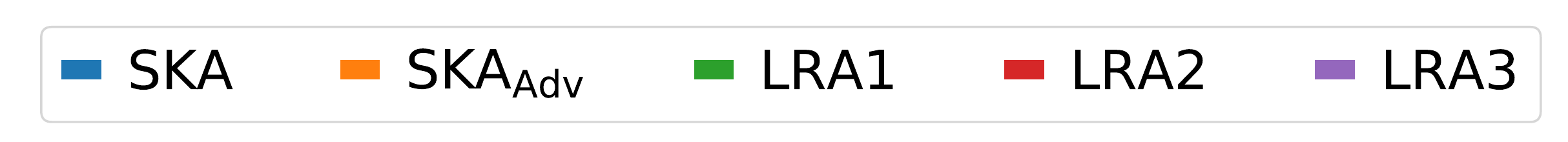}\\
		\includegraphics[scale=0.4]{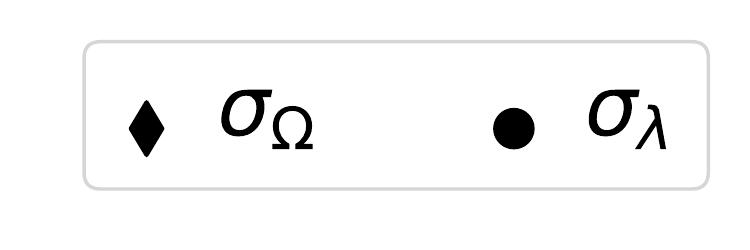}
		
		\caption{1$\sigma$ forecasted uncertainties on the abundance of PBHs, $\Omega_{\rm PBH}$, ($\sigma_\Omega$) and the Eddington ration, $\lambda$ ($\sigma_\lambda$) for different fiducial cases and experiments using Fisher matrices.
		}
		\label{tab:forecast}
	\end{table*}

The PBH signal will be barely detected by SKA, since  only for extreme cases in which $n_{\rm PBH}$ is very large,  the signal-to-noise ratio, $S/N$, for  $\Omega_{\rm PBH}$ and $\lambda$ is larger than unity.

As the amplitude of the power spectrum increases greatly at small scales, being able to resolve very small scales (i.e.,~ large $D_{\rm base}$, which implies large $\ell_{\rm cover}$) will be key to detect the PBH signal and constrain the parameters. This is why the forecast uncertainties for SKA$_{\rm Adv}$ are much smaller than for SKA 
(and similar considerations apply to LRA3 vs. LRA1).

\begin{figure}[hbtp!]
\centering
\includegraphics[width=\columnwidth]{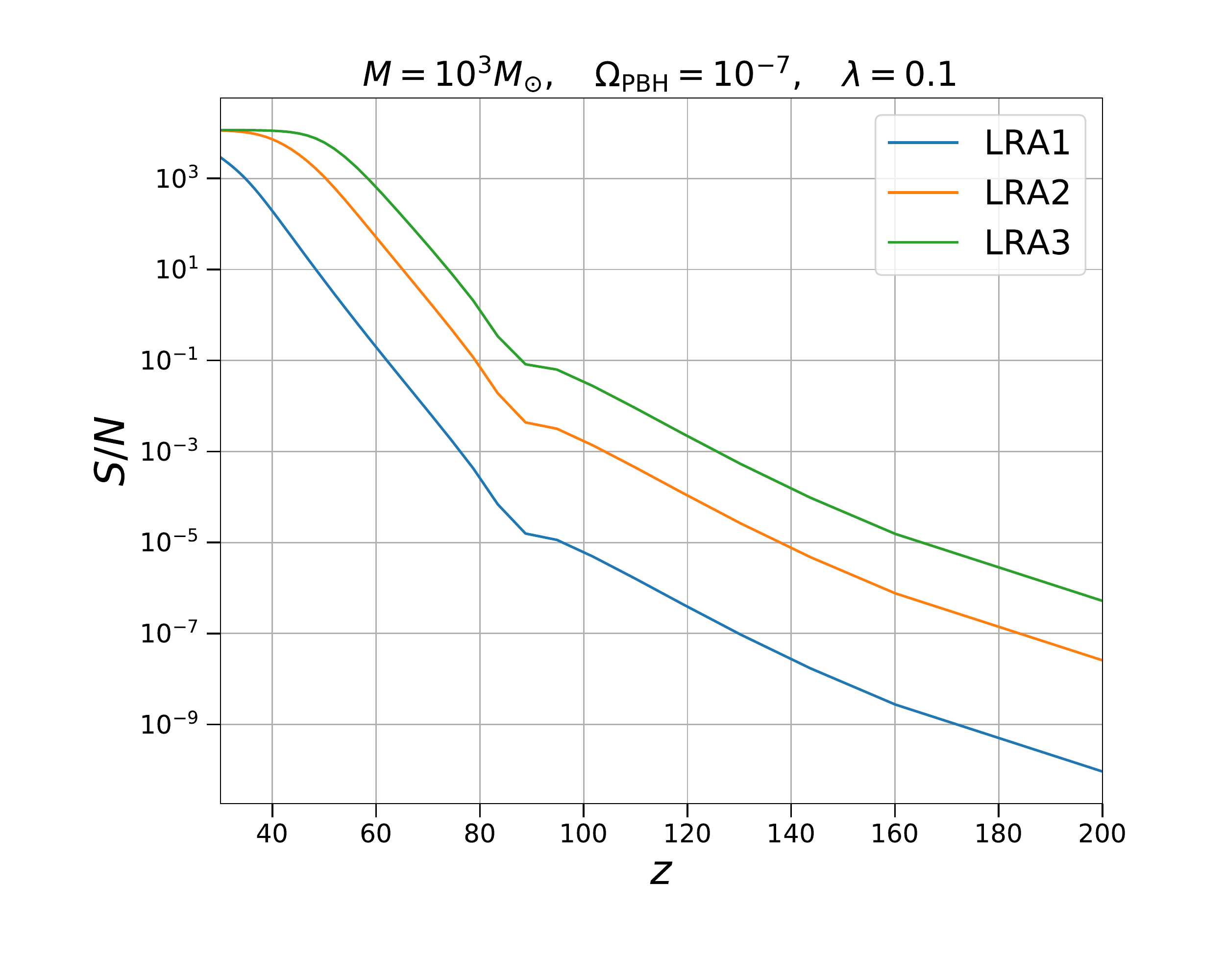}
\caption{\label{fig:SN_z}
Evolution of the signal-to-noise ratio between the power spectrum accounting for the PBH contribution and the standard one with respect to redshift. We consider $M = 10^3\Msun$, $\Omega_{\rm PBH}=10^{-7}$ and $\lambda=0.1$, and show the results for the three realizations of the LRA.}
\end{figure}

On the other hand, the contribution of PBHs to the power spectrum decays with redshift (\reffig{cells_z}), hence the $S/N$ between the case with PBHs and the standard one decreases fast with redshift, as shown in \reffig{SN_z} for the three realizations of LRA and $M = 10^3\Msun$, $\Omega_{\rm PBH}=10^{-7}$ and $\lambda=0.1$. As tomography does not add much information, $\ell_{\rm cover}$ has more impact in the final $S/N$.   Generally, forecast errors for LRA1 are larger than for SKA$_{\rm Adv}$; however, they are smaller for LRA2 than for SKA$_{\rm Adv}$, both having the same $D_{\rm base}$. This is true always except when both $\mathcal{M}$ and $n_{\rm PBH}$ are large.

To summarize, although a detection of the PBH contribution in the dark ages might be achieved by SKA, in order to measure $\Omega_{\rm PBH}$ and $\lambda$ accurately, a more ambitious experiment with a larger baseline is needed. Such measurements will be more precise if tomography is possible, for which experiments such as LRA are needed.

\section{Discussion and conclusions}\label{sec:Conclusions}
The origin and formation mechanism of SMBHs remains largely  unknown. If the growth of the black holes happens only through (standard) accretion,  in order to grow fast enough and reach  $M\sim 10^9\Msun$ at $z\sim 7$~\cite{Pacucci_growth} (and thus match the observed quasar abundance), massive seeds of~$\sim 10^4-10^5\Msun$ need to be already present in regions with large gas densities  at $z\sim 20$. However, if mergers are also considered, the seeds can be lighter. Therefore, there are three candidates to be the seeds of SMBHs: remnants of Population III stars, DCBHs or intermediate mass PBHs.

In this work, we address the observational signatures that intermediate mass PBHs would have on 21 cm IM during the dark ages. We model this signal starting from the characterization of the radial profiles of $T_{21}$ around a single PBH to compute the contribution to the standard sky-averaged signal and to the angular power spectrum, using the halo model. This is the first time that the signature of PBHs accounting for its full scale dependence is modeled in the 21 cm IM power spectrum. 

The values of  the abundance of SMBHs (and therefore, of the seeds needed, $\Omega_{\rm PBH}$), the radiative efficiency (i.e., the Eddington ratio, $\lambda$) and the mass of the possible seeds, $M$, are largely unconstrained.  
Therefore, we consider several parameter configurations as fiducial cases. We forecast observational errors on $\lambda$ and $\Omega_{\rm PBH}$ for each fiducial case assuming future observations made with SKA, a futuristic improved SKA-like experiment, and three different realizations of a futuristic radio array on the far side of the Moon (LRA).

We find that, although we consider three parameters ($M$, $\lambda$ and $\Omega_{\rm PBH}$), the final power spectrum is only sensitive to two combinations of them: $M\lambda =~\mathcal{M}$ (i.e., horizontal shifts of the one- and two-halo terms) and  $\Omega_{\rm PBH}/M~\propto n_{\rm PBH}$ (i.e., changes in the amplitude of the PBH contribution to the power spectrum). As a consequence, there is a degeneracy between the parameters, which can be expressed as  $C_\ell (M,\lambda,\Omega_{\rm PBH}) =~ C_\ell (M/\beta,\lambda\beta,\Omega_{\rm PBH}/\beta)$ (with $\beta$ being an arbitrary positive constant), as  can be seen in~\reffig{Dchi2}. This perfect degeneracy is expected to be partially broken with more detailed modelling. 

We find that the presence of PBHs increases the sky-averaged signal of 21 cm IM in absorption at $z\lesssim 50$, but it is only appreciable when both $\mathcal{M}$ and $n_{\rm PBH}$ are large (\reffig{T_z}). With respect to the angular power spectrum, we find an enhancement of the signal for $\ell~\gtrsim 10^2-10^3$ (\reffig{cells_cases}), which decays with redshift (because the size of the bubble around the PBH, i.e., the gas cloud producing a  signal different from the standard 
sky-averaged value, is smaller for larger redshift,~\reffig{dtb}), as shown in~\reffig{cells_z}.
Although the enhancement is large, measuring $\lambda$ and $\Omega_{\rm PBH}$ will be very difficult with SKA, as the effect is large only on  small scales that can be reached only with a much longer baseline.
On the other hand, as the signal-to-noise ratio decays fast with redshift, tomography does not add much information. 

In this paper we have concentrated on the dark ages, given that they directly probe an epoch where the seeds should be present if they are primordial (and absent otherwise).
Extending the analysis to lower redshift ranges would be interesting for experiments happening on a shorter timescale, although added complications due to astrophysics and degeneracy with other signals would be involved.

Our modelling makes several assumptions and simplifications, which we recap and discuss their resulting implications here. 
First of all, we consider that there is
no overdensity surrounding the PBH (or that the profile around it does not affect drastically the signal) and also neglect radiative
transfer effects. For this initial exploration of the subject,  we assume that the effects of  these two assumptions
compensate, since accounting for density profiles would generate smaller bubbles but larger mean free
paths of X-rays (consequence of the radiative transfer) would make the bubbles larger. 

To compute the contribution of PBHs to the standard signal, we model a single PBH and afterwards we use the number density of PBHs, $n_{\rm PBH}\propto~\Omega_{\rm PBH}/M$, to account for the full population. Hence, our formalism breaks down for large number densities. In such scenario,  bubbles around different PBHs overlap, so PBHs can not be considered as isolated anymore. Besides, PBHs contribute significantly to cosmic reionization, advancing it if their number density is too high. In this case, more accurate  modelling is needed. Actually, the cases with the largest $n_{\rm PBH}$ considered here should be interpreted carefully due to the effects commented above. This caveat could also be more relevant if the actual bubbles are larger than considered here due to radiative transfer effects. 

  Moreover, we assume a simple power-law spectrum for the radiation emitted by the PBH accretion without modelling the full spectral  energy distribution (although see Appendix~\ref{sec:AppendixA}) and that all processes are in equilibrium (steady-state approximation, hence we are limited to $M\leq 10^4\Msun$).
Although the effect of supersonic relative velocities between baryons and dark matter is included in our computation of the 21 cm IM fluctuations (see \refsec{standard_fluctuations}), it is not included in the modelling the heating of the IGM due to the PBH emission. Relative streaming velocities between gas and PBHs leave an imprint on $T_{21}$ radial profiles at the corresponding baryon acoustic oscillation scales, imprinting the corresponding features in the total angular power spectrum. The interested reader can find a study of the effects of relative velocities in the 21 cm IM power spectrum in the pre-reionization era, but after the first stars formed (hence at lower redshifts that those we are focused on) in e.g., \cite{Visbal_streaming}. We expect a similar qualitative behaviour for the case of PBHs at larger redshifts.

Finally, we have assumed an average value for the bias between the seeds and the dark matter distribution, while in reality its value might change with redshift and the mass of the seeds. However, given that the value of the bias is strictly related to the height of the peaks in the density field, it is also directly connected to the PBH initial mass and number. In principle, given that it affects the two-halo term contribution but not the one-halo term, variations of the bias would cause a slightly different signal, but we do not expect the final result of this paper to be substantially different.

Nonetheless, the impact of these assumptions and simplifications on  the final power spectrum in the scenario under study are subdominant, given the magnitude of the uncertainties due to the PBH parameters. 
On top of this, we assume that a comprehensive characterization of the foregrounds which affect the detectability of the signal is possible, hence they do not affect the $S/N$ or the forecast uncertainties in the PBH parameters reported in \refsec{detect}. 
In addition, we have considered that there is no other exotic energy injection during the dark ages and that star formation begins at $z\lesssim 30$, although this might not be the case. In such cases, the identification of a signal as the product of the IGM heating due to the PBHs would be more difficult. A comprehensive study of these effects using simulations and radiative transfer codes to account for PBH distribution, clustering, relative velocities, gas accretion, mergers, and/or extended mass distributions of the PBHs as well as an estimation of how removing the foreground wedge or an early star formation affects the detectability is left for future work. 

There are previous proposals to identify the seeds of the SMBHs from their observational signatures, e.g., with the 21 cm IM sky-averaged signals at $10\lesssim z~\lesssim 30$~\cite{Tanaka_smbh21cm} to distinguish between black holes formed from remnants of Population III stars or DCBHs (although the signal at larger redshifts might also come from PBHs) or with spectral distortions, to ascertain if the seeds are primordial~\cite{Kohri_smbh}. DCBHs are also one of the preferred candidates to explain the power spectrum of the Near Infrared Background and its cross correlation with the cosmic X-ray radiation, both at large scales~\cite{Yue_dcbhNIRB}. The emerging spectrum from the DCBH environment is non-zero only in this window~\cite{Pacucci_spectrum}, which may be useful for identifying  them with Chandra\footnote{http://cxc.harvard.edu/} or Athena\footnote{http://www.the-athena-x-ray-observatory.eu/}. Moreover, while the growth of remnants of Population III stars may remain undetectable for JWST\footnote{https://www.jwst.nasa.gov/}, the evolved stages of DCBHs might be identifiable~\cite{Natarajan_dcbhJWST}. However, these signatures might also have been produced by massive PBHs. We leave the study of this scenario for future work.  With advanced  gravitational wave detectors, such as LISA\footnote{https://lisa.nasa.gov/},  it will be possible to measure gravitational waves created in mergers of SMBHs at large redshifts, which will offer insights on the environments and history of such black holes and help to discriminate among the different candidates for being the seeds.

The advent of new experiments and corresponding observations will shed light on how SMBHs reached such huge masses and on the nature of  the massive seeds needed to explain their existence.
It is also possible that the three kinds of seeds discussed above coexist and give different signatures. We eagerly await   observations that will open the window toward higher redshifts and will give us the opportunity to improve our understanding of some of  the most extreme structures in the Universe.

\acknowledgements
We thank Andrea Ferrara for carefully reading the manuscript and for insightful comments.
We also thank Ely Kovetz, Julian B. Mu\~noz, Yacine Ali-Ha\"imoud and Mark Kamionkowski for useful discussions during the development of this work and Alkistis Pourtsidou, Philip Bull and Guido d'Amico for comments on the final version of the manuscript.  
Funding for this work was partially provided by the Spanish MINECO under projects AYA2014-58747-P AEI/FEDER UE and MDM-2014-0369 of ICCUB (Unidad de Excelencia Maria de Maeztu).
JLB is supported by the Spanish MINECO under grant BES-2015-071307, co-funded by the ESF and thanks the Royal Observatory of Edinburgh for hospitality. AR has received funding from the People Programme (Marie Curie Actions) of the European Union H2020 Programme under REA grant agreement number 706896 (COSMOFLAGS). LV acknowledges support of  European Union's Horizon 2020 research and innovation programme ERC (BePreSySe, grant agreement 725327) and thanks the Radcliffe  Institute for Advanced Study at Harvard University for hospitality.

\bibliography{biblio}

\begin{thebibliography}{150}
\expandafter\ifx\csname natexlab\endcsname\relax\def\natexlab#1{#1}\fi
\expandafter\ifx\csname bibnamefont\endcsname\relax
  \def\bibnamefont#1{#1}\fi
\expandafter\ifx\csname bibfnamefont\endcsname\relax
  \def\bibfnamefont#1{#1}\fi
\expandafter\ifx\csname citenamefont\endcsname\relax
  \def\citenamefont#1{#1}\fi
\expandafter\ifx\csname url\endcsname\relax
  \def\url#1{\texttt{#1}}\fi
\expandafter\ifx\csname urlprefix\endcsname\relax\def\urlprefix{URL }\fi
\providecommand{\bibinfo}[2]{#2}
\providecommand{\eprint}[2][]{\url{#2}}

\bibitem[{\citenamefont{Silk}(1967)}]{Silk:1967a}
\bibinfo{author}{\bibfnamefont{J.}~\bibnamefont{Silk}},
  \bibinfo{journal}{Nature} \textbf{\bibinfo{volume}{215}},
  \bibinfo{pages}{1155} (\bibinfo{year}{1967}).

\bibitem[{\citenamefont{Silk}(1968)}]{Silk:1967b}
\bibinfo{author}{\bibfnamefont{J.}~\bibnamefont{Silk}},
  \bibinfo{journal}{Astrophys. J.} \textbf{\bibinfo{volume}{151}},
  \bibinfo{pages}{459} (\bibinfo{year}{1968}).

\bibitem[{\citenamefont{{Carr} and {Silk}}(1983)}]{Carr_graininess}
\bibinfo{author}{\bibfnamefont{B.~J.} \bibnamefont{{Carr}}} \bibnamefont{and}
  \bibinfo{author}{\bibfnamefont{J.}~\bibnamefont{{Silk}}},
  \bibinfo{journal}{\apj} \textbf{\bibinfo{volume}{268}}, \bibinfo{pages}{1}
  (\bibinfo{year}{1983}).

\bibitem[{\citenamefont{{Kormendy} and {Ho}}(2013)}]{Kormendy_smbh}
\bibinfo{author}{\bibfnamefont{J.}~\bibnamefont{{Kormendy}}} \bibnamefont{and}
  \bibinfo{author}{\bibfnamefont{L.~C.} \bibnamefont{{Ho}}},
  \bibinfo{journal}{\araa} \textbf{\bibinfo{volume}{51}}, \bibinfo{pages}{511}
  (\bibinfo{year}{2013}), \eprint{1304.7762}.

\bibitem[{\citenamefont{{Ba{\~n}ados} et~al.}(2017)\citenamefont{{Ba{\~n}ados},
  {Venemans}, {Mazzucchelli}, {Farina}, {Walter}, {Wang}, {Decarli}, {Stern},
  {Fan}, {Davies} et~al.}}]{Bañados_SMBH}
\bibinfo{author}{\bibfnamefont{E.}~\bibnamefont{{Ba{\~n}ados}}},
  \bibinfo{author}{\bibfnamefont{B.~P.} \bibnamefont{{Venemans}}},
  \bibinfo{author}{\bibfnamefont{C.}~\bibnamefont{{Mazzucchelli}}},
  \bibinfo{author}{\bibfnamefont{E.~P.} \bibnamefont{{Farina}}},
  \bibinfo{author}{\bibfnamefont{F.}~\bibnamefont{{Walter}}},
  \bibinfo{author}{\bibfnamefont{F.}~\bibnamefont{{Wang}}},
  \bibinfo{author}{\bibfnamefont{R.}~\bibnamefont{{Decarli}}},
  \bibinfo{author}{\bibfnamefont{D.}~\bibnamefont{{Stern}}},
  \bibinfo{author}{\bibfnamefont{X.}~\bibnamefont{{Fan}}},
  \bibinfo{author}{\bibfnamefont{F.}~\bibnamefont{{Davies}}},
  \bibnamefont{et~al.}, \bibinfo{journal}{ArXiv e-prints}
  (\bibinfo{year}{2017}), \eprint{1712.01860}.

\bibitem[{\citenamefont{{Fan} et~al.}(2001)\citenamefont{{Fan}, {Narayanan},
  {Lupton}, {Strauss}, {Knapp}, {Becker}, {White}, {Pentericci}, {Leggett},
  {Haiman} et~al.}}]{Fan_quasar}
\bibinfo{author}{\bibfnamefont{X.}~\bibnamefont{{Fan}}},
  \bibinfo{author}{\bibfnamefont{V.~K.} \bibnamefont{{Narayanan}}},
  \bibinfo{author}{\bibfnamefont{R.~H.} \bibnamefont{{Lupton}}},
  \bibinfo{author}{\bibfnamefont{M.~A.} \bibnamefont{{Strauss}}},
  \bibinfo{author}{\bibfnamefont{G.~R.} \bibnamefont{{Knapp}}},
  \bibinfo{author}{\bibfnamefont{R.~H.} \bibnamefont{{Becker}}},
  \bibinfo{author}{\bibfnamefont{R.~L.} \bibnamefont{{White}}},
  \bibinfo{author}{\bibfnamefont{L.}~\bibnamefont{{Pentericci}}},
  \bibinfo{author}{\bibfnamefont{S.~K.} \bibnamefont{{Leggett}}},
  \bibinfo{author}{\bibfnamefont{Z.}~\bibnamefont{{Haiman}}},
  \bibnamefont{et~al.}, \bibinfo{journal}{\aj} \textbf{\bibinfo{volume}{122}},
  \bibinfo{pages}{2833} (\bibinfo{year}{2001}), \eprint{astro-ph/0108063}.

\bibitem[{\citenamefont{{Wu} et~al.}(2015)\citenamefont{{Wu}, {Wang}, {Fan},
  {Yi}, {Zuo}, {Bian}, {Jiang}, {McGreer}, {Wang}, {Yang} et~al.}}]{Wu_quasar}
\bibinfo{author}{\bibfnamefont{X.-B.} \bibnamefont{{Wu}}},
  \bibinfo{author}{\bibfnamefont{F.}~\bibnamefont{{Wang}}},
  \bibinfo{author}{\bibfnamefont{X.}~\bibnamefont{{Fan}}},
  \bibinfo{author}{\bibfnamefont{W.}~\bibnamefont{{Yi}}},
  \bibinfo{author}{\bibfnamefont{W.}~\bibnamefont{{Zuo}}},
  \bibinfo{author}{\bibfnamefont{F.}~\bibnamefont{{Bian}}},
  \bibinfo{author}{\bibfnamefont{L.}~\bibnamefont{{Jiang}}},
  \bibinfo{author}{\bibfnamefont{I.~D.} \bibnamefont{{McGreer}}},
  \bibinfo{author}{\bibfnamefont{R.}~\bibnamefont{{Wang}}},
  \bibinfo{author}{\bibfnamefont{J.}~\bibnamefont{{Yang}}},
  \bibnamefont{et~al.}, \bibinfo{journal}{\nat} \textbf{\bibinfo{volume}{518}},
  \bibinfo{pages}{512} (\bibinfo{year}{2015}), \eprint{1502.07418}.

\bibitem[{\citenamefont{{Mortlock} et~al.}(2011)\citenamefont{{Mortlock},
  {Warren}, {Venemans}, {Patel}, {Hewett}, {McMahon}, {Simpson}, {Theuns},
  {Gonz{\'a}les-Solares}, {Adamson} et~al.}}]{Mortlock_quasar}
\bibinfo{author}{\bibfnamefont{D.~J.} \bibnamefont{{Mortlock}}},
  \bibinfo{author}{\bibfnamefont{S.~J.} \bibnamefont{{Warren}}},
  \bibinfo{author}{\bibfnamefont{B.~P.} \bibnamefont{{Venemans}}},
  \bibinfo{author}{\bibfnamefont{M.}~\bibnamefont{{Patel}}},
  \bibinfo{author}{\bibfnamefont{P.~C.} \bibnamefont{{Hewett}}},
  \bibinfo{author}{\bibfnamefont{R.~G.} \bibnamefont{{McMahon}}},
  \bibinfo{author}{\bibfnamefont{C.}~\bibnamefont{{Simpson}}},
  \bibinfo{author}{\bibfnamefont{T.}~\bibnamefont{{Theuns}}},
  \bibinfo{author}{\bibfnamefont{E.~A.} \bibnamefont{{Gonz{\'a}les-Solares}}},
  \bibinfo{author}{\bibfnamefont{A.}~\bibnamefont{{Adamson}}},
  \bibnamefont{et~al.}, \bibinfo{journal}{\nat} \textbf{\bibinfo{volume}{474}},
  \bibinfo{pages}{616} (\bibinfo{year}{2011}), \eprint{1106.6088}.

\bibitem[{\citenamefont{{Smith} et~al.}(2017)\citenamefont{{Smith}, {Bromm},
  and {Loeb}}}]{Smith_firstsmbh}
\bibinfo{author}{\bibfnamefont{A.}~\bibnamefont{{Smith}}},
  \bibinfo{author}{\bibfnamefont{V.}~\bibnamefont{{Bromm}}}, \bibnamefont{and}
  \bibinfo{author}{\bibfnamefont{A.}~\bibnamefont{{Loeb}}},
  \bibinfo{journal}{Astronomy and Geophysics} \textbf{\bibinfo{volume}{58}},
  \bibinfo{pages}{3.22} (\bibinfo{year}{2017}), \eprint{1703.03083}.

\bibitem[{\citenamefont{{Oka} et~al.}(2016)\citenamefont{{Oka}, {Mizuno},
  {Miura}, and {Takekawa}}}]{oka_imbh1}
\bibinfo{author}{\bibfnamefont{T.}~\bibnamefont{{Oka}}},
  \bibinfo{author}{\bibfnamefont{R.}~\bibnamefont{{Mizuno}}},
  \bibinfo{author}{\bibfnamefont{K.}~\bibnamefont{{Miura}}}, \bibnamefont{and}
  \bibinfo{author}{\bibfnamefont{S.}~\bibnamefont{{Takekawa}}},
  \bibinfo{journal}{\apjl} \textbf{\bibinfo{volume}{816}}, \bibinfo{eid}{L7}
  (\bibinfo{year}{2016}), \eprint{1512.04661}.

\bibitem[{\citenamefont{{Oka} et~al.}(2017)\citenamefont{{Oka}, {Tsujimoto},
  {Iwata}, {Nomura}, and {Takekawa}}}]{oka_imbh2}
\bibinfo{author}{\bibfnamefont{T.}~\bibnamefont{{Oka}}},
  \bibinfo{author}{\bibfnamefont{S.}~\bibnamefont{{Tsujimoto}}},
  \bibinfo{author}{\bibfnamefont{Y.}~\bibnamefont{{Iwata}}},
  \bibinfo{author}{\bibfnamefont{M.}~\bibnamefont{{Nomura}}}, \bibnamefont{and}
  \bibinfo{author}{\bibfnamefont{S.}~\bibnamefont{{Takekawa}}},
  \bibinfo{journal}{ArXiv e-prints}  (\bibinfo{year}{2017}),
  \eprint{1707.07603}.

\bibitem[{\citenamefont{{Silk}}(2017)}]{silk_dwarf}
\bibinfo{author}{\bibfnamefont{J.}~\bibnamefont{{Silk}}},
  \bibinfo{journal}{Astrophys. J. Letters} \textbf{\bibinfo{volume}{839}},
  \bibinfo{eid}{L13} (\bibinfo{year}{2017}), \eprint{1703.08553}.

\bibitem[{\citenamefont{{Pacucci}
  et~al.}(2017{\natexlab{a}})\citenamefont{{Pacucci}, {Natarajan}, {Volonteri},
  {Cappelluti}, and {Urry}}}]{Pacucci_growth}
\bibinfo{author}{\bibfnamefont{F.}~\bibnamefont{{Pacucci}}},
  \bibinfo{author}{\bibfnamefont{P.}~\bibnamefont{{Natarajan}}},
  \bibinfo{author}{\bibfnamefont{M.}~\bibnamefont{{Volonteri}}},
  \bibinfo{author}{\bibfnamefont{N.}~\bibnamefont{{Cappelluti}}},
  \bibnamefont{and} \bibinfo{author}{\bibfnamefont{C.~M.}
  \bibnamefont{{Urry}}}, \bibinfo{journal}{ArXiv e-prints}
  (\bibinfo{year}{2017}{\natexlab{a}}), \eprint{1710.09375}.

\bibitem[{\citenamefont{{Pacucci}
  et~al.}(2015{\natexlab{a}})\citenamefont{{Pacucci}, {Volonteri}, and
  {Ferrara}}}]{Pacucci_growth2}
\bibinfo{author}{\bibfnamefont{F.}~\bibnamefont{{Pacucci}}},
  \bibinfo{author}{\bibfnamefont{M.}~\bibnamefont{{Volonteri}}},
  \bibnamefont{and}
  \bibinfo{author}{\bibfnamefont{A.}~\bibnamefont{{Ferrara}}},
  \bibinfo{journal}{\mnras} \textbf{\bibinfo{volume}{452}},
  \bibinfo{pages}{1922} (\bibinfo{year}{2015}{\natexlab{a}}),
  \eprint{1506.04750}.

\bibitem[{\citenamefont{{Pacucci}
  et~al.}(2017{\natexlab{b}})\citenamefont{{Pacucci}, {Natarajan}, and
  {Ferrara}}}]{Pacucci_limit}
\bibinfo{author}{\bibfnamefont{F.}~\bibnamefont{{Pacucci}}},
  \bibinfo{author}{\bibfnamefont{P.}~\bibnamefont{{Natarajan}}},
  \bibnamefont{and}
  \bibinfo{author}{\bibfnamefont{A.}~\bibnamefont{{Ferrara}}},
  \bibinfo{journal}{\apjl} \textbf{\bibinfo{volume}{835}}, \bibinfo{eid}{L36}
  (\bibinfo{year}{2017}{\natexlab{b}}), \eprint{1701.06565}.

\bibitem[{\citenamefont{{Volonteri}}(2010)}]{Volonteri_smbhorigin}
\bibinfo{author}{\bibfnamefont{M.}~\bibnamefont{{Volonteri}}},
  \bibinfo{journal}{\aapr} \textbf{\bibinfo{volume}{18}}, \bibinfo{pages}{279}
  (\bibinfo{year}{2010}), \eprint{1003.4404}.

\bibitem[{\citenamefont{{Latif} and {Ferrara}}(2016)}]{Latif_seeds}
\bibinfo{author}{\bibfnamefont{M.~A.} \bibnamefont{{Latif}}} \bibnamefont{and}
  \bibinfo{author}{\bibfnamefont{A.}~\bibnamefont{{Ferrara}}},
  \bibinfo{journal}{Pub. of the Astron. Soc. of Australia}
  \textbf{\bibinfo{volume}{33}}, \bibinfo{eid}{e051} (\bibinfo{year}{2016}),
  \eprint{1605.07391}.

\bibitem[{\citenamefont{{D'Amico} et~al.}(2018)\citenamefont{{D'Amico},
  {Panci}, {Lupi}, {Bovino}, and {Silk}}}]{dAmico_dissDM}
\bibinfo{author}{\bibfnamefont{G.}~\bibnamefont{{D'Amico}}},
  \bibinfo{author}{\bibfnamefont{P.}~\bibnamefont{{Panci}}},
  \bibinfo{author}{\bibfnamefont{A.}~\bibnamefont{{Lupi}}},
  \bibinfo{author}{\bibfnamefont{S.}~\bibnamefont{{Bovino}}}, \bibnamefont{and}
  \bibinfo{author}{\bibfnamefont{J.}~\bibnamefont{{Silk}}},
  \bibinfo{journal}{\mnras} \textbf{\bibinfo{volume}{473}},
  \bibinfo{pages}{328} (\bibinfo{year}{2018}), \eprint{1707.03419}.

\bibitem[{\citenamefont{{Madau} and {Rees}}(2001)}]{Madau_pop3}
\bibinfo{author}{\bibfnamefont{P.}~\bibnamefont{{Madau}}} \bibnamefont{and}
  \bibinfo{author}{\bibfnamefont{M.~J.} \bibnamefont{{Rees}}},
  \bibinfo{journal}{\apjl} \textbf{\bibinfo{volume}{551}}, \bibinfo{pages}{L27}
  (\bibinfo{year}{2001}), \eprint{astro-ph/0101223}.

\bibitem[{\citenamefont{{Tanaka} and {Haiman}}(2009)}]{Tanaka_seedsaccretion}
\bibinfo{author}{\bibfnamefont{T.}~\bibnamefont{{Tanaka}}} \bibnamefont{and}
  \bibinfo{author}{\bibfnamefont{Z.}~\bibnamefont{{Haiman}}},
  \bibinfo{journal}{\apj} \textbf{\bibinfo{volume}{696}}, \bibinfo{pages}{1798}
  (\bibinfo{year}{2009}), \eprint{0807.4702}.

\bibitem[{\citenamefont{{Li} et~al.}(2007)\citenamefont{{Li}, {Hernquist},
  {Robertson}, {Cox}, {Hopkins}, {Springel}, {Gao}, {Di Matteo}, {Zentner},
  {Jenkins} et~al.}}]{Li_bhmergers}
\bibinfo{author}{\bibfnamefont{Y.}~\bibnamefont{{Li}}},
  \bibinfo{author}{\bibfnamefont{L.}~\bibnamefont{{Hernquist}}},
  \bibinfo{author}{\bibfnamefont{B.}~\bibnamefont{{Robertson}}},
  \bibinfo{author}{\bibfnamefont{T.~J.} \bibnamefont{{Cox}}},
  \bibinfo{author}{\bibfnamefont{P.~F.} \bibnamefont{{Hopkins}}},
  \bibinfo{author}{\bibfnamefont{V.}~\bibnamefont{{Springel}}},
  \bibinfo{author}{\bibfnamefont{L.}~\bibnamefont{{Gao}}},
  \bibinfo{author}{\bibfnamefont{T.}~\bibnamefont{{Di Matteo}}},
  \bibinfo{author}{\bibfnamefont{A.~R.} \bibnamefont{{Zentner}}},
  \bibinfo{author}{\bibfnamefont{A.}~\bibnamefont{{Jenkins}}},
  \bibnamefont{et~al.}, \bibinfo{journal}{\apj} \textbf{\bibinfo{volume}{665}},
  \bibinfo{pages}{187} (\bibinfo{year}{2007}), \eprint{astro-ph/0608190}.

\bibitem[{\citenamefont{{Madau} et~al.}(2014)\citenamefont{{Madau}, {Haardt},
  and {Dotti}}}]{Madau_superedd}
\bibinfo{author}{\bibfnamefont{P.}~\bibnamefont{{Madau}}},
  \bibinfo{author}{\bibfnamefont{F.}~\bibnamefont{{Haardt}}}, \bibnamefont{and}
  \bibinfo{author}{\bibfnamefont{M.}~\bibnamefont{{Dotti}}},
  \bibinfo{journal}{\apjl} \textbf{\bibinfo{volume}{784}}, \bibinfo{eid}{L38}
  (\bibinfo{year}{2014}), \eprint{1402.6995}.

\bibitem[{\citenamefont{{Lupi} et~al.}(2016)\citenamefont{{Lupi}, {Haardt},
  {Dotti}, {Fiacconi}, {Mayer}, and {Madau}}}]{Lupi_superedd}
\bibinfo{author}{\bibfnamefont{A.}~\bibnamefont{{Lupi}}},
  \bibinfo{author}{\bibfnamefont{F.}~\bibnamefont{{Haardt}}},
  \bibinfo{author}{\bibfnamefont{M.}~\bibnamefont{{Dotti}}},
  \bibinfo{author}{\bibfnamefont{D.}~\bibnamefont{{Fiacconi}}},
  \bibinfo{author}{\bibfnamefont{L.}~\bibnamefont{{Mayer}}}, \bibnamefont{and}
  \bibinfo{author}{\bibfnamefont{P.}~\bibnamefont{{Madau}}},
  \bibinfo{journal}{\mnras} \textbf{\bibinfo{volume}{456}},
  \bibinfo{pages}{2993} (\bibinfo{year}{2016}), \eprint{1512.02651}.

\bibitem[{\citenamefont{{Inayoshi} et~al.}(2016)\citenamefont{{Inayoshi},
  {Haiman}, and {Ostriker}}}]{Inayoshi_superedd}
\bibinfo{author}{\bibfnamefont{K.}~\bibnamefont{{Inayoshi}}},
  \bibinfo{author}{\bibfnamefont{Z.}~\bibnamefont{{Haiman}}}, \bibnamefont{and}
  \bibinfo{author}{\bibfnamefont{J.~P.} \bibnamefont{{Ostriker}}},
  \bibinfo{journal}{\mnras} \textbf{\bibinfo{volume}{459}},
  \bibinfo{pages}{3738} (\bibinfo{year}{2016}), \eprint{1511.02116}.

\bibitem[{\citenamefont{{Baldassare} et~al.}(2015)\citenamefont{{Baldassare},
  {Reines}, {Gallo}, and {Greene}}}]{Baldassare_imbh}
\bibinfo{author}{\bibfnamefont{V.~F.} \bibnamefont{{Baldassare}}},
  \bibinfo{author}{\bibfnamefont{A.~E.} \bibnamefont{{Reines}}},
  \bibinfo{author}{\bibfnamefont{E.}~\bibnamefont{{Gallo}}}, \bibnamefont{and}
  \bibinfo{author}{\bibfnamefont{J.~E.} \bibnamefont{{Greene}}},
  \bibinfo{journal}{\apjl} \textbf{\bibinfo{volume}{809}}, \bibinfo{eid}{L14}
  (\bibinfo{year}{2015}), \eprint{1506.07531}.

\bibitem[{\citenamefont{{Mezcua} et~al.}(2016)\citenamefont{{Mezcua}, {Civano},
  {Fabbiano}, {Miyaji}, and {Marchesi}}}]{Mezcua_imbh}
\bibinfo{author}{\bibfnamefont{M.}~\bibnamefont{{Mezcua}}},
  \bibinfo{author}{\bibfnamefont{F.}~\bibnamefont{{Civano}}},
  \bibinfo{author}{\bibfnamefont{G.}~\bibnamefont{{Fabbiano}}},
  \bibinfo{author}{\bibfnamefont{T.}~\bibnamefont{{Miyaji}}}, \bibnamefont{and}
  \bibinfo{author}{\bibfnamefont{S.}~\bibnamefont{{Marchesi}}},
  \bibinfo{journal}{\apj} \textbf{\bibinfo{volume}{817}}, \bibinfo{eid}{20}
  (\bibinfo{year}{2016}), \eprint{1511.05844}.

\bibitem[{\citenamefont{{Salvaterra} et~al.}(2012)\citenamefont{{Salvaterra},
  {Haardt}, {Volonteri}, and {Moretti}}}]{Salvaterra_xray}
\bibinfo{author}{\bibfnamefont{R.}~\bibnamefont{{Salvaterra}}},
  \bibinfo{author}{\bibfnamefont{F.}~\bibnamefont{{Haardt}}},
  \bibinfo{author}{\bibfnamefont{M.}~\bibnamefont{{Volonteri}}},
  \bibnamefont{and}
  \bibinfo{author}{\bibfnamefont{A.}~\bibnamefont{{Moretti}}},
  \bibinfo{journal}{\aap} \textbf{\bibinfo{volume}{545}}, \bibinfo{eid}{L6}
  (\bibinfo{year}{2012}), \eprint{1209.1095}.

\bibitem[{\citenamefont{{Dijkstra} et~al.}(2004)\citenamefont{{Dijkstra},
  {Haiman}, and {Loeb}}}]{Dijkstra04}
\bibinfo{author}{\bibfnamefont{M.}~\bibnamefont{{Dijkstra}}},
  \bibinfo{author}{\bibfnamefont{Z.}~\bibnamefont{{Haiman}}}, \bibnamefont{and}
  \bibinfo{author}{\bibfnamefont{A.}~\bibnamefont{{Loeb}}},
  \bibinfo{journal}{\apj} \textbf{\bibinfo{volume}{613}}, \bibinfo{pages}{646}
  (\bibinfo{year}{2004}), \eprint{astro-ph/0403078}.

\bibitem[{\citenamefont{{Bromm} and {Loeb}}(2003)}]{Bromm_dcbh}
\bibinfo{author}{\bibfnamefont{V.}~\bibnamefont{{Bromm}}} \bibnamefont{and}
  \bibinfo{author}{\bibfnamefont{A.}~\bibnamefont{{Loeb}}},
  \bibinfo{journal}{\apj} \textbf{\bibinfo{volume}{596}}, \bibinfo{pages}{34}
  (\bibinfo{year}{2003}), \eprint{astro-ph/0212400}.

\bibitem[{\citenamefont{{Begelman} et~al.}(2006)\citenamefont{{Begelman},
  {Volonteri}, and {Rees}}}]{Begelman_dcbh}
\bibinfo{author}{\bibfnamefont{M.~C.} \bibnamefont{{Begelman}}},
  \bibinfo{author}{\bibfnamefont{M.}~\bibnamefont{{Volonteri}}},
  \bibnamefont{and} \bibinfo{author}{\bibfnamefont{M.~J.}
  \bibnamefont{{Rees}}}, \bibinfo{journal}{\mnras}
  \textbf{\bibinfo{volume}{370}}, \bibinfo{pages}{289} (\bibinfo{year}{2006}),
  \eprint{astro-ph/0602363}.

\bibitem[{\citenamefont{{Lodato} and {Natarajan}}(2006)}]{Lodato_dcbh}
\bibinfo{author}{\bibfnamefont{G.}~\bibnamefont{{Lodato}}} \bibnamefont{and}
  \bibinfo{author}{\bibfnamefont{P.}~\bibnamefont{{Natarajan}}},
  \bibinfo{journal}{\mnras} \textbf{\bibinfo{volume}{371}},
  \bibinfo{pages}{1813} (\bibinfo{year}{2006}), \eprint{astro-ph/0606159}.

\bibitem[{\citenamefont{{Choi} et~al.}(2015)\citenamefont{{Choi}, {Shlosman},
  and {Begelman}}}]{Choi_dcbh}
\bibinfo{author}{\bibfnamefont{J.-H.} \bibnamefont{{Choi}}},
  \bibinfo{author}{\bibfnamefont{I.}~\bibnamefont{{Shlosman}}},
  \bibnamefont{and} \bibinfo{author}{\bibfnamefont{M.~C.}
  \bibnamefont{{Begelman}}}, \bibinfo{journal}{\mnras}
  \textbf{\bibinfo{volume}{450}}, \bibinfo{pages}{4411} (\bibinfo{year}{2015}),
  \eprint{1412.2761}.

\bibitem[{\citenamefont{{Volonteri} et~al.}(2008)\citenamefont{{Volonteri},
  {Lodato}, and {Natarajan}}}]{Volonteri_evolseeds}
\bibinfo{author}{\bibfnamefont{M.}~\bibnamefont{{Volonteri}}},
  \bibinfo{author}{\bibfnamefont{G.}~\bibnamefont{{Lodato}}}, \bibnamefont{and}
  \bibinfo{author}{\bibfnamefont{P.}~\bibnamefont{{Natarajan}}},
  \bibinfo{journal}{\mnras} \textbf{\bibinfo{volume}{383}},
  \bibinfo{pages}{1079} (\bibinfo{year}{2008}), \eprint{0709.0529}.

\bibitem[{\citenamefont{{Johnson} et~al.}(2012)\citenamefont{{Johnson},
  {Whalen}, {Fryer}, and {Li}}}]{Johnson_dcbh}
\bibinfo{author}{\bibfnamefont{J.~L.} \bibnamefont{{Johnson}}},
  \bibinfo{author}{\bibfnamefont{D.~J.} \bibnamefont{{Whalen}}},
  \bibinfo{author}{\bibfnamefont{C.~L.} \bibnamefont{{Fryer}}},
  \bibnamefont{and} \bibinfo{author}{\bibfnamefont{H.}~\bibnamefont{{Li}}},
  \bibinfo{journal}{\apj} \textbf{\bibinfo{volume}{750}}, \bibinfo{eid}{66}
  (\bibinfo{year}{2012}), \eprint{1112.2726}.

\bibitem[{\citenamefont{{Agarwal} et~al.}(2014)\citenamefont{{Agarwal}, {Dalla
  Vecchia}, {Johnson}, {Khochfar}, and {Paardekooper}}}]{Agarwal_dcbh}
\bibinfo{author}{\bibfnamefont{B.}~\bibnamefont{{Agarwal}}},
  \bibinfo{author}{\bibfnamefont{C.}~\bibnamefont{{Dalla Vecchia}}},
  \bibinfo{author}{\bibfnamefont{J.~L.} \bibnamefont{{Johnson}}},
  \bibinfo{author}{\bibfnamefont{S.}~\bibnamefont{{Khochfar}}},
  \bibnamefont{and} \bibinfo{author}{\bibfnamefont{J.-P.}
  \bibnamefont{{Paardekooper}}}, \bibinfo{journal}{\mnras}
  \textbf{\bibinfo{volume}{443}}, \bibinfo{pages}{648} (\bibinfo{year}{2014}),
  \eprint{1403.5267}.

\bibitem[{\citenamefont{{Shang} et~al.}(2010)\citenamefont{{Shang}, {Bryan},
  and {Haiman}}}]{Shang_dcbh}
\bibinfo{author}{\bibfnamefont{C.}~\bibnamefont{{Shang}}},
  \bibinfo{author}{\bibfnamefont{G.~L.} \bibnamefont{{Bryan}}},
  \bibnamefont{and} \bibinfo{author}{\bibfnamefont{Z.}~\bibnamefont{{Haiman}}},
  \bibinfo{journal}{\mnras} \textbf{\bibinfo{volume}{402}},
  \bibinfo{pages}{1249} (\bibinfo{year}{2010}), \eprint{0906.4773}.

\bibitem[{\citenamefont{{Regan} et~al.}(2017)\citenamefont{{Regan}, {Visbal},
  {Wise}, {Haiman}, {Johansson}, and {Bryan}}}]{Regan_dcbh}
\bibinfo{author}{\bibfnamefont{J.~A.} \bibnamefont{{Regan}}},
  \bibinfo{author}{\bibfnamefont{E.}~\bibnamefont{{Visbal}}},
  \bibinfo{author}{\bibfnamefont{J.~H.} \bibnamefont{{Wise}}},
  \bibinfo{author}{\bibfnamefont{Z.}~\bibnamefont{{Haiman}}},
  \bibinfo{author}{\bibfnamefont{P.~H.} \bibnamefont{{Johansson}}},
  \bibnamefont{and} \bibinfo{author}{\bibfnamefont{G.~L.}
  \bibnamefont{{Bryan}}}, \bibinfo{journal}{Nature Astronomy}
  \textbf{\bibinfo{volume}{1}}, \bibinfo{eid}{0075} (\bibinfo{year}{2017}),
  \eprint{1703.03805}.

\bibitem[{\citenamefont{{Yue} et~al.}(2017)\citenamefont{{Yue}, {Ferrara},
  {Pacucci}, and {Omukai}}}]{Yue_dcbh}
\bibinfo{author}{\bibfnamefont{B.}~\bibnamefont{{Yue}}},
  \bibinfo{author}{\bibfnamefont{A.}~\bibnamefont{{Ferrara}}},
  \bibinfo{author}{\bibfnamefont{F.}~\bibnamefont{{Pacucci}}},
  \bibnamefont{and} \bibinfo{author}{\bibfnamefont{K.}~\bibnamefont{{Omukai}}},
  \bibinfo{journal}{\apj} \textbf{\bibinfo{volume}{838}}, \bibinfo{eid}{111}
  (\bibinfo{year}{2017}), \eprint{1612.07885}.

\bibitem[{\citenamefont{{Yue} et~al.}(2013)\citenamefont{{Yue}, {Ferrara},
  {Salvaterra}, {Xu}, and {Chen}}}]{Yue_dcbhNIRB}
\bibinfo{author}{\bibfnamefont{B.}~\bibnamefont{{Yue}}},
  \bibinfo{author}{\bibfnamefont{A.}~\bibnamefont{{Ferrara}}},
  \bibinfo{author}{\bibfnamefont{R.}~\bibnamefont{{Salvaterra}}},
  \bibinfo{author}{\bibfnamefont{Y.}~\bibnamefont{{Xu}}}, \bibnamefont{and}
  \bibinfo{author}{\bibfnamefont{X.}~\bibnamefont{{Chen}}},
  \bibinfo{journal}{\mnras} \textbf{\bibinfo{volume}{433}},
  \bibinfo{pages}{1556} (\bibinfo{year}{2013}), \eprint{1305.5177}.

\bibitem[{\citenamefont{{Pacucci} and {Ferrara}}(2015)}]{Pacucci_dcbh}
\bibinfo{author}{\bibfnamefont{F.}~\bibnamefont{{Pacucci}}} \bibnamefont{and}
  \bibinfo{author}{\bibfnamefont{A.}~\bibnamefont{{Ferrara}}},
  \bibinfo{journal}{\mnras} \textbf{\bibinfo{volume}{448}},
  \bibinfo{pages}{104} (\bibinfo{year}{2015}), \eprint{1501.00989}.

\bibitem[{\citenamefont{{Pacucci} et~al.}(2016)\citenamefont{{Pacucci},
  {Ferrara}, {Grazian}, {Fiore}, {Giallongo}, and {Puccetti}}}]{Pacucci_seeds}
\bibinfo{author}{\bibfnamefont{F.}~\bibnamefont{{Pacucci}}},
  \bibinfo{author}{\bibfnamefont{A.}~\bibnamefont{{Ferrara}}},
  \bibinfo{author}{\bibfnamefont{A.}~\bibnamefont{{Grazian}}},
  \bibinfo{author}{\bibfnamefont{F.}~\bibnamefont{{Fiore}}},
  \bibinfo{author}{\bibfnamefont{E.}~\bibnamefont{{Giallongo}}},
  \bibnamefont{and}
  \bibinfo{author}{\bibfnamefont{S.}~\bibnamefont{{Puccetti}}},
  \bibinfo{journal}{\mnras} \textbf{\bibinfo{volume}{459}},
  \bibinfo{pages}{1432} (\bibinfo{year}{2016}), \eprint{1603.08522}.

\bibitem[{\citenamefont{{Dijkstra} et~al.}(2014)\citenamefont{{Dijkstra},
  {Ferrara}, and {Mesinger}}}]{Dijkstra_dcbh}
\bibinfo{author}{\bibfnamefont{M.}~\bibnamefont{{Dijkstra}}},
  \bibinfo{author}{\bibfnamefont{A.}~\bibnamefont{{Ferrara}}},
  \bibnamefont{and}
  \bibinfo{author}{\bibfnamefont{A.}~\bibnamefont{{Mesinger}}},
  \bibinfo{journal}{\mnras} \textbf{\bibinfo{volume}{442}},
  \bibinfo{pages}{2036} (\bibinfo{year}{2014}), \eprint{1405.6743}.

\bibitem[{\citenamefont{{Latif} et~al.}(2015)\citenamefont{{Latif}, {Bovino},
  {Grassi}, {Schleicher}, and {Spaans}}}]{Latif_dcbh}
\bibinfo{author}{\bibfnamefont{M.~A.} \bibnamefont{{Latif}}},
  \bibinfo{author}{\bibfnamefont{S.}~\bibnamefont{{Bovino}}},
  \bibinfo{author}{\bibfnamefont{T.}~\bibnamefont{{Grassi}}},
  \bibinfo{author}{\bibfnamefont{D.~R.~G.} \bibnamefont{{Schleicher}}},
  \bibnamefont{and} \bibinfo{author}{\bibfnamefont{M.}~\bibnamefont{{Spaans}}},
  \bibinfo{journal}{\mnras} \textbf{\bibinfo{volume}{446}},
  \bibinfo{pages}{3163} (\bibinfo{year}{2015}), \eprint{1408.3061}.

\bibitem[{\citenamefont{{Habouzit} et~al.}(2016)\citenamefont{{Habouzit},
  {Volonteri}, {Latif}, {Dubois}, and {Peirani}}}]{Habouzit_dcbh}
\bibinfo{author}{\bibfnamefont{M.}~\bibnamefont{{Habouzit}}},
  \bibinfo{author}{\bibfnamefont{M.}~\bibnamefont{{Volonteri}}},
  \bibinfo{author}{\bibfnamefont{M.}~\bibnamefont{{Latif}}},
  \bibinfo{author}{\bibfnamefont{Y.}~\bibnamefont{{Dubois}}}, \bibnamefont{and}
  \bibinfo{author}{\bibfnamefont{S.}~\bibnamefont{{Peirani}}},
  \bibinfo{journal}{\mnras} \textbf{\bibinfo{volume}{463}},
  \bibinfo{pages}{529} (\bibinfo{year}{2016}), \eprint{1601.00557}.

\bibitem[{\citenamefont{Dolgov and Silk}(1993)}]{Dolgov_smbh}
\bibinfo{author}{\bibfnamefont{A.}~\bibnamefont{Dolgov}} \bibnamefont{and}
  \bibinfo{author}{\bibfnamefont{J.}~\bibnamefont{Silk}},
  \bibinfo{journal}{Phys. Rev.} \textbf{\bibinfo{volume}{D47}},
  \bibinfo{pages}{4244} (\bibinfo{year}{1993}).

\bibitem[{\citenamefont{{Kohri} et~al.}(2014)\citenamefont{{Kohri}, {Nakama},
  and {Suyama}}}]{Kohri_smbh}
\bibinfo{author}{\bibfnamefont{K.}~\bibnamefont{{Kohri}}},
  \bibinfo{author}{\bibfnamefont{T.}~\bibnamefont{{Nakama}}}, \bibnamefont{and}
  \bibinfo{author}{\bibfnamefont{T.}~\bibnamefont{{Suyama}}},
  \bibinfo{journal}{Phys. Rev. D} \textbf{\bibinfo{volume}{90}},
  \bibinfo{eid}{083514} (\bibinfo{year}{2014}), \eprint{1405.5999}.

\bibitem[{\citenamefont{{Clesse} and
  {Garc{\'{\i}}a-Bellido}}(2015)}]{Clesse_pbh}
\bibinfo{author}{\bibfnamefont{S.}~\bibnamefont{{Clesse}}} \bibnamefont{and}
  \bibinfo{author}{\bibfnamefont{J.}~\bibnamefont{{Garc{\'{\i}}a-Bellido}}},
  \bibinfo{journal}{\prd} \textbf{\bibinfo{volume}{92}}, \bibinfo{eid}{023524}
  (\bibinfo{year}{2015}), \eprint{1501.07565}.

\bibitem[{\citenamefont{{Kawasaki} et~al.}(2012)\citenamefont{{Kawasaki},
  {Kusenko}, and {Yanagida}}}]{Kawasaki_seeds}
\bibinfo{author}{\bibfnamefont{M.}~\bibnamefont{{Kawasaki}}},
  \bibinfo{author}{\bibfnamefont{A.}~\bibnamefont{{Kusenko}}},
  \bibnamefont{and} \bibinfo{author}{\bibfnamefont{T.~T.}
  \bibnamefont{{Yanagida}}}, \bibinfo{journal}{Physics Letters B}
  \textbf{\bibinfo{volume}{711}}, \bibinfo{pages}{1} (\bibinfo{year}{2012}),
  \eprint{1202.3848}.

\bibitem[{\citenamefont{{Khlopov} et~al.}(2005)\citenamefont{{Khlopov},
  {Rubin}, and {Sakharov}}}]{Khlopov_formmech}
\bibinfo{author}{\bibfnamefont{M.~Y.} \bibnamefont{{Khlopov}}},
  \bibinfo{author}{\bibfnamefont{S.~G.} \bibnamefont{{Rubin}}},
  \bibnamefont{and} \bibinfo{author}{\bibfnamefont{A.~S.}
  \bibnamefont{{Sakharov}}}, \bibinfo{journal}{Astroparticle Physics}
  \textbf{\bibinfo{volume}{23}}, \bibinfo{pages}{265} (\bibinfo{year}{2005}),
  \eprint{astro-ph/0401532}.

\bibitem[{\citenamefont{{Dolgov} et~al.}(2009)\citenamefont{{Dolgov},
  {Kawasaki}, and {Kevlishvili}}}]{Dolgov_formmech}
\bibinfo{author}{\bibfnamefont{A.~D.} \bibnamefont{{Dolgov}}},
  \bibinfo{author}{\bibfnamefont{M.}~\bibnamefont{{Kawasaki}}},
  \bibnamefont{and}
  \bibinfo{author}{\bibfnamefont{N.}~\bibnamefont{{Kevlishvili}}},
  \bibinfo{journal}{Nuclear Physics B} \textbf{\bibinfo{volume}{807}},
  \bibinfo{pages}{229} (\bibinfo{year}{2009}), \eprint{0806.2986}.

\bibitem[{\citenamefont{{Zel'dovich} and {Novikov}}(1967)}]{Zeldovich_pbh}
\bibinfo{author}{\bibfnamefont{Y.~B.} \bibnamefont{{Zel'dovich}}}
  \bibnamefont{and} \bibinfo{author}{\bibfnamefont{I.~D.}
  \bibnamefont{{Novikov}}}, \bibinfo{journal}{Soviet Astronomy}
  \textbf{\bibinfo{volume}{10}}, \bibinfo{pages}{602} (\bibinfo{year}{1967}).

\bibitem[{\citenamefont{{Bird} et~al.}(2016)\citenamefont{{Bird}, {Cholis},
  {Mu{\~n}oz}, {Ali-Ha{\"\i}moud}, {Kamionkowski}, {Kovetz}, {Raccanelli}, and
  {Riess}}}]{Bird_pbh}
\bibinfo{author}{\bibfnamefont{S.}~\bibnamefont{{Bird}}},
  \bibinfo{author}{\bibfnamefont{I.}~\bibnamefont{{Cholis}}},
  \bibinfo{author}{\bibfnamefont{J.~B.} \bibnamefont{{Mu{\~n}oz}}},
  \bibinfo{author}{\bibfnamefont{Y.}~\bibnamefont{{Ali-Ha{\"\i}moud}}},
  \bibinfo{author}{\bibfnamefont{M.}~\bibnamefont{{Kamionkowski}}},
  \bibinfo{author}{\bibfnamefont{E.~D.} \bibnamefont{{Kovetz}}},
  \bibinfo{author}{\bibfnamefont{A.}~\bibnamefont{{Raccanelli}}},
  \bibnamefont{and} \bibinfo{author}{\bibfnamefont{A.~G.}
  \bibnamefont{{Riess}}}, \bibinfo{journal}{Physical Review Letters}
  \textbf{\bibinfo{volume}{116}}, \bibinfo{eid}{201301} (\bibinfo{year}{2016}),
  \eprint{1603.00464}.

\bibitem[{\citenamefont{Abbott et~al.}(2016)}]{abbott:ligo}
\bibinfo{author}{\bibfnamefont{B.~P.} \bibnamefont{Abbott}}
  \bibnamefont{et~al.} (\bibinfo{collaboration}{LIGO Scientific Collaboration
  and Virgo Collaboration}), \bibinfo{journal}{Phys. Rev. Lett.}
  \textbf{\bibinfo{volume}{116}}, \bibinfo{pages}{061102}
  (\bibinfo{year}{2016}), \eprint{1602.03837}.

\bibitem[{\citenamefont{Carr et~al.}(2010)\citenamefont{Carr, Kohri, Sendouda,
  and Yokoyama}}]{Carr:evaporation}
\bibinfo{author}{\bibfnamefont{B.~J.} \bibnamefont{Carr}},
  \bibinfo{author}{\bibfnamefont{K.}~\bibnamefont{Kohri}},
  \bibinfo{author}{\bibfnamefont{Y.}~\bibnamefont{Sendouda}}, \bibnamefont{and}
  \bibinfo{author}{\bibfnamefont{J.}~\bibnamefont{Yokoyama}},
  \bibinfo{journal}{Phys. Rev. D} \textbf{\bibinfo{volume}{81}},
  \bibinfo{pages}{104019} (\bibinfo{year}{2010}), \eprint{0912.5297}.

\bibitem[{\citenamefont{Griest et~al.}(2014)\citenamefont{Griest, Cieplak, and
  Lehner}}]{griest:keplerconstraint}
\bibinfo{author}{\bibfnamefont{K.}~\bibnamefont{Griest}},
  \bibinfo{author}{\bibfnamefont{A.~M.} \bibnamefont{Cieplak}},
  \bibnamefont{and} \bibinfo{author}{\bibfnamefont{M.~J.}
  \bibnamefont{Lehner}}, \bibinfo{journal}{The Astrophysical Journal}
  \textbf{\bibinfo{volume}{786}}, \bibinfo{pages}{158} (\bibinfo{year}{2014}),
  \eprint{1307.5798}.

\bibitem[{\citenamefont{Niikura et~al.}(2017)\citenamefont{Niikura, Takada,
  Yasuda, Lupton, Sumi, More, More, Oguri, and
  Chiba}}]{niikura:microlensingconstraint}
\bibinfo{author}{\bibfnamefont{H.}~\bibnamefont{Niikura}},
  \bibinfo{author}{\bibfnamefont{M.}~\bibnamefont{Takada}},
  \bibinfo{author}{\bibfnamefont{N.}~\bibnamefont{Yasuda}},
  \bibinfo{author}{\bibfnamefont{R.~H.} \bibnamefont{Lupton}},
  \bibinfo{author}{\bibfnamefont{T.}~\bibnamefont{Sumi}},
  \bibinfo{author}{\bibfnamefont{S.}~\bibnamefont{More}},
  \bibinfo{author}{\bibfnamefont{A.}~\bibnamefont{More}},
  \bibinfo{author}{\bibfnamefont{M.}~\bibnamefont{Oguri}}, \bibnamefont{and}
  \bibinfo{author}{\bibfnamefont{M.}~\bibnamefont{Chiba}}
  (\bibinfo{year}{2017}), \eprint{1701.02151}.

\bibitem[{\citenamefont{Gaggero et~al.}(2017)\citenamefont{Gaggero, Bertone,
  Calore, Connors, Lovell, Markoff, and Storm}}]{gaggero}
\bibinfo{author}{\bibfnamefont{D.}~\bibnamefont{Gaggero}},
  \bibinfo{author}{\bibfnamefont{G.}~\bibnamefont{Bertone}},
  \bibinfo{author}{\bibfnamefont{F.}~\bibnamefont{Calore}},
  \bibinfo{author}{\bibfnamefont{R.~M.~T.} \bibnamefont{Connors}},
  \bibinfo{author}{\bibfnamefont{M.}~\bibnamefont{Lovell}},
  \bibinfo{author}{\bibfnamefont{S.}~\bibnamefont{Markoff}}, \bibnamefont{and}
  \bibinfo{author}{\bibfnamefont{E.}~\bibnamefont{Storm}},
  \bibinfo{journal}{Phys. Rev. Lett.} \textbf{\bibinfo{volume}{118}},
  \bibinfo{pages}{241101} (\bibinfo{year}{2017}), \eprint{1612.00457}.

\bibitem[{\citenamefont{Quinn et~al.}(2009)\citenamefont{Quinn, Wilkinson,
  Irwin, Marshall, Koch, and Belokurov}}]{quinn:widebinaryconstraint}
\bibinfo{author}{\bibfnamefont{D.~P.} \bibnamefont{Quinn}},
  \bibinfo{author}{\bibfnamefont{M.~I.} \bibnamefont{Wilkinson}},
  \bibinfo{author}{\bibfnamefont{M.~J.} \bibnamefont{Irwin}},
  \bibinfo{author}{\bibfnamefont{J.}~\bibnamefont{Marshall}},
  \bibinfo{author}{\bibfnamefont{A.}~\bibnamefont{Koch}}, \bibnamefont{and}
  \bibinfo{author}{\bibfnamefont{V.}~\bibnamefont{Belokurov}},
  \bibinfo{journal}{Monthly Notices of the Royal Astronomical Society: Letters}
  \textbf{\bibinfo{volume}{396}}, \bibinfo{pages}{L11} (\bibinfo{year}{2009}),
  \eprint{0903.1644}.

\bibitem[{\citenamefont{{Ali-Ha{\"\i}moud} and
  {Kamionkowski}}(2017)}]{Ali-Haimoud_PBH}
\bibinfo{author}{\bibfnamefont{Y.}~\bibnamefont{{Ali-Ha{\"\i}moud}}}
  \bibnamefont{and}
  \bibinfo{author}{\bibfnamefont{M.}~\bibnamefont{{Kamionkowski}}},
  \bibinfo{journal}{Phys. Rev. D} \textbf{\bibinfo{volume}{95}},
  \bibinfo{eid}{043534} (\bibinfo{year}{2017}), \eprint{1612.05644}.

\bibitem[{\citenamefont{Poulin et~al.}(2017)\citenamefont{Poulin, Serpico,
  Calore, Clesse, and Kohri}}]{poulin:cmbconstraint}
\bibinfo{author}{\bibfnamefont{V.}~\bibnamefont{Poulin}},
  \bibinfo{author}{\bibfnamefont{P.~D.} \bibnamefont{Serpico}},
  \bibinfo{author}{\bibfnamefont{F.}~\bibnamefont{Calore}},
  \bibinfo{author}{\bibfnamefont{S.}~\bibnamefont{Clesse}}, \bibnamefont{and}
  \bibinfo{author}{\bibfnamefont{K.}~\bibnamefont{Kohri}}
  (\bibinfo{year}{2017}), \eprint{1707.04206}.

\bibitem[{\citenamefont{{Bernal} et~al.}(2017)\citenamefont{{Bernal},
  {Bellomo}, {Raccanelli}, and {Verde}}}]{Bernal_pbh}
\bibinfo{author}{\bibfnamefont{J.~L.} \bibnamefont{{Bernal}}},
  \bibinfo{author}{\bibfnamefont{N.}~\bibnamefont{{Bellomo}}},
  \bibinfo{author}{\bibfnamefont{A.}~\bibnamefont{{Raccanelli}}},
  \bibnamefont{and} \bibinfo{author}{\bibfnamefont{L.}~\bibnamefont{{Verde}}},
  \bibinfo{journal}{\jcap} \textbf{\bibinfo{volume}{10}}, \bibinfo{eid}{052}
  (\bibinfo{year}{2017}), \eprint{1709.07465}.

\bibitem[{\citenamefont{Raccanelli et~al.}(2017)\citenamefont{Raccanelli,
  Vidotto, and Verde}}]{Raccanelli:QG}
\bibinfo{author}{\bibfnamefont{A.}~\bibnamefont{Raccanelli}},
  \bibinfo{author}{\bibfnamefont{F.}~\bibnamefont{Vidotto}}, \bibnamefont{and}
  \bibinfo{author}{\bibfnamefont{L.}~\bibnamefont{Verde}}
  (\bibinfo{year}{2017}), \eprint{1708.02588}.

\bibitem[{\citenamefont{Mu{\~n}oz et~al.}(2016)\citenamefont{Mu{\~n}oz, Kovetz,
  Dai, and Kamionkowski}}]{Munoz:2016FRB}
\bibinfo{author}{\bibfnamefont{J.~B.} \bibnamefont{Mu{\~n}oz}},
  \bibinfo{author}{\bibfnamefont{E.~D.} \bibnamefont{Kovetz}},
  \bibinfo{author}{\bibfnamefont{L.}~\bibnamefont{Dai}}, \bibnamefont{and}
  \bibinfo{author}{\bibfnamefont{M.}~\bibnamefont{Kamionkowski}},
  \bibinfo{journal}{Phys. Rev. Lett.} \textbf{\bibinfo{volume}{117}},
  \bibinfo{pages}{091301} (\bibinfo{year}{2016}), \eprint{1605.00008}.

\bibitem[{\citenamefont{Raccanelli et~al.}(2016)\citenamefont{Raccanelli,
  Kovetz, Bird, Cholis, and Munoz}}]{raccanelli:cross}
\bibinfo{author}{\bibfnamefont{A.}~\bibnamefont{Raccanelli}},
  \bibinfo{author}{\bibfnamefont{E.~D.} \bibnamefont{Kovetz}},
  \bibinfo{author}{\bibfnamefont{S.}~\bibnamefont{Bird}},
  \bibinfo{author}{\bibfnamefont{I.}~\bibnamefont{Cholis}}, \bibnamefont{and}
  \bibinfo{author}{\bibfnamefont{J.~B.} \bibnamefont{Munoz}},
  \bibinfo{journal}{Phys. Rev.} \textbf{\bibinfo{volume}{D94}},
  \bibinfo{pages}{023516} (\bibinfo{year}{2016}), \eprint{1605.01405}.

\bibitem[{\citenamefont{Raccanelli}(2017)}]{raccanelli:radio}
\bibinfo{author}{\bibfnamefont{A.}~\bibnamefont{Raccanelli}},
  \bibinfo{journal}{Mon. Not. Roy. Astron. Soc.}
  \textbf{\bibinfo{volume}{469}}, \bibinfo{pages}{656} (\bibinfo{year}{2017}),
  \eprint{1609.09377}.

\bibitem[{\citenamefont{Nishikawa et~al.}(2017)\citenamefont{Nishikawa, Kovetz,
  Kamionkowski, and Silk}}]{Nishikawa:2017}
\bibinfo{author}{\bibfnamefont{H.}~\bibnamefont{Nishikawa}},
  \bibinfo{author}{\bibfnamefont{E.~D.} \bibnamefont{Kovetz}},
  \bibinfo{author}{\bibfnamefont{M.}~\bibnamefont{Kamionkowski}},
  \bibnamefont{and} \bibinfo{author}{\bibfnamefont{J.}~\bibnamefont{Silk}}
  (\bibinfo{year}{2017}), \eprint{1708.08449}.

\bibitem[{\citenamefont{Cholis et~al.}(2016)\citenamefont{Cholis, Kovetz,
  Ali-Ha{\"\i}moud, Bird, Kamionkowski, Mu{\~n}oz, and Raccanelli}}]{cholis}
\bibinfo{author}{\bibfnamefont{I.}~\bibnamefont{Cholis}},
  \bibinfo{author}{\bibfnamefont{E.~D.} \bibnamefont{Kovetz}},
  \bibinfo{author}{\bibfnamefont{Y.}~\bibnamefont{Ali-Ha{\"\i}moud}},
  \bibinfo{author}{\bibfnamefont{S.}~\bibnamefont{Bird}},
  \bibinfo{author}{\bibfnamefont{M.}~\bibnamefont{Kamionkowski}},
  \bibinfo{author}{\bibfnamefont{J.~B.} \bibnamefont{Mu{\~n}oz}},
  \bibnamefont{and}
  \bibinfo{author}{\bibfnamefont{A.}~\bibnamefont{Raccanelli}},
  \bibinfo{journal}{Phys. Rev.} \textbf{\bibinfo{volume}{D94}},
  \bibinfo{pages}{084013} (\bibinfo{year}{2016}), \eprint{1606.07437}.

\bibitem[{\citenamefont{Kovetz et~al.}(2017{\natexlab{a}})\citenamefont{Kovetz,
  Cholis, Breysse, and Kamionkowski}}]{Kovetz:2017BHMF}
\bibinfo{author}{\bibfnamefont{E.~D.} \bibnamefont{Kovetz}},
  \bibinfo{author}{\bibfnamefont{I.}~\bibnamefont{Cholis}},
  \bibinfo{author}{\bibfnamefont{P.~C.} \bibnamefont{Breysse}},
  \bibnamefont{and}
  \bibinfo{author}{\bibfnamefont{M.}~\bibnamefont{Kamionkowski}},
  \bibinfo{journal}{Phys. Rev.} \textbf{\bibinfo{volume}{D95}},
  \bibinfo{pages}{103010} (\bibinfo{year}{2017}{\natexlab{a}}),
  \eprint{1611.01157}.

\bibitem[{\citenamefont{Kovetz}(2017)}]{Kovetz:2017GW}
\bibinfo{author}{\bibfnamefont{E.~D.} \bibnamefont{Kovetz}},
  \bibinfo{journal}{Phys. Rev. Lett.} \textbf{\bibinfo{volume}{119}},
  \bibinfo{pages}{131301} (\bibinfo{year}{2017}), \eprint{1705.09182}.

\bibitem[{\citenamefont{Ali-Ha\"imoud et~al.}(2017)\citenamefont{Ali-Ha\"imoud,
  Kovetz, and Kamionkowski}}]{alihaimoud:pbhmergerrate}
\bibinfo{author}{\bibfnamefont{Y.}~\bibnamefont{Ali-Ha\"imoud}},
  \bibinfo{author}{\bibfnamefont{E.~D.} \bibnamefont{Kovetz}},
  \bibnamefont{and}
  \bibinfo{author}{\bibfnamefont{M.}~\bibnamefont{Kamionkowski}}
  (\bibinfo{year}{2017}), \eprint{1709.06576}.

\bibitem[{\citenamefont{Mandic et~al.}(2016)\citenamefont{Mandic, Bird, and
  Cholis}}]{Mandic:2016}
\bibinfo{author}{\bibfnamefont{V.}~\bibnamefont{Mandic}},
  \bibinfo{author}{\bibfnamefont{S.}~\bibnamefont{Bird}}, \bibnamefont{and}
  \bibinfo{author}{\bibfnamefont{I.}~\bibnamefont{Cholis}},
  \bibinfo{journal}{Phys. Rev. Lett.} \textbf{\bibinfo{volume}{117}},
  \bibinfo{pages}{201102} (\bibinfo{year}{2016}), \eprint{1608.06699}.

\bibitem[{\citenamefont{Cholis}(2017)}]{Cholis:2016}
\bibinfo{author}{\bibfnamefont{I.}~\bibnamefont{Cholis}},
  \bibinfo{journal}{JCAP} \textbf{\bibinfo{volume}{1706}}, \bibinfo{pages}{037}
  (\bibinfo{year}{2017}), \eprint{1609.03565}.

\bibitem[{\citenamefont{Nakama et~al.}(2017)\citenamefont{Nakama, Silk, and
  Kamionkowski}}]{Nakama:2016}
\bibinfo{author}{\bibfnamefont{T.}~\bibnamefont{Nakama}},
  \bibinfo{author}{\bibfnamefont{J.}~\bibnamefont{Silk}}, \bibnamefont{and}
  \bibinfo{author}{\bibfnamefont{M.}~\bibnamefont{Kamionkowski}},
  \bibinfo{journal}{Phys. Rev.} \textbf{\bibinfo{volume}{D95}},
  \bibinfo{pages}{043511} (\bibinfo{year}{2017}), \eprint{1612.06264}.

\bibitem[{\citenamefont{Raidal et~al.}(2017)\citenamefont{Raidal, Vaskonen, and
  Veerm\"{a}e}}]{marti:gwpbhmergers}
\bibinfo{author}{\bibfnamefont{M.}~\bibnamefont{Raidal}},
  \bibinfo{author}{\bibfnamefont{V.}~\bibnamefont{Vaskonen}}, \bibnamefont{and}
  \bibinfo{author}{\bibfnamefont{H.}~\bibnamefont{Veerm\"{a}e}},
  \bibinfo{journal}{Journal of Cosmology and Astroparticle Physics}
  \textbf{\bibinfo{volume}{2017}}, \bibinfo{pages}{037} (\bibinfo{year}{2017}),
  \eprint{1707.01480}.

\bibitem[{\citenamefont{Brandt}(2016)}]{brandt:ufdgconstraint}
\bibinfo{author}{\bibfnamefont{T.~D.} \bibnamefont{Brandt}},
  \bibinfo{journal}{The Astrophysical Journal Letters}
  \textbf{\bibinfo{volume}{824}}, \bibinfo{pages}{L31} (\bibinfo{year}{2016}),
  \eprint{1605.03665}.

\bibitem[{\citenamefont{{Bellomo} et~al.}(2017)\citenamefont{{Bellomo},
  {Bernal}, {Raccanelli}, and {Verde}}}]{bellomo:pbhemfconstraints}
\bibinfo{author}{\bibfnamefont{N.}~\bibnamefont{{Bellomo}}},
  \bibinfo{author}{\bibfnamefont{J.~L.} \bibnamefont{{Bernal}}},
  \bibinfo{author}{\bibfnamefont{A.}~\bibnamefont{{Raccanelli}}},
  \bibnamefont{and} \bibinfo{author}{\bibfnamefont{L.}~\bibnamefont{{Verde}}},
  \bibinfo{journal}{ArXiv e-prints}  (\bibinfo{year}{2017}),
  \eprint{1709.07467}.

\bibitem[{\citenamefont{Kovetz et~al.}(2017{\natexlab{b}})}]{Kovetz:2017}
\bibinfo{author}{\bibfnamefont{E.~D.} \bibnamefont{Kovetz}}
  \bibnamefont{et~al.} (\bibinfo{year}{2017}{\natexlab{b}}),
  \eprint{1709.09066}.

\bibitem[{\citenamefont{{Raccanelli} et~al.}(2016)\citenamefont{{Raccanelli},
  {Kovetz}, {Dai}, and {Kamionkowski}}}]{Raccanelli_isw21cm}
\bibinfo{author}{\bibfnamefont{A.}~\bibnamefont{{Raccanelli}}},
  \bibinfo{author}{\bibfnamefont{E.}~\bibnamefont{{Kovetz}}},
  \bibinfo{author}{\bibfnamefont{L.}~\bibnamefont{{Dai}}}, \bibnamefont{and}
  \bibinfo{author}{\bibfnamefont{M.}~\bibnamefont{{Kamionkowski}}},
  \bibinfo{journal}{\prd} \textbf{\bibinfo{volume}{93}}, \bibinfo{eid}{083512}
  (\bibinfo{year}{2016}), \eprint{1502.03107}.

\bibitem[{\citenamefont{{Kovetz} and
  {Kamionkowski}}(2013)}]{Kovetz_lensing21cm}
\bibinfo{author}{\bibfnamefont{E.~D.} \bibnamefont{{Kovetz}}} \bibnamefont{and}
  \bibinfo{author}{\bibfnamefont{M.}~\bibnamefont{{Kamionkowski}}},
  \bibinfo{journal}{\prd} \textbf{\bibinfo{volume}{87}}, \bibinfo{eid}{063516}
  (\bibinfo{year}{2013}), \eprint{1210.3041}.

\bibitem[{\citenamefont{{Mu{\~n}oz} et~al.}(2015)\citenamefont{{Mu{\~n}oz},
  {Ali-Ha{\"\i}moud}, and {Kamionkowski}}}]{Munoz_21cmbispec}
\bibinfo{author}{\bibfnamefont{J.~B.} \bibnamefont{{Mu{\~n}oz}}},
  \bibinfo{author}{\bibfnamefont{Y.}~\bibnamefont{{Ali-Ha{\"\i}moud}}},
  \bibnamefont{and}
  \bibinfo{author}{\bibfnamefont{M.}~\bibnamefont{{Kamionkowski}}},
  \bibinfo{journal}{\prd} \textbf{\bibinfo{volume}{92}}, \bibinfo{eid}{083508}
  (\bibinfo{year}{2015}), \eprint{1506.04152}.

\bibitem[{\citenamefont{Mu{\~n}oz et~al.}(2017)\citenamefont{Mu{\~n}oz, Kovetz,
  Raccanelli, Kamionkowski, and Silk}}]{Munoz:2016}
\bibinfo{author}{\bibfnamefont{J.~B.} \bibnamefont{Mu{\~n}oz}},
  \bibinfo{author}{\bibfnamefont{E.~D.} \bibnamefont{Kovetz}},
  \bibinfo{author}{\bibfnamefont{A.}~\bibnamefont{Raccanelli}},
  \bibinfo{author}{\bibfnamefont{M.}~\bibnamefont{Kamionkowski}},
  \bibnamefont{and} \bibinfo{author}{\bibfnamefont{J.}~\bibnamefont{Silk}},
  \bibinfo{journal}{JCAP} \textbf{\bibinfo{volume}{1705}}, \bibinfo{pages}{032}
  (\bibinfo{year}{2017}), \eprint{1611.05883}.

\bibitem[{\citenamefont{Pourtsidou}(2016)}]{Pourtsidou:2016}
\bibinfo{author}{\bibfnamefont{A.}~\bibnamefont{Pourtsidou}}
  (\bibinfo{year}{2016}), \eprint{1612.05138}.

\bibitem[{\citenamefont{Sekiguchi et~al.}(2017)\citenamefont{Sekiguchi,
  Takahashi, Tashiro, and Yokoyama}}]{Sekiguchi:2017}
\bibinfo{author}{\bibfnamefont{T.}~\bibnamefont{Sekiguchi}},
  \bibinfo{author}{\bibfnamefont{T.}~\bibnamefont{Takahashi}},
  \bibinfo{author}{\bibfnamefont{H.}~\bibnamefont{Tashiro}}, \bibnamefont{and}
  \bibinfo{author}{\bibfnamefont{S.}~\bibnamefont{Yokoyama}}
  (\bibinfo{year}{2017}), \eprint{1705.00405}.

\bibitem[{\citenamefont{Mu{\~n}oz et~al.}(2015)\citenamefont{Mu{\~n}oz, Kovetz,
  and Ali-Ha{\"\i}moud}}]{Munoz:2015}
\bibinfo{author}{\bibfnamefont{J.~B.} \bibnamefont{Mu{\~n}oz}},
  \bibinfo{author}{\bibfnamefont{E.~D.} \bibnamefont{Kovetz}},
  \bibnamefont{and}
  \bibinfo{author}{\bibfnamefont{Y.}~\bibnamefont{Ali-Ha{\"\i}moud}},
  \bibinfo{journal}{Phys. Rev.} \textbf{\bibinfo{volume}{D92}},
  \bibinfo{pages}{083528} (\bibinfo{year}{2015}), \eprint{1509.00029}.

\bibitem[{\citenamefont{{Shiraishi} et~al.}(2016)\citenamefont{{Shiraishi},
  {Mu{\~n}oz}, {Kamionkowski}, and {Raccanelli}}}]{Shiraishi_iso21cm}
\bibinfo{author}{\bibfnamefont{M.}~\bibnamefont{{Shiraishi}}},
  \bibinfo{author}{\bibfnamefont{J.~B.} \bibnamefont{{Mu{\~n}oz}}},
  \bibinfo{author}{\bibfnamefont{M.}~\bibnamefont{{Kamionkowski}}},
  \bibnamefont{and}
  \bibinfo{author}{\bibfnamefont{A.}~\bibnamefont{{Raccanelli}}},
  \bibinfo{journal}{\prd} \textbf{\bibinfo{volume}{93}}, \bibinfo{eid}{103506}
  (\bibinfo{year}{2016}), \eprint{1603.01206}.

\bibitem[{\citenamefont{{Evoli} et~al.}(2014)\citenamefont{{Evoli}, {Mesinger},
  and {Ferrara}}}]{Evoli_dcdm21}
\bibinfo{author}{\bibfnamefont{C.}~\bibnamefont{{Evoli}}},
  \bibinfo{author}{\bibfnamefont{A.}~\bibnamefont{{Mesinger}}},
  \bibnamefont{and}
  \bibinfo{author}{\bibfnamefont{A.}~\bibnamefont{{Ferrara}}},
  \bibinfo{journal}{\jcap} \textbf{\bibinfo{volume}{11}}, \bibinfo{eid}{024}
  (\bibinfo{year}{2014}), \eprint{1408.1109}.

\bibitem[{\citenamefont{{Gong} and {Kitajima}}(2017)}]{Gong_pbh21cm}
\bibinfo{author}{\bibfnamefont{J.-O.} \bibnamefont{{Gong}}} \bibnamefont{and}
  \bibinfo{author}{\bibfnamefont{N.}~\bibnamefont{{Kitajima}}},
  \bibinfo{journal}{\jcap} \textbf{\bibinfo{volume}{8}}, \bibinfo{eid}{017}
  (\bibinfo{year}{2017}), \eprint{1704.04132}.

\bibitem[{\citenamefont{{Mack} and {Wesley}}(2008)}]{Mack_lightpbh}
\bibinfo{author}{\bibfnamefont{K.~J.} \bibnamefont{{Mack}}} \bibnamefont{and}
  \bibinfo{author}{\bibfnamefont{D.~H.} \bibnamefont{{Wesley}}},
  \bibinfo{journal}{ArXiv e-prints}  (\bibinfo{year}{2008}),
  \eprint{0805.1531}.

\bibitem[{\citenamefont{{Iliev} et~al.}(2002)\citenamefont{{Iliev}, {Shapiro},
  {Ferrara}, and {Martel}}}]{Iliev_minihalo_02}
\bibinfo{author}{\bibfnamefont{I.~T.} \bibnamefont{{Iliev}}},
  \bibinfo{author}{\bibfnamefont{P.~R.} \bibnamefont{{Shapiro}}},
  \bibinfo{author}{\bibfnamefont{A.}~\bibnamefont{{Ferrara}}},
  \bibnamefont{and} \bibinfo{author}{\bibfnamefont{H.}~\bibnamefont{{Martel}}},
  \bibinfo{journal}{\apjl} \textbf{\bibinfo{volume}{572}},
  \bibinfo{pages}{L123} (\bibinfo{year}{2002}), \eprint{astro-ph/0202410}.

\bibitem[{\citenamefont{{Furlanetto} and {Oh}}(2006)}]{Furlanetto06}
\bibinfo{author}{\bibfnamefont{S.~R.} \bibnamefont{{Furlanetto}}}
  \bibnamefont{and} \bibinfo{author}{\bibfnamefont{S.~P.} \bibnamefont{{Oh}}},
  \bibinfo{journal}{\apj} \textbf{\bibinfo{volume}{652}}, \bibinfo{pages}{849}
  (\bibinfo{year}{2006}), \eprint{astro-ph/0604080}.

\bibitem[{\citenamefont{{Tanaka} et~al.}(2016)\citenamefont{{Tanaka},
  {O'Leary}, and {Perna}}}]{Tanaka_smbh21cm}
\bibinfo{author}{\bibfnamefont{T.~L.} \bibnamefont{{Tanaka}}},
  \bibinfo{author}{\bibfnamefont{R.~M.} \bibnamefont{{O'Leary}}},
  \bibnamefont{and} \bibinfo{author}{\bibfnamefont{R.}~\bibnamefont{{Perna}}},
  \bibinfo{journal}{\mnras} \textbf{\bibinfo{volume}{455}},
  \bibinfo{pages}{2619} (\bibinfo{year}{2016}), \eprint{1509.05406}.

\bibitem[{\citenamefont{{Cohen} et~al.}(2016)\citenamefont{{Cohen}, {Fialkov},
  {Barkana}, and {Lotem}}}]{Cohen_21cm}
\bibinfo{author}{\bibfnamefont{A.}~\bibnamefont{{Cohen}}},
  \bibinfo{author}{\bibfnamefont{A.}~\bibnamefont{{Fialkov}}},
  \bibinfo{author}{\bibfnamefont{R.}~\bibnamefont{{Barkana}}},
  \bibnamefont{and} \bibinfo{author}{\bibfnamefont{M.}~\bibnamefont{{Lotem}}},
  \bibinfo{journal}{ArXiv e-prints}  (\bibinfo{year}{2016}),
  \eprint{1609.02312}.

\bibitem[{\citenamefont{{Cohen} et~al.}(2017)\citenamefont{{Cohen}, {Fialkov},
  and {Barkana}}}]{Cohen_21cmPSparams}
\bibinfo{author}{\bibfnamefont{A.}~\bibnamefont{{Cohen}}},
  \bibinfo{author}{\bibfnamefont{A.}~\bibnamefont{{Fialkov}}},
  \bibnamefont{and}
  \bibinfo{author}{\bibfnamefont{R.}~\bibnamefont{{Barkana}}},
  \bibinfo{journal}{ArXiv e-prints}  (\bibinfo{year}{2017}),
  \eprint{1709.02122}.

\bibitem[{\citenamefont{{Vald{\'e}s} et~al.}(2013)\citenamefont{{Vald{\'e}s},
  {Evoli}, {Mesinger}, {Ferrara}, and {Yoshida}}}]{Valdes_anni-dm21cm}
\bibinfo{author}{\bibfnamefont{M.}~\bibnamefont{{Vald{\'e}s}}},
  \bibinfo{author}{\bibfnamefont{C.}~\bibnamefont{{Evoli}}},
  \bibinfo{author}{\bibfnamefont{A.}~\bibnamefont{{Mesinger}}},
  \bibinfo{author}{\bibfnamefont{A.}~\bibnamefont{{Ferrara}}},
  \bibnamefont{and}
  \bibinfo{author}{\bibfnamefont{N.}~\bibnamefont{{Yoshida}}},
  \bibinfo{journal}{\mnras} \textbf{\bibinfo{volume}{429}},
  \bibinfo{pages}{1705} (\bibinfo{year}{2013}), \eprint{1209.2120}.

\bibitem[{\citenamefont{Santos et~al.}(2015)}]{SKA_IM}
\bibinfo{author}{\bibfnamefont{M.}~\bibnamefont{Santos}} \bibnamefont{et~al.},
  \bibinfo{journal}{PoS} \textbf{\bibinfo{volume}{AASKA14}},
  \bibinfo{pages}{019} (\bibinfo{year}{2015}).

\bibitem[{\citenamefont{{Silk} et~al.}(in preparation)\citenamefont{{Silk},
  {Raccanelli}, S., J., and {Kovetz}}}]{Silk:telescope}
\bibinfo{author}{\bibfnamefont{J.}~\bibnamefont{{Silk}}},
  \bibinfo{author}{\bibfnamefont{A.}~\bibnamefont{{Raccanelli}}},
  \bibinfo{author}{\bibfnamefont{C.}~\bibnamefont{S.}},
  \bibinfo{author}{\bibfnamefont{M.}~\bibnamefont{J.}}, \bibnamefont{and}
  \bibinfo{author}{\bibfnamefont{E.~D.} \bibnamefont{{Kovetz}}}
  (\bibinfo{year}{in preparation}).

\bibitem[{\citenamefont{Ade and others
  (Planck~Collaboration)}(2016)}]{Planckparameterspaper}
\bibinfo{author}{\bibfnamefont{P.~A.~R.} \bibnamefont{Ade}} \bibnamefont{and}
  \bibinfo{author}{\bibnamefont{others (Planck~Collaboration)}},
  \bibinfo{journal}{\aap} \textbf{\bibinfo{volume}{594}}, \bibinfo{eid}{A13}
  (\bibinfo{year}{2016}), \eprint{1502.01589}.

\bibitem[{\citenamefont{{Field}}(1959{\natexlab{a}})}]{Field_59}
\bibinfo{author}{\bibfnamefont{G.~B.} \bibnamefont{{Field}}},
  \bibinfo{journal}{\apj} \textbf{\bibinfo{volume}{129}}, \bibinfo{pages}{525}
  (\bibinfo{year}{1959}{\natexlab{a}}).

\bibitem[{\citenamefont{{Field}}(1958)}]{Field58}
\bibinfo{author}{\bibfnamefont{G.~B.} \bibnamefont{{Field}}},
  \bibinfo{journal}{Proceedings of the IRE} \textbf{\bibinfo{volume}{46}},
  \bibinfo{pages}{240} (\bibinfo{year}{1958}).

\bibitem[{\citenamefont{{Field}}(1959{\natexlab{b}})}]{Field_59B}
\bibinfo{author}{\bibfnamefont{G.~B.} \bibnamefont{{Field}}},
  \bibinfo{journal}{\apj} \textbf{\bibinfo{volume}{129}}, \bibinfo{pages}{551}
  (\bibinfo{year}{1959}{\natexlab{b}}).

\bibitem[{\citenamefont{{Kuhlen} et~al.}(2006)\citenamefont{{Kuhlen}, {Madau},
  and {Montgomery}}}]{Kuhlen06}
\bibinfo{author}{\bibfnamefont{M.}~\bibnamefont{{Kuhlen}}},
  \bibinfo{author}{\bibfnamefont{P.}~\bibnamefont{{Madau}}}, \bibnamefont{and}
  \bibinfo{author}{\bibfnamefont{R.}~\bibnamefont{{Montgomery}}},
  \bibinfo{journal}{\apjl} \textbf{\bibinfo{volume}{637}}, \bibinfo{pages}{L1}
  (\bibinfo{year}{2006}), \eprint{astro-ph/0510814}.

\bibitem[{\citenamefont{{Shull} and {van Steenberg}}(1985)}]{Shull85}
\bibinfo{author}{\bibfnamefont{J.~M.} \bibnamefont{{Shull}}} \bibnamefont{and}
  \bibinfo{author}{\bibfnamefont{M.~E.} \bibnamefont{{van Steenberg}}},
  \bibinfo{journal}{\apj} \textbf{\bibinfo{volume}{298}}, \bibinfo{pages}{268}
  (\bibinfo{year}{1985}).

\bibitem[{\citenamefont{{Loeb} and {Zaldarriaga}}(2004)}]{Loeb_21cm}
\bibinfo{author}{\bibfnamefont{A.}~\bibnamefont{{Loeb}}} \bibnamefont{and}
  \bibinfo{author}{\bibfnamefont{M.}~\bibnamefont{{Zaldarriaga}}},
  \bibinfo{journal}{Physical Review Letters} \textbf{\bibinfo{volume}{92}},
  \bibinfo{eid}{211301} (\bibinfo{year}{2004}), \eprint{astro-ph/0312134}.

\bibitem[{\citenamefont{{Bharadwaj} and {Ali}}(2004)}]{Bharadwaj_jlpm2}
\bibinfo{author}{\bibfnamefont{S.}~\bibnamefont{{Bharadwaj}}} \bibnamefont{and}
  \bibinfo{author}{\bibfnamefont{S.~S.} \bibnamefont{{Ali}}},
  \bibinfo{journal}{\mnras} \textbf{\bibinfo{volume}{352}},
  \bibinfo{pages}{142} (\bibinfo{year}{2004}), \eprint{astro-ph/0401206}.

\bibitem[{\citenamefont{{Ali-Ha{\"\i}moud}
  et~al.}(2014)\citenamefont{{Ali-Ha{\"\i}moud}, {Meerburg}, and
  {Yuan}}}]{Ali-Haimoud_21cmpert}
\bibinfo{author}{\bibfnamefont{Y.}~\bibnamefont{{Ali-Ha{\"\i}moud}}},
  \bibinfo{author}{\bibfnamefont{P.~D.} \bibnamefont{{Meerburg}}},
  \bibnamefont{and} \bibinfo{author}{\bibfnamefont{S.}~\bibnamefont{{Yuan}}},
  \bibinfo{journal}{\prd} \textbf{\bibinfo{volume}{89}}, \bibinfo{eid}{083506}
  (\bibinfo{year}{2014}), \eprint{1312.4948}.

\bibitem[{\citenamefont{{Tseliakhovich} and
  {Hirata}}(2010)}]{Tseliakovich_supersonic}
\bibinfo{author}{\bibfnamefont{D.}~\bibnamefont{{Tseliakhovich}}}
  \bibnamefont{and} \bibinfo{author}{\bibfnamefont{C.}~\bibnamefont{{Hirata}}},
  \bibinfo{journal}{\prd} \textbf{\bibinfo{volume}{82}}, \bibinfo{eid}{083520}
  (\bibinfo{year}{2010}), \eprint{1005.2416}.

\bibitem[{\citenamefont{{Hirano} et~al.}(2017)\citenamefont{{Hirano},
  {Hosokawa}, {Yoshida}, and {Kuiper}}}]{Hirano_streaming}
\bibinfo{author}{\bibfnamefont{S.}~\bibnamefont{{Hirano}}},
  \bibinfo{author}{\bibfnamefont{T.}~\bibnamefont{{Hosokawa}}},
  \bibinfo{author}{\bibfnamefont{N.}~\bibnamefont{{Yoshida}}},
  \bibnamefont{and} \bibinfo{author}{\bibfnamefont{R.}~\bibnamefont{{Kuiper}}},
  \bibinfo{journal}{Science} \textbf{\bibinfo{volume}{357}},
  \bibinfo{pages}{1375} (\bibinfo{year}{2017}), \eprint{1709.09863}.

\bibitem[{\citenamefont{{Tanaka} et~al.}(2013)\citenamefont{{Tanaka}, {Li}, and
  {Haiman}}}]{Tanaka_streaming}
\bibinfo{author}{\bibfnamefont{T.~L.} \bibnamefont{{Tanaka}}},
  \bibinfo{author}{\bibfnamefont{M.}~\bibnamefont{{Li}}}, \bibnamefont{and}
  \bibinfo{author}{\bibfnamefont{Z.}~\bibnamefont{{Haiman}}},
  \bibinfo{journal}{\mnras} \textbf{\bibinfo{volume}{435}},
  \bibinfo{pages}{3559} (\bibinfo{year}{2013}), \eprint{1309.2301}.

\bibitem[{\citenamefont{{Pillepich} et~al.}(2007)\citenamefont{{Pillepich},
  {Porciani}, and {Matarrese}}}]{Pillepich_21cm}
\bibinfo{author}{\bibfnamefont{A.}~\bibnamefont{{Pillepich}}},
  \bibinfo{author}{\bibfnamefont{C.}~\bibnamefont{{Porciani}}},
  \bibnamefont{and}
  \bibinfo{author}{\bibfnamefont{S.}~\bibnamefont{{Matarrese}}},
  \bibinfo{journal}{\apj} \textbf{\bibinfo{volume}{662}}, \bibinfo{pages}{1}
  (\bibinfo{year}{2007}), \eprint{astro-ph/0611126}.

\bibitem[{\citenamefont{{Lewis} and {Challinor}}(2007)}]{Lewis_21cm}
\bibinfo{author}{\bibfnamefont{A.}~\bibnamefont{{Lewis}}} \bibnamefont{and}
  \bibinfo{author}{\bibfnamefont{A.}~\bibnamefont{{Challinor}}},
  \bibinfo{journal}{\prd} \textbf{\bibinfo{volume}{76}}, \bibinfo{eid}{083005}
  (\bibinfo{year}{2007}), \eprint{astro-ph/0702600}.

\bibitem[{\citenamefont{{Limber}}(1953)}]{Limber:1953}
\bibinfo{author}{\bibfnamefont{D.~N.} \bibnamefont{{Limber}}},
  \bibinfo{journal}{\apj} \textbf{\bibinfo{volume}{117}}, \bibinfo{pages}{134}
  (\bibinfo{year}{1953}).

\bibitem[{\citenamefont{{Hu}}(2000)}]{Hu_cmblensing_flatsky}
\bibinfo{author}{\bibfnamefont{W.}~\bibnamefont{{Hu}}}, \bibinfo{journal}{\prd}
  \textbf{\bibinfo{volume}{62}}, \bibinfo{eid}{043007} (\bibinfo{year}{2000}),
  \eprint{astro-ph/0001303}.

\bibitem[{\citenamefont{{White} et~al.}(1999)\citenamefont{{White},
  {Carlstrom}, {Dragovan}, and {Holzapfel}}}]{White_flatsky}
\bibinfo{author}{\bibfnamefont{M.}~\bibnamefont{{White}}},
  \bibinfo{author}{\bibfnamefont{J.}~\bibnamefont{{Carlstrom}}},
  \bibinfo{author}{\bibfnamefont{M.}~\bibnamefont{{Dragovan}}},
  \bibnamefont{and} \bibinfo{author}{\bibfnamefont{S.~W.~L.}
  \bibnamefont{{Holzapfel}}}, \bibinfo{journal}{ArXiv Astrophysics e-prints}
  (\bibinfo{year}{1999}), \eprint{astro-ph/9912422}.

\bibitem[{\citenamefont{{Jaffe} et~al.}(2000)\citenamefont{{Jaffe},
  {Kamionkowski}, and {Wang}}}]{Jaffe_noise}
\bibinfo{author}{\bibfnamefont{A.~H.} \bibnamefont{{Jaffe}}},
  \bibinfo{author}{\bibfnamefont{M.}~\bibnamefont{{Kamionkowski}}},
  \bibnamefont{and} \bibinfo{author}{\bibfnamefont{L.}~\bibnamefont{{Wang}}},
  \bibinfo{journal}{\prd} \textbf{\bibinfo{volume}{61}}, \bibinfo{eid}{083501}
  (\bibinfo{year}{2000}), \eprint{astro-ph/9909281}.

\bibitem[{\citenamefont{{Knox} and {Song}}(2002)}]{Knox_inflation}
\bibinfo{author}{\bibfnamefont{L.}~\bibnamefont{{Knox}}} \bibnamefont{and}
  \bibinfo{author}{\bibfnamefont{Y.-S.} \bibnamefont{{Song}}},
  \bibinfo{journal}{Physical Review Letters} \textbf{\bibinfo{volume}{89}},
  \bibinfo{pages}{011303} (\bibinfo{year}{2002}), \eprint{astro-ph/0202286}.

\bibitem[{\citenamefont{{Kesden} et~al.}(2002)\citenamefont{{Kesden}, {Cooray},
  and {Kamionkowski}}}]{Kesden_gw}
\bibinfo{author}{\bibfnamefont{M.}~\bibnamefont{{Kesden}}},
  \bibinfo{author}{\bibfnamefont{A.}~\bibnamefont{{Cooray}}}, \bibnamefont{and}
  \bibinfo{author}{\bibfnamefont{M.}~\bibnamefont{{Kamionkowski}}},
  \bibinfo{journal}{Physical Review Letters} \textbf{\bibinfo{volume}{89}},
  \bibinfo{pages}{011304} (\bibinfo{year}{2002}), \eprint{astro-ph/0202434}.

\bibitem[{\citenamefont{{Zaldarriaga} et~al.}(2004)\citenamefont{{Zaldarriaga},
  {Furlanetto}, and {Hernquist}}}]{Zaldarriaga_21cm}
\bibinfo{author}{\bibfnamefont{M.}~\bibnamefont{{Zaldarriaga}}},
  \bibinfo{author}{\bibfnamefont{S.~R.} \bibnamefont{{Furlanetto}}},
  \bibnamefont{and}
  \bibinfo{author}{\bibfnamefont{L.}~\bibnamefont{{Hernquist}}},
  \bibinfo{journal}{\apj} \textbf{\bibinfo{volume}{608}}, \bibinfo{pages}{622}
  (\bibinfo{year}{2004}), \eprint{astro-ph/0311514}.

\bibitem[{\citenamefont{{Mozdzen} et~al.}(2017)\citenamefont{{Mozdzen},
  {Bowman}, {Monsalve}, and {Rogers}}}]{Mozdzen_21cmskytemp}
\bibinfo{author}{\bibfnamefont{T.~J.} \bibnamefont{{Mozdzen}}},
  \bibinfo{author}{\bibfnamefont{J.~D.} \bibnamefont{{Bowman}}},
  \bibinfo{author}{\bibfnamefont{R.~A.} \bibnamefont{{Monsalve}}},
  \bibnamefont{and} \bibinfo{author}{\bibfnamefont{A.~E.~E.}
  \bibnamefont{{Rogers}}}, \bibinfo{journal}{\mnras}
  \textbf{\bibinfo{volume}{464}}, \bibinfo{pages}{4995} (\bibinfo{year}{2017}),
  \eprint{1609.08705}.

\bibitem[{\citenamefont{{Ricotti}}(2007)}]{Ricotti_accretion}
\bibinfo{author}{\bibfnamefont{M.}~\bibnamefont{{Ricotti}}},
  \bibinfo{journal}{Astrophys. J.} \textbf{\bibinfo{volume}{662}},
  \bibinfo{pages}{53} (\bibinfo{year}{2007}), \eprint{0706.0864}.

\bibitem[{\citenamefont{Carr et~al.}(2017)\citenamefont{Carr, Raidal, Tenkanen,
  Vaskonen, and Veerm\"ae}}]{carr:comparison2}
\bibinfo{author}{\bibfnamefont{B.}~\bibnamefont{Carr}},
  \bibinfo{author}{\bibfnamefont{M.}~\bibnamefont{Raidal}},
  \bibinfo{author}{\bibfnamefont{T.}~\bibnamefont{Tenkanen}},
  \bibinfo{author}{\bibfnamefont{V.}~\bibnamefont{Vaskonen}}, \bibnamefont{and}
  \bibinfo{author}{\bibfnamefont{H.}~\bibnamefont{Veerm\"ae}},
  \bibinfo{journal}{Phys. Rev. D} \textbf{\bibinfo{volume}{96}},
  \bibinfo{pages}{023514} (\bibinfo{year}{2017}), \urlprefix\url{1705.05567}.

\bibitem[{\citenamefont{{Thomas} and {Zaroubi}}(2008)}]{Thomas_mqcode}
\bibinfo{author}{\bibfnamefont{R.~M.} \bibnamefont{{Thomas}}} \bibnamefont{and}
  \bibinfo{author}{\bibfnamefont{S.}~\bibnamefont{{Zaroubi}}},
  \bibinfo{journal}{\mnras} \textbf{\bibinfo{volume}{384}},
  \bibinfo{pages}{1080} (\bibinfo{year}{2008}), \eprint{0709.1657}.

\bibitem[{\citenamefont{{Tanaka} et~al.}(2012)\citenamefont{{Tanaka}, {Perna},
  and {Haiman}}}]{Tanaka_sed}
\bibinfo{author}{\bibfnamefont{T.}~\bibnamefont{{Tanaka}}},
  \bibinfo{author}{\bibfnamefont{R.}~\bibnamefont{{Perna}}}, \bibnamefont{and}
  \bibinfo{author}{\bibfnamefont{Z.}~\bibnamefont{{Haiman}}},
  \bibinfo{journal}{\mnras} \textbf{\bibinfo{volume}{425}},
  \bibinfo{pages}{2974} (\bibinfo{year}{2012}), \eprint{1205.6467}.

\bibitem[{\citenamefont{{Pacucci}
  et~al.}(2015{\natexlab{b}})\citenamefont{{Pacucci}, {Ferrara}, {Volonteri},
  and {Dubus}}}]{Pacucci_spectrum}
\bibinfo{author}{\bibfnamefont{F.}~\bibnamefont{{Pacucci}}},
  \bibinfo{author}{\bibfnamefont{A.}~\bibnamefont{{Ferrara}}},
  \bibinfo{author}{\bibfnamefont{M.}~\bibnamefont{{Volonteri}}},
  \bibnamefont{and} \bibinfo{author}{\bibfnamefont{G.}~\bibnamefont{{Dubus}}},
  \bibinfo{journal}{\mnras} \textbf{\bibinfo{volume}{454}},
  \bibinfo{pages}{3771} (\bibinfo{year}{2015}{\natexlab{b}}),
  \eprint{1506.05299}.

\bibitem[{\citenamefont{{Zaroubi} et~al.}(2007)\citenamefont{{Zaroubi},
  {Thomas}, {Sugiyama}, and {Silk}}}]{Zaroubi07}
\bibinfo{author}{\bibfnamefont{S.}~\bibnamefont{{Zaroubi}}},
  \bibinfo{author}{\bibfnamefont{R.~M.} \bibnamefont{{Thomas}}},
  \bibinfo{author}{\bibfnamefont{N.}~\bibnamefont{{Sugiyama}}},
  \bibnamefont{and} \bibinfo{author}{\bibfnamefont{J.}~\bibnamefont{{Silk}}},
  \bibinfo{journal}{\mnras} \textbf{\bibinfo{volume}{375}},
  \bibinfo{pages}{1269} (\bibinfo{year}{2007}), \eprint{astro-ph/0609151}.

\bibitem[{\citenamefont{{Fialkov} et~al.}(2014)\citenamefont{{Fialkov},
  {Barkana}, and {Visbal}}}]{Fialkov_sed}
\bibinfo{author}{\bibfnamefont{A.}~\bibnamefont{{Fialkov}}},
  \bibinfo{author}{\bibfnamefont{R.}~\bibnamefont{{Barkana}}},
  \bibnamefont{and} \bibinfo{author}{\bibfnamefont{E.}~\bibnamefont{{Visbal}}},
  \bibinfo{journal}{\nat} \textbf{\bibinfo{volume}{506}}, \bibinfo{pages}{197}
  (\bibinfo{year}{2014}), \eprint{1402.0940}.

\bibitem[{\citenamefont{{Pacucci} et~al.}(2014)\citenamefont{{Pacucci},
  {Mesinger}, {Mineo}, and {Ferrara}}}]{Pacucci_sed}
\bibinfo{author}{\bibfnamefont{F.}~\bibnamefont{{Pacucci}}},
  \bibinfo{author}{\bibfnamefont{A.}~\bibnamefont{{Mesinger}}},
  \bibinfo{author}{\bibfnamefont{S.}~\bibnamefont{{Mineo}}}, \bibnamefont{and}
  \bibinfo{author}{\bibfnamefont{A.}~\bibnamefont{{Ferrara}}},
  \bibinfo{journal}{\mnras} \textbf{\bibinfo{volume}{443}},
  \bibinfo{pages}{678} (\bibinfo{year}{2014}), \eprint{1403.6125}.

\bibitem[{\citenamefont{{Sazonov} et~al.}(2004)\citenamefont{{Sazonov},
  {Ostriker}, and {Sunyaev}}}]{Sazonov_spectrum}
\bibinfo{author}{\bibfnamefont{S.~Y.} \bibnamefont{{Sazonov}}},
  \bibinfo{author}{\bibfnamefont{J.~P.} \bibnamefont{{Ostriker}}},
  \bibnamefont{and} \bibinfo{author}{\bibfnamefont{R.~A.}
  \bibnamefont{{Sunyaev}}}, \bibinfo{journal}{\mnras}
  \textbf{\bibinfo{volume}{347}}, \bibinfo{pages}{144} (\bibinfo{year}{2004}),
  \eprint{astro-ph/0305233}.

\bibitem[{\citenamefont{{Zdziarski} and {Svensson}}(1989)}]{Zdziarski_sigma}
\bibinfo{author}{\bibfnamefont{A.~A.} \bibnamefont{{Zdziarski}}}
  \bibnamefont{and}
  \bibinfo{author}{\bibfnamefont{R.}~\bibnamefont{{Svensson}}},
  \bibinfo{journal}{\apj} \textbf{\bibinfo{volume}{344}}, \bibinfo{pages}{551}
  (\bibinfo{year}{1989}).

\bibitem[{\citenamefont{Ali-Ha\"{\i}moud and Hirata}(2010)}]{alihamoud:hyrec1}
\bibinfo{author}{\bibfnamefont{Y.}~\bibnamefont{Ali-Ha\"{\i}moud}}
  \bibnamefont{and} \bibinfo{author}{\bibfnamefont{C.~M.}
  \bibnamefont{Hirata}}, \bibinfo{journal}{Phys. Rev. D}
  \textbf{\bibinfo{volume}{82}}, \bibinfo{pages}{063521}
  (\bibinfo{year}{2010}), \eprint{https://arxiv.org/abs/1006.1355}.

\bibitem[{\citenamefont{Ali-Ha\"{\i}moud and Hirata}(2011)}]{alihamoud:hyrec2}
\bibinfo{author}{\bibfnamefont{Y.}~\bibnamefont{Ali-Ha\"{\i}moud}}
  \bibnamefont{and} \bibinfo{author}{\bibfnamefont{C.~M.}
  \bibnamefont{Hirata}}, \bibinfo{journal}{Phys. Rev. D}
  \textbf{\bibinfo{volume}{83}}, \bibinfo{pages}{043513}
  (\bibinfo{year}{2011}), \eprint{https://arxiv.org/abs/1011.3758}.

\bibitem[{\citenamefont{{Conselice} et~al.}(2016)\citenamefont{{Conselice},
  {Wilkinson}, {Duncan}, and {Mortlock}}}]{Conselice_numbergalaxies}
\bibinfo{author}{\bibfnamefont{C.~J.} \bibnamefont{{Conselice}}},
  \bibinfo{author}{\bibfnamefont{A.}~\bibnamefont{{Wilkinson}}},
  \bibinfo{author}{\bibfnamefont{K.}~\bibnamefont{{Duncan}}}, \bibnamefont{and}
  \bibinfo{author}{\bibfnamefont{A.}~\bibnamefont{{Mortlock}}},
  \bibinfo{journal}{\apj} \textbf{\bibinfo{volume}{830}}, \bibinfo{eid}{83}
  (\bibinfo{year}{2016}), \eprint{1607.03909}.

\bibitem[{\citenamefont{{Shen} and {Kelly}}(2012)}]{Shen_lambda}
\bibinfo{author}{\bibfnamefont{Y.}~\bibnamefont{{Shen}}} \bibnamefont{and}
  \bibinfo{author}{\bibfnamefont{B.~C.} \bibnamefont{{Kelly}}},
  \bibinfo{journal}{\apj} \textbf{\bibinfo{volume}{746}}, \bibinfo{eid}{169}
  (\bibinfo{year}{2012}), \eprint{1107.4372}.

\bibitem[{\citenamefont{{Kelly} and {Shen}}(2013)}]{Kelly_lambda}
\bibinfo{author}{\bibfnamefont{B.~C.} \bibnamefont{{Kelly}}} \bibnamefont{and}
  \bibinfo{author}{\bibfnamefont{Y.}~\bibnamefont{{Shen}}},
  \bibinfo{journal}{\apj} \textbf{\bibinfo{volume}{764}}, \bibinfo{eid}{45}
  (\bibinfo{year}{2013}), \eprint{1209.0477}.

\bibitem[{\citenamefont{{Hopkins} and {Hernquist}}(2009)}]{Hopkins_lambda}
\bibinfo{author}{\bibfnamefont{P.~F.} \bibnamefont{{Hopkins}}}
  \bibnamefont{and}
  \bibinfo{author}{\bibfnamefont{L.}~\bibnamefont{{Hernquist}}},
  \bibinfo{journal}{\apj} \textbf{\bibinfo{volume}{698}}, \bibinfo{pages}{1550}
  (\bibinfo{year}{2009}), \eprint{0809.3789}.

\bibitem[{\citenamefont{{Panessa} et~al.}(2006)\citenamefont{{Panessa},
  {Bassani}, {Cappi}, {Dadina}, {Barcons}, {Carrera}, {Ho}, and
  {Iwasawa}}}]{Panessa_Plambda}
\bibinfo{author}{\bibfnamefont{F.}~\bibnamefont{{Panessa}}},
  \bibinfo{author}{\bibfnamefont{L.}~\bibnamefont{{Bassani}}},
  \bibinfo{author}{\bibfnamefont{M.}~\bibnamefont{{Cappi}}},
  \bibinfo{author}{\bibfnamefont{M.}~\bibnamefont{{Dadina}}},
  \bibinfo{author}{\bibfnamefont{X.}~\bibnamefont{{Barcons}}},
  \bibinfo{author}{\bibfnamefont{F.~J.} \bibnamefont{{Carrera}}},
  \bibinfo{author}{\bibfnamefont{L.~C.} \bibnamefont{{Ho}}}, \bibnamefont{and}
  \bibinfo{author}{\bibfnamefont{K.}~\bibnamefont{{Iwasawa}}},
  \bibinfo{journal}{\aap} \textbf{\bibinfo{volume}{455}}, \bibinfo{pages}{173}
  (\bibinfo{year}{2006}), \eprint{astro-ph/0605236}.

\bibitem[{\citenamefont{{Hopkins} et~al.}(2006)\citenamefont{{Hopkins},
  {Narayan}, and {Hernquist}}}]{Hopkins_Plambda}
\bibinfo{author}{\bibfnamefont{P.~F.} \bibnamefont{{Hopkins}}},
  \bibinfo{author}{\bibfnamefont{R.}~\bibnamefont{{Narayan}}},
  \bibnamefont{and}
  \bibinfo{author}{\bibfnamefont{L.}~\bibnamefont{{Hernquist}}},
  \bibinfo{journal}{\apj} \textbf{\bibinfo{volume}{643}}, \bibinfo{pages}{641}
  (\bibinfo{year}{2006}), \eprint{astro-ph/0510369}.

\bibitem[{\citenamefont{{Babi{\'c}} et~al.}(2007)\citenamefont{{Babi{\'c}},
  {Miller}, {Jarvis}, {Turner}, {Alexander}, and {Croom}}}]{Babic_Plambda}
\bibinfo{author}{\bibfnamefont{A.}~\bibnamefont{{Babi{\'c}}}},
  \bibinfo{author}{\bibfnamefont{L.}~\bibnamefont{{Miller}}},
  \bibinfo{author}{\bibfnamefont{M.~J.} \bibnamefont{{Jarvis}}},
  \bibinfo{author}{\bibfnamefont{T.~J.} \bibnamefont{{Turner}}},
  \bibinfo{author}{\bibfnamefont{D.~M.} \bibnamefont{{Alexander}}},
  \bibnamefont{and} \bibinfo{author}{\bibfnamefont{S.~M.}
  \bibnamefont{{Croom}}}, \bibinfo{journal}{\aap}
  \textbf{\bibinfo{volume}{474}}, \bibinfo{pages}{755} (\bibinfo{year}{2007}),
  \eprint{0709.0786}.

\bibitem[{\citenamefont{{Merloni} and {Heinz}}(2008)}]{Merloni_lambda}
\bibinfo{author}{\bibfnamefont{A.}~\bibnamefont{{Merloni}}} \bibnamefont{and}
  \bibinfo{author}{\bibfnamefont{S.}~\bibnamefont{{Heinz}}},
  \bibinfo{journal}{\mnras} \textbf{\bibinfo{volume}{388}},
  \bibinfo{pages}{1011} (\bibinfo{year}{2008}), \eprint{0805.2499}.

\bibitem[{\citenamefont{{Komatsu} and {Kitayama}}(1999)}]{Komatsu99_SZ}
\bibinfo{author}{\bibfnamefont{E.}~\bibnamefont{{Komatsu}}} \bibnamefont{and}
  \bibinfo{author}{\bibfnamefont{T.}~\bibnamefont{{Kitayama}}},
  \bibinfo{journal}{\apjl} \textbf{\bibinfo{volume}{526}}, \bibinfo{pages}{L1}
  (\bibinfo{year}{1999}), \eprint{astro-ph/9908087}.

\bibitem[{\citenamefont{{Cole} and {Kaiser}}(1988)}]{Cole88}
\bibinfo{author}{\bibfnamefont{S.}~\bibnamefont{{Cole}}} \bibnamefont{and}
  \bibinfo{author}{\bibfnamefont{N.}~\bibnamefont{{Kaiser}}},
  \bibinfo{journal}{\mnras} \textbf{\bibinfo{volume}{233}},
  \bibinfo{pages}{637} (\bibinfo{year}{1988}).

\bibitem[{\citenamefont{{Cooray} and {Sheth}}(2002)}]{Cooray_halomodel}
\bibinfo{author}{\bibfnamefont{A.}~\bibnamefont{{Cooray}}} \bibnamefont{and}
  \bibinfo{author}{\bibfnamefont{R.}~\bibnamefont{{Sheth}}},
  \bibinfo{journal}{\physrep} \textbf{\bibinfo{volume}{372}},
  \bibinfo{pages}{1} (\bibinfo{year}{2002}), \eprint{astro-ph/0206508}.

\bibitem[{\citenamefont{{Afshordi} et~al.}(2003)\citenamefont{{Afshordi},
  {McDonald}, and {Spergel}}}]{Afshordi_pbh}
\bibinfo{author}{\bibfnamefont{N.}~\bibnamefont{{Afshordi}}},
  \bibinfo{author}{\bibfnamefont{P.}~\bibnamefont{{McDonald}}},
  \bibnamefont{and} \bibinfo{author}{\bibfnamefont{D.~N.}
  \bibnamefont{{Spergel}}}, \bibinfo{journal}{\apjl}
  \textbf{\bibinfo{volume}{594}}, \bibinfo{pages}{L71} (\bibinfo{year}{2003}),
  \eprint{astro-ph/0302035}.

\bibitem[{\citenamefont{{Kashlinsky}}(2016)}]{Kashilinsky16}
\bibinfo{author}{\bibfnamefont{A.}~\bibnamefont{{Kashlinsky}}},
  \bibinfo{journal}{\apjl} \textbf{\bibinfo{volume}{823}}, \bibinfo{eid}{L25}
  (\bibinfo{year}{2016}), \eprint{1605.04023}.

\bibitem[{\citenamefont{{Bowman} and {Rogers}}(2010)}]{EDGES}
\bibinfo{author}{\bibfnamefont{J.~D.} \bibnamefont{{Bowman}}} \bibnamefont{and}
  \bibinfo{author}{\bibfnamefont{A.~E.~E.} \bibnamefont{{Rogers}}},
  \bibinfo{journal}{\nat} \textbf{\bibinfo{volume}{468}}, \bibinfo{pages}{796}
  (\bibinfo{year}{2010}), \eprint{1209.1117}.

\bibitem[{\citenamefont{{Bernardi} et~al.}(2016)\citenamefont{{Bernardi},
  {Zwart}, {Price}, {Greenhill}, {Mesinger}, {Dowell}, {Eftekhari},
  {Ellingson}, {Kocz}, and {Schinzel}}}]{LEDA}
\bibinfo{author}{\bibfnamefont{G.}~\bibnamefont{{Bernardi}}},
  \bibinfo{author}{\bibfnamefont{J.~T.~L.} \bibnamefont{{Zwart}}},
  \bibinfo{author}{\bibfnamefont{D.}~\bibnamefont{{Price}}},
  \bibinfo{author}{\bibfnamefont{L.~J.} \bibnamefont{{Greenhill}}},
  \bibinfo{author}{\bibfnamefont{A.}~\bibnamefont{{Mesinger}}},
  \bibinfo{author}{\bibfnamefont{J.}~\bibnamefont{{Dowell}}},
  \bibinfo{author}{\bibfnamefont{T.}~\bibnamefont{{Eftekhari}}},
  \bibinfo{author}{\bibfnamefont{S.~W.} \bibnamefont{{Ellingson}}},
  \bibinfo{author}{\bibfnamefont{J.}~\bibnamefont{{Kocz}}}, \bibnamefont{and}
  \bibinfo{author}{\bibfnamefont{F.}~\bibnamefont{{Schinzel}}},
  \bibinfo{journal}{\mnras} \textbf{\bibinfo{volume}{461}},
  \bibinfo{pages}{2847} (\bibinfo{year}{2016}), \eprint{1606.06006}.

\bibitem[{\citenamefont{{Singh} et~al.}(2017)\citenamefont{{Singh},
  {Subrahmanyan}, {Udaya Shankar}, {Sathyanarayana Rao}, {Fialkov}, {Cohen},
  {Barkana}, {Girish}, {Raghunathan}, {Somashekar} et~al.}}]{SARAS}
\bibinfo{author}{\bibfnamefont{S.}~\bibnamefont{{Singh}}},
  \bibinfo{author}{\bibfnamefont{R.}~\bibnamefont{{Subrahmanyan}}},
  \bibinfo{author}{\bibfnamefont{N.}~\bibnamefont{{Udaya Shankar}}},
  \bibinfo{author}{\bibfnamefont{M.}~\bibnamefont{{Sathyanarayana Rao}}},
  \bibinfo{author}{\bibfnamefont{A.}~\bibnamefont{{Fialkov}}},
  \bibinfo{author}{\bibfnamefont{A.}~\bibnamefont{{Cohen}}},
  \bibinfo{author}{\bibfnamefont{R.}~\bibnamefont{{Barkana}}},
  \bibinfo{author}{\bibfnamefont{B.~S.} \bibnamefont{{Girish}}},
  \bibinfo{author}{\bibfnamefont{A.}~\bibnamefont{{Raghunathan}}},
  \bibinfo{author}{\bibfnamefont{R.}~\bibnamefont{{Somashekar}}},
  \bibnamefont{et~al.}, \bibinfo{journal}{ArXiv e-prints}
  (\bibinfo{year}{2017}), \eprint{1711.11281}.

\bibitem[{\citenamefont{{Jester} and {Falcke}}(2009)}]{Jester_darkmoon}
\bibinfo{author}{\bibfnamefont{S.}~\bibnamefont{{Jester}}} \bibnamefont{and}
  \bibinfo{author}{\bibfnamefont{H.}~\bibnamefont{{Falcke}}},
  \bibinfo{journal}{New Astron. Rev.} \textbf{\bibinfo{volume}{53}},
  \bibinfo{pages}{1} (\bibinfo{year}{2009}), \eprint{0902.0493}.

\bibitem[{\citenamefont{{Burns} et~al.}(2012)\citenamefont{{Burns}, {Lazio},
  and {Bottke}}}]{Burns_darkmoon}
\bibinfo{author}{\bibfnamefont{J.~O.} \bibnamefont{{Burns}}},
  \bibinfo{author}{\bibfnamefont{T.~J.~W.} \bibnamefont{{Lazio}}},
  \bibnamefont{and} \bibinfo{author}{\bibfnamefont{W.}~\bibnamefont{{Bottke}}},
  \bibinfo{journal}{ArXiv e-prints}  (\bibinfo{year}{2012}),
  \eprint{1209.2233}.

\bibitem[{\citenamefont{{Visbal} et~al.}(2012)\citenamefont{{Visbal},
  {Barkana}, {Fialkov}, {Tseliakhovich}, and {Hirata}}}]{Visbal_streaming}
\bibinfo{author}{\bibfnamefont{E.}~\bibnamefont{{Visbal}}},
  \bibinfo{author}{\bibfnamefont{R.}~\bibnamefont{{Barkana}}},
  \bibinfo{author}{\bibfnamefont{A.}~\bibnamefont{{Fialkov}}},
  \bibinfo{author}{\bibfnamefont{D.}~\bibnamefont{{Tseliakhovich}}},
  \bibnamefont{and} \bibinfo{author}{\bibfnamefont{C.~M.}
  \bibnamefont{{Hirata}}}, \bibinfo{journal}{\nat}
  \textbf{\bibinfo{volume}{487}}, \bibinfo{pages}{70} (\bibinfo{year}{2012}),
  \eprint{1201.1005}.

\bibitem[{\citenamefont{{Natarajan} et~al.}(2017)\citenamefont{{Natarajan},
  {Pacucci}, {Ferrara}, {Agarwal}, {Ricarte}, {Zackrisson}, and
  {Cappelluti}}}]{Natarajan_dcbhJWST}
\bibinfo{author}{\bibfnamefont{P.}~\bibnamefont{{Natarajan}}},
  \bibinfo{author}{\bibfnamefont{F.}~\bibnamefont{{Pacucci}}},
  \bibinfo{author}{\bibfnamefont{A.}~\bibnamefont{{Ferrara}}},
  \bibinfo{author}{\bibfnamefont{B.}~\bibnamefont{{Agarwal}}},
  \bibinfo{author}{\bibfnamefont{A.}~\bibnamefont{{Ricarte}}},
  \bibinfo{author}{\bibfnamefont{E.}~\bibnamefont{{Zackrisson}}},
  \bibnamefont{and}
  \bibinfo{author}{\bibfnamefont{N.}~\bibnamefont{{Cappelluti}}},
  \bibinfo{journal}{\apj} \textbf{\bibinfo{volume}{838}}, \bibinfo{eid}{117}
  (\bibinfo{year}{2017}), \eprint{1610.05312}.

\end{thebibliography}

\appendix
\section{Dependence on the flux}\label{sec:AppendixA}
In this appendix we discuss the dependence of the final $T_{21}$ radial profile and power spectrum on the emitted spectrum assumed. We referred to the spectrum used in the main text (\refeq{spectrum}) as Power-Law (PL) in opposition to a Power-Law with Low Energies (PL LE) in which the energy range is extended at the low energy limit ($10.4\; {\rm eV}~\leq E~\leq 100 \;{\rm keV}$) and a Power-Law with High Energies (PL HE) in which the energy range is extended at the high energy limit ($200 \;{\rm eV}~\leq E~\leq 300 \;{\rm keV}$). All of them have the same exponent: $-1$.

\begin{figure}[t!]
\centering
\includegraphics[width=\columnwidth]{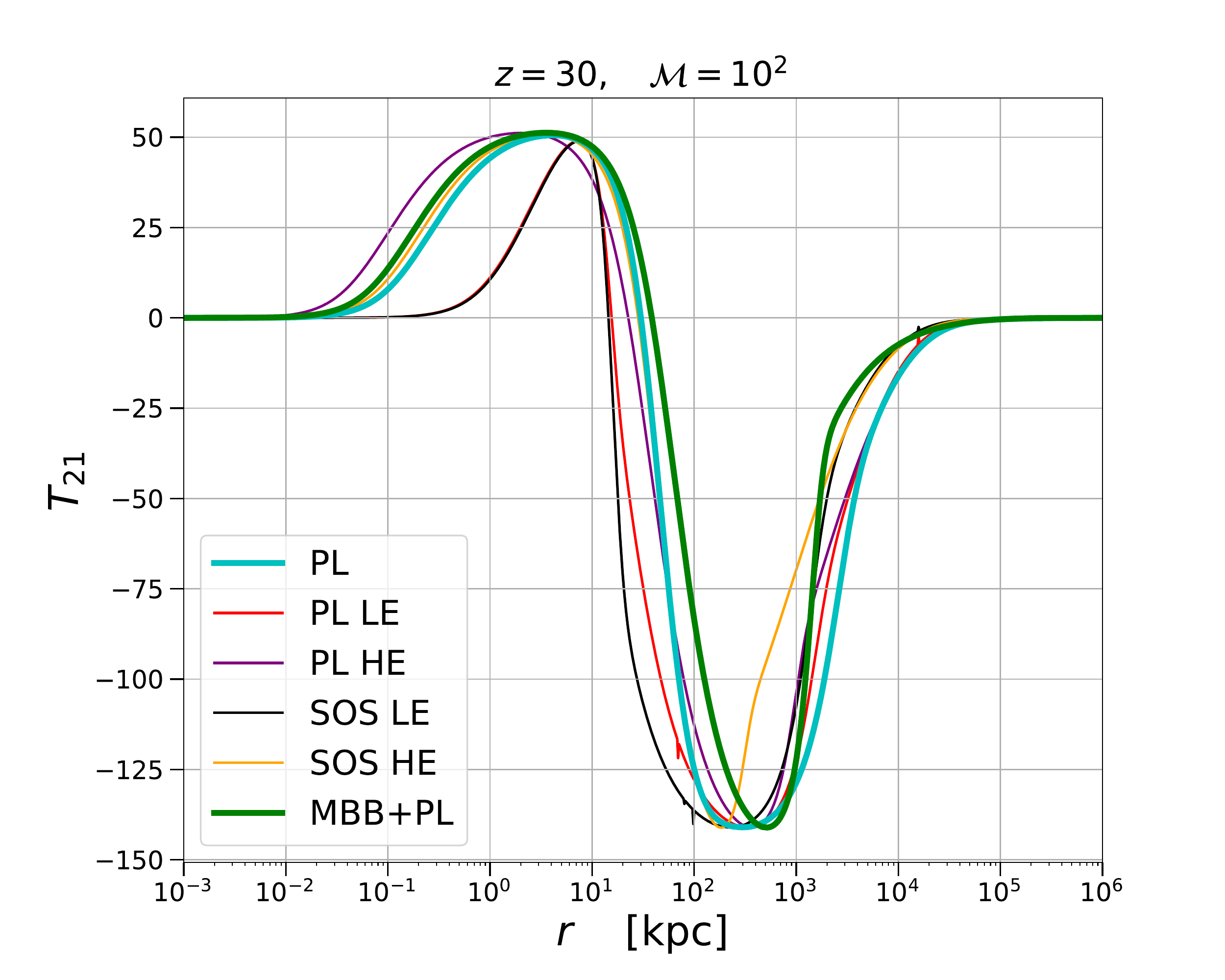}
\caption{\label{fig:T21_spectra}
Differential brightness temperature profile $T_{21}$ for a PBH with $\mathcal{M}=100$ at $z=30$ for different spectra for the emitted radiation.}
\end{figure}

\begin{figure}[t!]
\centering
\includegraphics[width=\columnwidth]{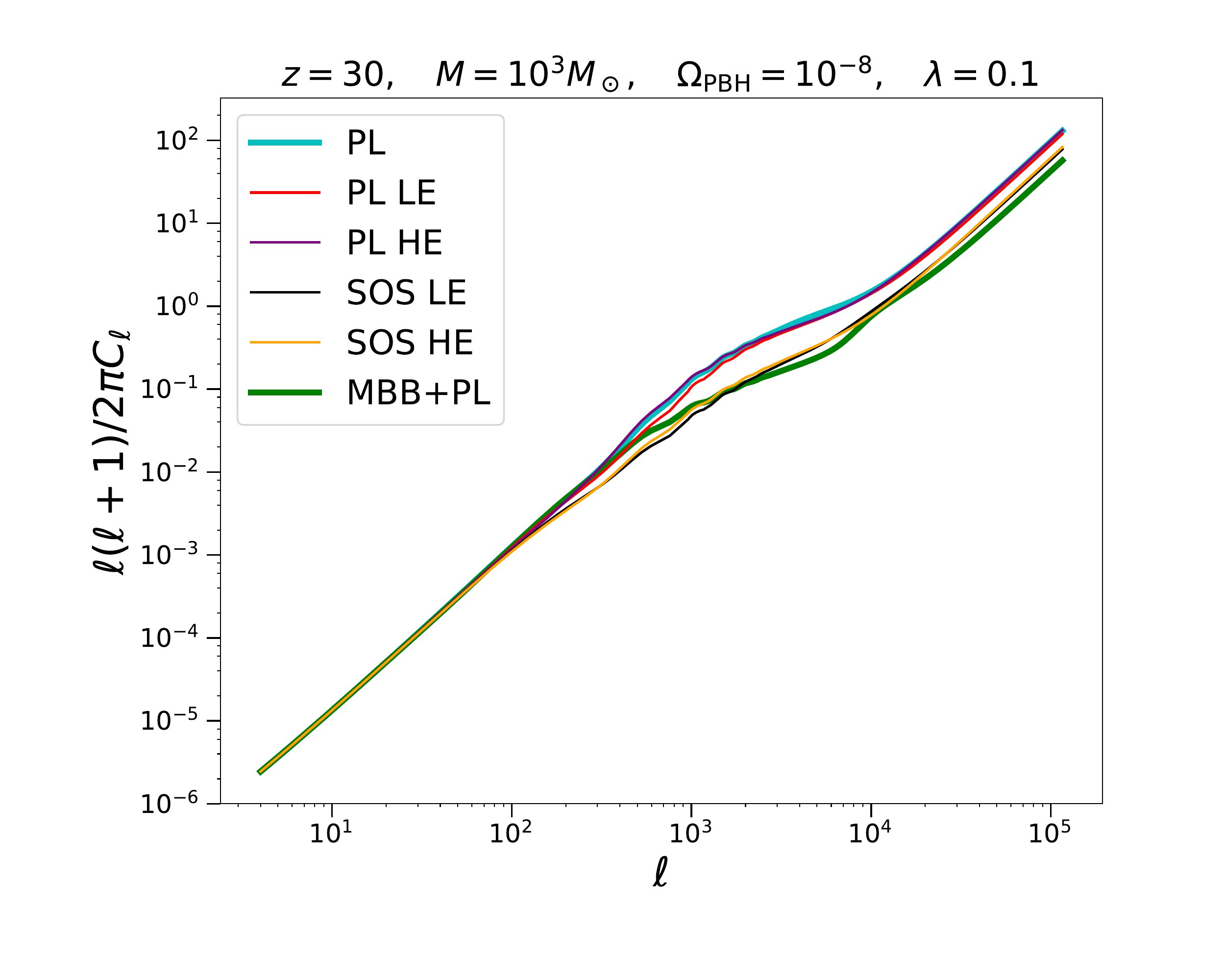}
\caption{\label{fig:Cells_spectra}
Angular power spectrum comparing the total signal in 21 cm IM  for $M=10^3\Msun$, $\Omega_{\rm PBH} = 10^{-8}$, $\lambda=0.1$, $z=30$ and $\Delta\nu = 1$ MHz for different spectra for the emitted radiation.}
\end{figure}

We also consider a more elaborated spectrum, as the one introduced by Sazonov, Ostriker~\& Sunyaev (2004)~\cite{Sazonov_spectrum}:
\begin{equation}
F(E) =~\mathcal{A}(\mathcal{M})\left\lbrace
\begin{aligned}
&E^{-1.7}, & 10.4\, {\rm eV} < E < 1\,{\rm keV},	\\
&E^{-1}, & 1~\,{\rm keV} < E < 100\,{\rm keV},~\\
&E^{-1.6}, & 100\, {\rm keV} < E,
\end{aligned}\right.
\label{eq:SOS_spectrum}
\end{equation}
with $\mathcal{A}(\mathcal{M})$ computed as in~\refeq{A_Ml}. We consider one case with a high energy cut of 100 keV (SOS LE) and another with the cut at 300 keV (SOS HE).
 Finally, we consider a more realistic spectrum which includes the contribution of the disk as a multicolor black body spectrum, added to a power-law with  index $-1$ for energies larger than $\sim 3T_{\rm max}$ (where $k_BT_{\rm max}=\left(M/\Msun  \right)^{-0.25}$ KeV) and with a high energy exponential cut off at 300 KeV, modelling the emission of the hot corona.  We follow \cite{Yue_dcbhNIRB} and normalize each contribution to the total emission to have the same luminosity.
 The emission from the disk can be expressed as:
\begin{equation}
F_{MBB}(E) = \mathcal{A}_{\rm MBB}(\mathcal{M})\int_0^{T_{\rm max}}B(E,T)\left(\frac{T}{T_{\rm max}}\right)^{-11/3}\frac{dT}{T_{\rm max}} 
\end{equation}
Again, we set a low energy limit at 10.4 eV.

We show the resulting $T_{21}(r)$ and $C_\ell$ for all the emission models explained above  in~\reffig{T21_spectra} and~\reffig{Cells_spectra}, respectively.  Although the radial profiles of $T_{21}$ are different, the effect on the final power spectrum is small, compared with the uncertainties in the PBH parameters (i.e., $\Omega_{\rm PBH}$, $M$ and $n_{\rm PBH}$). Moreover, as the dependence on the PBH parameters is the same for all the different emission spectra, significant changes on the forecasts reported on~\reftab{forecast} or on the two dimensional confidence levels shown in~\reffig{Dchi2} are not expected.

\end{document}